\documentclass[%
aps,
prl,
reprint,
groupedaddress,
]{revtex4-2}

\usepackage{lipsum}
\usepackage{amsmath}
\usepackage{cleveref}
\usepackage{graphicx}
\usepackage{xcolor}
\usepackage{siunitx}

\begin{document}


\title{Unresolved-Sideband Optomechanics with Hexagonal Boron Nitride:\\ Induced Transparency, Gain, and Frequency Combs}


\author{Francesco Fogliano}
\affiliation{Physics Department, University of Basel}
\author{Thibaud Ruelle}
\affiliation{Physics Department, University of Basel}
\author{David Jaeger}
\affiliation{Physics Department, University of Basel}
\author{Martino Poggio}
\email[]{martino.poggio@unibas.ch}
\affiliation{Physics Department, University of Basel}


\date{\today}

\begin{abstract}
Optomechanically induced transparency (OMIT) is usually modeled and studied in the resolved-sideband regime, but many compact microcavity platforms operate in the unresolved-sideband limit $(\kappa \gg \Omega_m)$. Here we investigate OMIT in this regime using a tunable fiber-based Fabry--Perot microcavity coupled to a suspended hexagonal boron nitride (hBN) drum resonator in a membrane-in-the-middle geometry. The system achieves a large single-photon coupling rate of $g_0/2\pi \sim \qty{180}{\kilo\hertz}$ and exhibits strong radiation-pressure backaction. By measuring OMIT spectra as a function of pump power and cavity detuning, we observe a crossover from a transparency-like dip to a gain feature in the reflected response. These maps are quantitatively reproduced by the full linearized optomechanical response, demonstrating the breakdown of the standard rotating-wave approximation used in the resolved-sideband limit. Finally, we drive the system into a nonlinear regime to generate optomechanical frequency combs. These results establish hBN fiber-cavities as a versatile architecture for unresolved-sideband optomechanics, nonlinear dynamics, and hybrid device integration.
\end{abstract}


\maketitle

\begin{figure*}[!t]
	\includegraphics[width=1\linewidth]{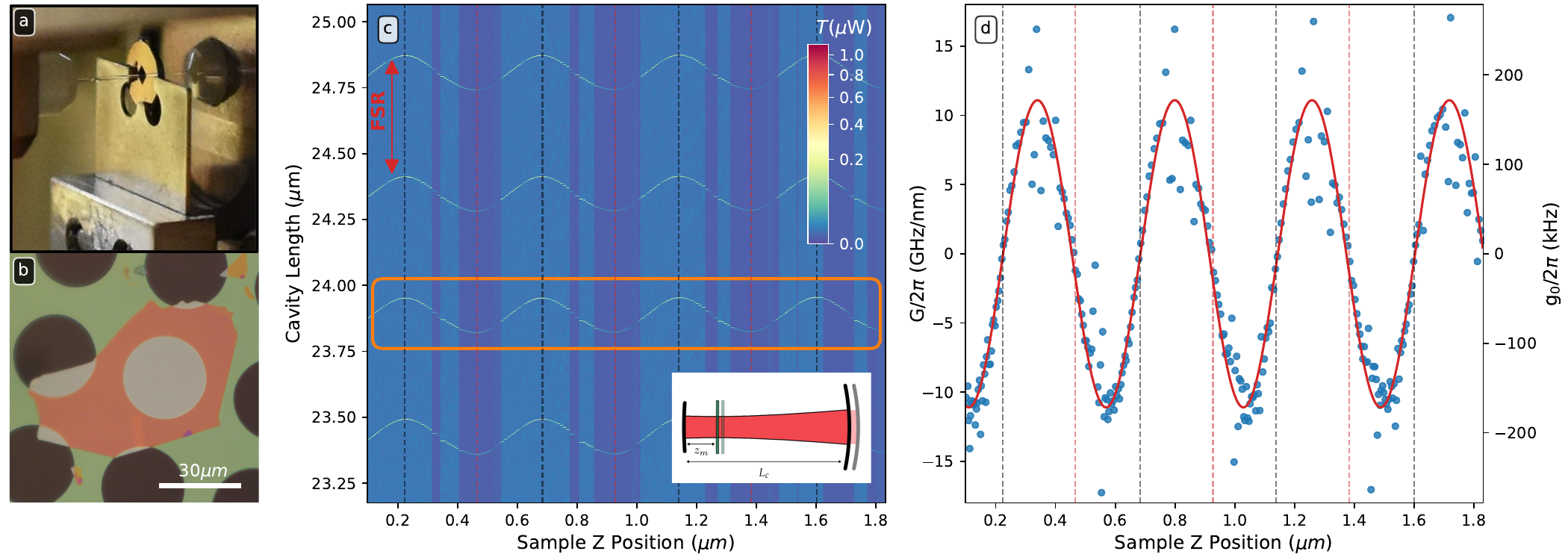}
	\caption{Membrane-in-the-middle fiber cavity and static optomechanical coupling. (a) Photograph of the assembled fiber-cavity and sample-frame system. (b) Optical image of the suspended hBN drum used in the measurements. (c) Cavity-frequency shift of several optical modes as a function of membrane position along the cavity axis. (d) Corresponding dispersive frequency pull $G$ and equivalent single-photon coupling rate $g_0$; blue points: extracted data; red curve: derivative of a sinusoidal fit. Black (red) dashed lines: cavity nodes (antinodes).}
	\label{fig:Fig1_sys_overview}
\end{figure*}

Cavity optomechanics provides a versatile framework to control and read out mechanical motion through radiation pressure, enabling coherent interactions between optical and mechanical degrees of freedom in systems ranging from macroscopic Fabry--Perot cavities to chip-scale nanophotonic resonators \cite{aspelmeyerCavityOptomechanics2014,bodiyaSubhertzOptomechanicallyInduced2019,thompsonStrongDispersiveCoupling2008,safavi-naeiniTwoDimensionalPhononicPhotonicBand2014}. A central interference effect in this context is optomechanically induced transparency (OMIT), which arises when a strong control tone and a weak probe tone simultaneously drive the cavity: the beating between the two tones excites the mechanical resonator, and the sidebands generated by this motion interfere with the intracavity probe field, producing a narrow modification of the probe response around the mechanical resonance \cite{agarwalElectromagneticallyInducedTransparency2010,weisOptomechanicallyInducedTransparency2010,xiongFundamentalsApplicationsOptomechanically2018}. Beyond its analogy to atomic electromagnetically induced transparency, OMIT has become an important tool for probing coherent optomechanical coupling and radiation-pressure backaction, and for applications including optical group-delay control (slow and fast light), wavelength conversion, and multimode interference control \cite{xiongFundamentalsApplicationsOptomechanically2018,fanCascadedOpticalTransparency2015,lakeTwocolourInterferometrySwitching2020,laiTunableOptomechanicallyInduced2020,yanOptomechanicallyInducedUltraslow2021}.

A particularly attractive direction for extending optomechanics is the use of low-dimensional mechanical resonators, where small effective mass can strongly enhance the single-photon interaction. Among these, atomically thin and layered materials offer a particularly appealing combination of low mass, high mechanical compliance, and compatibility with compact optical architectures \cite{falinMechanicalPropertiesAtomically2017,tavernarakisOptomechanicsHybridCarbon2018,singhOptomechanicalCouplingMultilayer2014}. Hexagonal boron nitride (hBN) is especially attractive here: suspended hBN drums exhibit low mechanical dissipation and wide-bandgap optical transparency compatible with high-finesse cavities, while optically active hBN defects offer a route to hybrid emitter--photon--phonon devices \cite{wangCavityQuantumElectrodynamics2021,qianEmitterOptomechanicalInteraction2025}. Coupling such resonators to open-access fiber microcavities is therefore a promising route toward strong optomechanical interactions in a membrane-in-the-middle (MIM) geometry \cite{thompsonStrongDispersiveCoupling2008,sankeyStrongTunableNonlinear2010,karuzaOptomechanicallyInducedTransparency2013,steinmetzStableFiberbasedFabryPerot2006,hungerFiberFabryPerot2010,ruelleTunableFiberFabry2022,jaegerMechanicalModeImaging2023}. In this context, the reconfigurable cavity architecture of \cite{ruelleTunableFiberFabry2022} provides a practical way to combine suspended 2D-material resonators with small optical mode volumes. Recent fiber-cavity experiments have demonstrated radiation-pressure backaction in SiN- and hBN-based resonators \cite{rochauDynamicalBackactionUltrahighFinesse2021,sanchezarribasRadiationPressureBackaction2023}, while modern membrane-in-the-middle platforms continue to pursue room-temperature quantum control and collective photon-mediated dynamics \cite{saarinenLaserCoolingMembraneinthemiddle2023,huangRoomtemperatureQuantumOptomechanics2024,yaoLongRangeOptomechanicalInteractions2025}.

The use of compact microcavity devices also changes which optomechanical regime is naturally relevant. Most standard OMIT experiments and their interpretation operate in or near the resolved-sideband regime, where the cavity filters the mechanical sidebands strongly enough that a rotating-wave approximation (RWA) is accurate \cite{weisOptomechanicallyInducedTransparency2010}. By contrast, many compact cavities and microdevices naturally tend toward the unresolved-sideband regime (USR), where the optical linewidth remains large compared with the mechanical frequency \cite{monselDissipativeDispersiveCavity2024,fitzgeraldCavityOptomechanicsPhotonic2021a}. In this regime, both Stokes and anti-Stokes processes remain inside the cavity linewidth, and the usual RWA description fails. Theoretical work predicts that retaining the full linearized response leads to a modified OMIT lineshape, including a crossover from a conventional transparency-like dip to a probe-power enhancement feature \cite{ojanenGroundstateCoolingMechanical2014,yanOptomechanicallyInducedTransparency2020,yanOptomechanicallyInducedOptical2021}. Although the USR is often viewed as less favorable than the resolved-sideband regime for standard quantum optomechanics, it is directly relevant to compact microcavity platforms and has attracted renewed interest in connection with engineered cooling, state transfer, and hybrid or auxiliary-nonlinearity-assisted schemes \cite{lauGroundStateCoolingHighFidelity2020,navarathnaContinuousOpticaltoMechanicalQuantum2023,zoepflKerrEnhancedBackaction2023,monselDissipativeDispersiveCavity2024,kumarOptomechanicallyInducedTransparency2024,sarmaGroundstateCoolingMicromechanical2016,zhangStrongMechanicalSqueezing2019,chuSimultaneousCoolingDegenerate2024,diaz-naufalKerrenhancedOptomechanicalCooling2025}.

In this work we combine a tunable open-access fiber Fabry--Perot microcavity with a suspended hBN drum resonator and realize an optomechanical MIM platform with large single-photon coupling. We first characterize the static dispersive coupling and benchmark the device through radiation-pressure dynamical backaction. We then investigate OMIT deep in the unresolved-sideband regime and resolve the predicted power-driven transition from a transparency-like dip to a probe gain feature in the normalized reflected-probe response, in quantitative agreement with the full linearized USR model. Finally, we show that the same system can also be driven into a nonlinear regime of optomechanical frequency-comb generation \cite{miriOptomechanicalFrequencyCombs2018,guOpticalFrequencyComb2024}. Together, these results establish hBN fiber-cavity devices as a high-finesse, strong-coupling 2D-material platform for exploring optomechanical interference and nonlinear dynamics beyond the resolved-sideband approximation.

\begin{figure*}[!t]
	\includegraphics[width=1.0\linewidth]{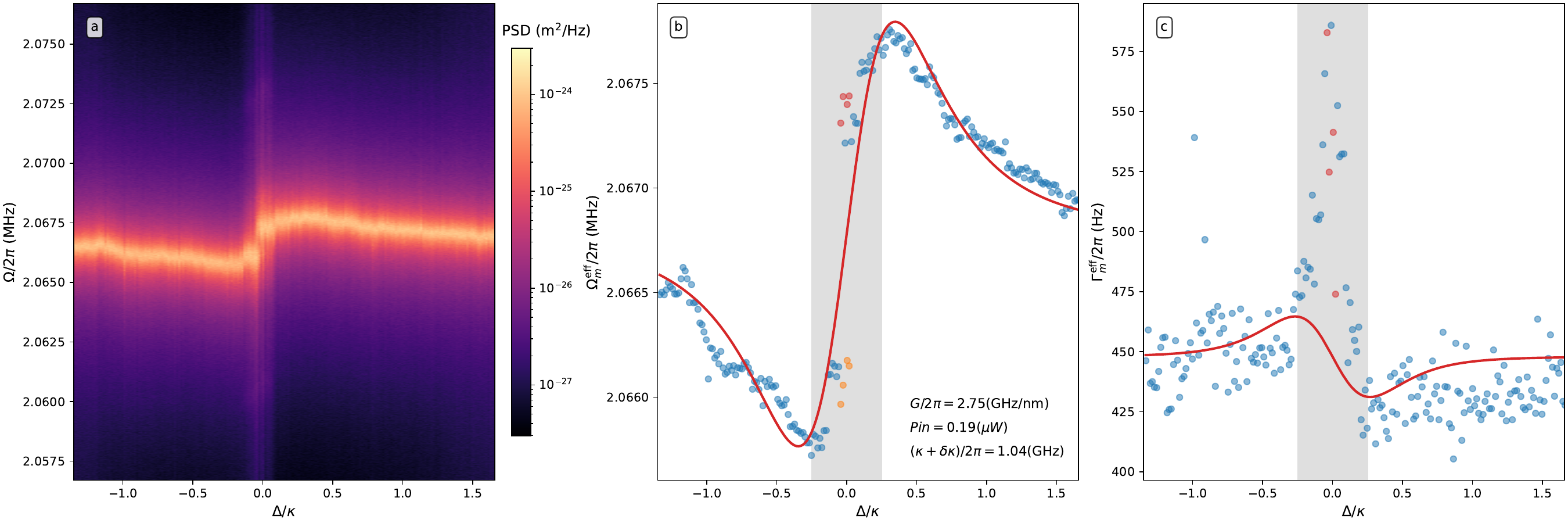}
	\caption{Radiation-pressure dynamical backaction of the hBN mode. (a) Calibrated displacement power spectral density versus normalized pump detuning $\Delta/\kappa$ and mechanical frequency. (b) Extracted effective resonance frequency $\Omega_m^{\mathrm{eff}}/2\pi$ versus detuning. Blue points correspond to single-Lorentzian fits, while orange and red points indicate the lower and upper branches from a two-Lorentzian fit. The red curve is the optical-spring fit. (c) Extracted apparent effective linewidth $\Gamma_m^{\mathrm{eff}}/2\pi$ together with the damping prediction computed from the frequency-fit parameters. The gray band marks the unstable near-resonant region.}
	\label{fig:Fig3_Backaction}
\end{figure*}

\subparagraph{Platform and static coupling}
Our experiment combines a tunable open-access fiber Fabry--Perot microcavity with a suspended hBN drum resonator in a MIM geometry, as summarized in \cref{fig:Fig1_sys_overview}. The platform builds on the cryo-compatible, widely tunable fiber-cavity architecture of \cite{ruelleTunableFiberFabry2022} and on hBN drum resonators fabricated on a hybrid hBN/SiN platform \cite{jaegerMechanicalModeImaging2023}. At the operating wavelength near \qty{920}{\nano\meter} and cavity length $L_{\mathrm{cav}}\sim\qty{24}{\micro\meter}$, the cavity has a finesse of order $10^4$ and a total linewidth $\kappa/2\pi \sim \qty{1.0}{\giga\hertz}$. In contrast to many fixed microcavity implementations, the present system allows independent positioning of the cavity and membrane, exchange of samples, and tuning over many free spectral ranges while maintaining a small optical mode volume ($\sim 250\,\lambda^3$). The reflected signal is detected on a common photodiode used both for cavity locking and the OMIT readout; detailed tone layout and calibration procedures are given in the End Matter and SI.

\Cref{fig:Fig1_sys_overview}(a) shows the assembled cavity and sample frame, while \cref{fig:Fig1_sys_overview}(b) shows the hBN drum used in the measurements, with diameter $\qty{30}{\um}$ and thickness $\qty{48}{\nm}$. The selected mode has mechanical resonance frequency $\Omega_m/2\pi \sim \qty{2}{\mega\hertz}$, effective mass $M_{\mathrm{eff}}=\qty{17.1}{\pico\gram}$, and room-temperature quality factor of order $5\times 10^3$. The small mass of the suspended hBN resonator strongly enhances the achievable single-photon coupling.

To characterize the static optomechanical interaction, we scan the sample position $z$ along the intracavity standing wave while sweeping the cavity length through several optical resonances. The resulting shift of the tracked cavity resonance frequency $\omega_{\mathrm{cav}}(z)$ is shown in \cref{fig:Fig1_sys_overview}(c), and the corresponding dispersive frequency-pull parameter, $G(z) = -\frac{\partial \omega_{\mathrm{cav}}}{\partial z}$, is shown in \cref{fig:Fig1_sys_overview}(d) together with the equivalent single-photon coupling rate $g_0(z)=G(z)x_{\mathrm{zpf}}$, with $x_{\mathrm{zpf}}=\sqrt{\frac{\hbar}{2M_{\mathrm{eff}}\Omega_m}}$ the resonator zero-point motion. The calibration and coordinate convention used to express these data in the standard symmetric membrane-in-the-middle picture are described in the End Matter and in more detail in the SI. The extracted maximum reaches $g_0/2\pi \sim \qty{180}{\kilo\hertz}$ for the selected hBN mode and cavity configuration.

This value is unusually large for an open-cavity MIM platform, well above the Hz or low kHz-scale couplings reported in related cavity-membrane systems and approaching the reported values of integrated optomechanical-crystal benchmarks \cite{rochauDynamicalBackactionUltrahighFinesse2021,sanchezarribasRadiationPressureBackaction2023,tenbrakeDirectLaserwrittenOptomechanical2024,safavi-naeiniTwoDimensionalPhononicPhotonicBand2014,marzioniAmplitudePhaseNoise2023,mayorHighPhotonphononPair2025a,guoIntegratedOpticalreadoutHighQ2022a,chenLownoiseOptomechanicalSingle2026,planzMembraneinthemiddleOptomechanicsSoftclamped2023,fedoseevStimulatedRamanAdiabatic2021}. At the same time, the cavity linewidth increases strongly away from the standing-wave nodes because of enhanced scattering, clipping, and local coupling to higher-order optical modes. For this reason, the dynamical measurements reported below are performed close to a cavity node, where the local dispersive coupling is reduced to about $G/2\pi \sim \qty{1}{\giga\hertz\per\nano\meter}$, corresponding to $g_0/2\pi \sim \qty{16}{\kilo\hertz}$ for the selected mode, but the cavity remains stable and the effective linewidth broadening is strongly reduced. This static characterization therefore sets both the accessible coupling and the practical operating window for the backaction, OMIT, and nonlinear measurements discussed below.

\begin{figure*}[!t]
	\includegraphics[width=1\linewidth]{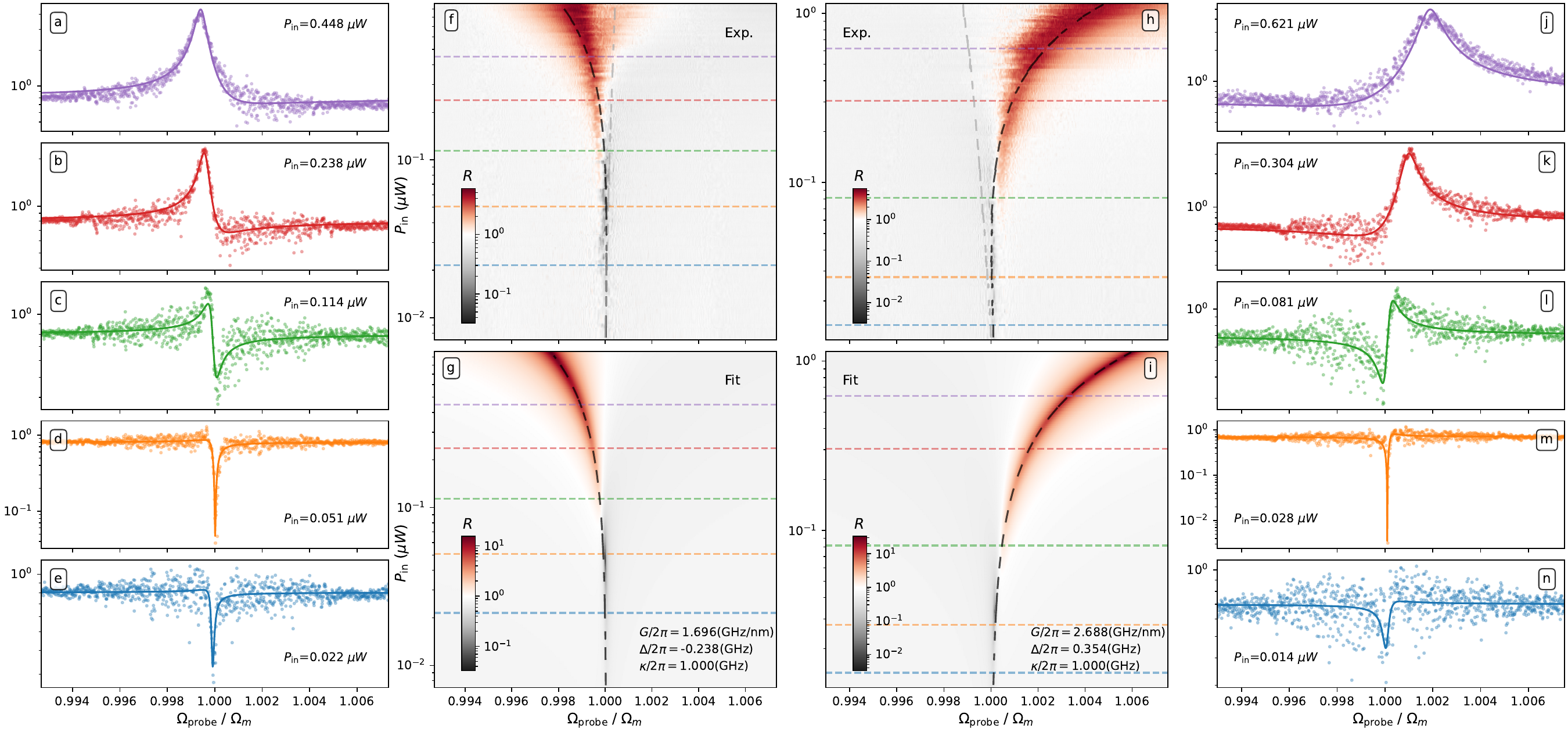}
	\caption{OMIT response versus input power at fixed cavity detuning. (a--e) Red-detuned line cuts; (f,g) corresponding measured and model maps. (h,i) Blue-detuned measured and model maps; (j--n) corresponding line cuts. Maps show $R$ versus probe offset $\Omega_{\mathrm{probe}}/\Omega_m$ and input power $P_{\mathrm{in}}$; colored dashed lines mark the cut powers. Points are data and solid lines are the model. At low power, the response shows a narrow OMIT dip near $\Omega_{\mathrm{probe}}/\Omega_m \simeq 1$; at higher power it evolves into a feature with $R>1$. The lower panels list the fitted effective local coupling and detuning; the cavity linewidth is fixed to $\kappa/2\pi=\qty{1.0}{\giga\hertz}$.}
	\label{fig:Fig4_OMIT_vs_Power}
\end{figure*}

\subparagraph{OM Backaction}
Radiation pressure modifies the effective mechanical susceptibility, producing dynamical backaction in the form of an optical spring and optomechanical damping \cite{aspelmeyerCavityOptomechanics2014}. Here we denote by $\omega_l$ the optical frequency of the pump tone driving the cavity and define its detuning as $\Delta=\omega_l-\omega_{\mathrm{cav}}$. We quantify the backaction by sweeping $\Delta$ across the cavity resonance at fixed input power and recording the thermal displacement spectrum of the selected hBN mechanical mode from the transmitted-light readout.

\Cref{fig:Fig3_Backaction}(a) displays the calibrated displacement power spectral density as a function of normalized detuning $\Delta/\kappa$ and mechanical frequency. For each detuning, the thermomechanical resonance is fitted with either a single Lorentzian or, when a clearly resolved split structure appears, a sum of two Lorentzians. The extracted effective resonance frequency and apparent linewidth are shown in \cref{fig:Fig3_Backaction}(b,c). The red curve in panel (b) is a fit to the standard linearized backaction expression for the optical spring, from which we infer a local coupling $G/2\pi \simeq \qty{2.75}{\giga\hertz\per\nano\meter}$ for an input power of $P_{\mathrm{in}}\simeq \qty{0.19}{\micro\watt}$, with an effective linewidth $(\kappa+\delta\kappa)/2\pi \simeq \qty{1.05}{\giga\hertz}$, where $\delta\kappa$ is the optical broadening induced by the drum thermal motion \cite{leijssenNonlinearCavityOptomechanics2017}. The curve in panel (c) is the corresponding damping computed from the same fitted parameters.

The optical-spring shift is well reproduced over most of the detuning range and provides an independent dynamical benchmark of the strong radiation-pressure coupling inferred from the static scan. The linewidth data follow the expected trend away from cavity resonance, but close to resonance they are broadened and occasionally split by lock instability and thermally driven detuning fluctuations. We therefore use the optical spring as the robust quantitative observable and interpret the near-resonant linewidth only qualitatively.

\subparagraph{OMIT in the USR}
We describe the OMIT response of our MIM fiber cavity in the USR, $\kappa \gg \Omega_m$, where both Stokes and anti-Stokes scattering must be retained. In a frame rotating at the optical carrier frequency $\omega_l$, the linearized dynamics of a driven cavity mode ($\hat{a}$) dispersively coupled to a mechanical mode ($\hat{b}$) follow from the standard radiation-pressure
Hamiltonian \cite{aspelmeyerCavityOptomechanics2014}. The static radiation-pressure displacement $\bar{q}$ shifts the detuning to $\bar{\Delta}=\Delta+G\bar{q}$, with $G=-\partial\omega_{\mathrm{cav}}/\partial q$. The mechanical response is described by the susceptibility
\begin{equation}
	\chi_m(\Omega)=\frac{1}{M_{\mathrm{eff}} \left(\Omega_m^2-\Omega^2-i\Gamma_m\Omega\right)} ,
\end{equation}
where $M_{\mathrm{eff}}$ and $\Gamma_m$ are the effective mass and mechanical damping rate. 

A strong pump tone establishes a coherent intracavity amplitude $\bar{a}$, taken real, giving the field-enhanced coupling $g=g_0\sqrt{\bar{n}_{\mathrm{cav}}}$ with $\bar{n}_{\mathrm{cav}}=|\bar{a}|^2$, and the parameter $\alpha=\hbar G^2|\bar{a}|^2$. We probe the system with a weak sideband at frequency offset $\Omega$ from the pump and solve the linearized Langevin equations without applying the RWA. The full derivation is given in the SI. With $\kappa_{\mathrm{ext}}$ the external coupling rate, the intracavity probe component is
\begin{equation}
	\label{eq:Aminus3}
	A^-(\Omega)	=	\frac{\sqrt{\kappa_{\mathrm{ext}}} a_p}
	{\left[ \frac{\kappa}{2} - i(\bar{\Delta}+\Omega) \right] 
		+ \frac{\alpha}{\frac{i}{\chi_m} - \frac{\alpha}{\left[ \frac{\kappa}{2} + i(\bar{\Delta}-\Omega) \right]}} }
\end{equation}
where $a_p$ is the probe photon-flux amplitude.

\Cref{eq:Aminus3} makes explicit how the usual expression in the RWA is recovered by neglecting the counter-rotating Stokes pathway. In the full USR result, the mechanical response is dressed by the additional term $\alpha/[\frac{\kappa}{2}+i(\bar{\Delta}-\Omega)]$, which vanishes only when the cavity filters the Stokes sideband ($|\bar{\Delta}-\Omega|\gg\kappa/2$). The reflected probe field follows from the input--output relation $a^R_{\rm out}=a_{\rm in}-\sqrt{\kappa_{\rm ext}} a$, where $a_{\rm in}$ is the incident field.

\begin{figure*}[!t]
	\includegraphics[width=1\linewidth]{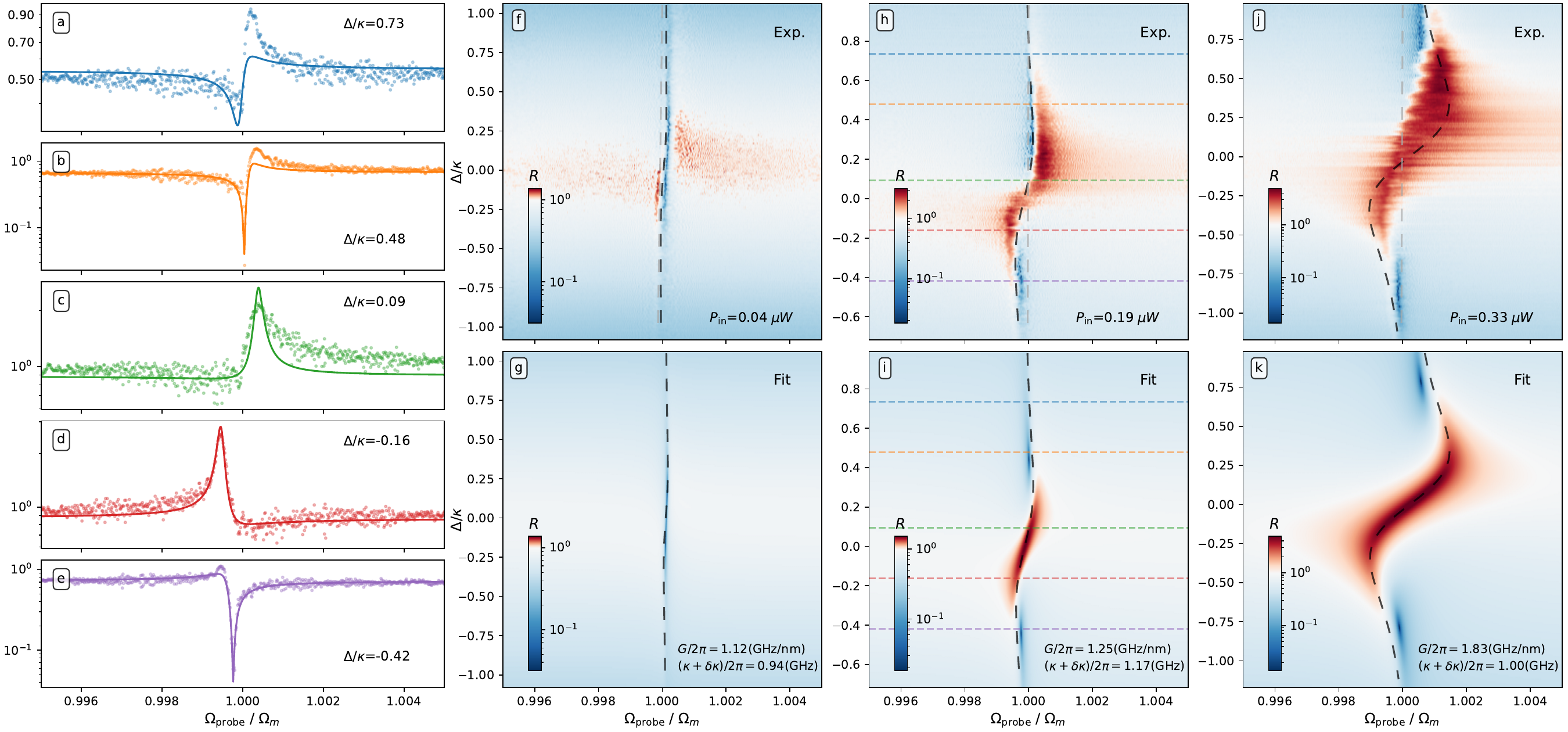}
	\caption{OMIT response versus cavity detuning at fixed input power. (a--e) Representative line cuts at the detunings marked in (h,i); points are data and solid lines are the model. (f,g), (h,i), and (j,k) Measured and model responses for three input powers, plotted versus probe offset $\Omega_{\mathrm{probe}}/\Omega_m$ and normalized pump detuning $\Delta/\kappa$. At low power, the response is dominated by a narrow OMIT dip near $\Omega_{\mathrm{probe}}/\Omega_m\simeq1$; at higher power a region with $R>1$ develops. The dip/peak position shifts consistently with the optical spring. Lower panels give the fitted effective local coupling and effective cavity linewidth. Black dashed lines indicate the optical-spring shift; light-gray dashed lines in the experimental panels show the measured slow drift of the mechanical resonance.}
	\label{fig:Fig5_OMIT_vs_Detuning}
\end{figure*}

\subparagraph{OMIT and Gain}
We now turn to the OMIT response of the hBN optomechanical system in the USR, where the usual RWA fails and counter-rotating scattering contributions modify the probe response both qualitatively and quantitatively \cite{yanOptomechanicallyInducedTransparency2020,yanOptomechanicallyInducedOptical2021}. The measured observable is
the normalized demodulated probe response $R(\Omega)=\frac{|Z(\Omega)|}{|Z_{\mathrm{off}}|}$, where $\Omega=\omega_p-\omega_l$ is the probe offset from the pump. The response is directly proportional to the full intracavity probe amplitude $|A^-(\Omega)|$ derived in \cref{eq:Aminus3} (see End Matter and SI). With this convention, the OMIT feature appears either as a dip ($R<1$) or, at higher power, as a probe-power enhancement feature ($R>1$) around $\Omega \simeq \Omega_m$.

\textit{Power sweeps at fixed detuning---}
\Cref{fig:Fig4_OMIT_vs_Power} shows two representative datasets acquired at fixed cavity detuning while increasing the input power. In each case, the probe offset $\Omega$ is swept across the mechanical resonance for a sequence of pump powers. The central top panels display the measured normalized response $R(\Omega,P_\mathrm{in})$, the central bottom panels show the corresponding response from the full USR model, and the outer columns show representative line cuts.

The left half of \cref{fig:Fig4_OMIT_vs_Power} corresponds to a fixed red-detuned pump, while the right half corresponds to a fixed blue-detuned pump. In both cases, the low-power response shows the familiar narrow OMIT dip near $\Omega/\Omega_m \simeq 1$, consistent with destructive interference between the directly reflected probe field and the field scattered from the pump through the mechanical susceptibility. As the pump power is increased, this dip becomes progressively distorted and eventually evolves into a peak. Within our normalization convention, this marks the onset of a probe-power enhancement feature. For our experimental parameters, this power-driven dip-to-peak transition is observed for both signs of detuning, although its shape depends strongly on the sideband resolution $\kappa/\Omega_m$ (see SI).

These trends are reproduced by the full expression for the intracavity probe sideband $A^-(\Omega)$ in \cref{eq:Aminus3}, which retains both the resonant and counter-rotating contributions. In the USR, the cavity does not strongly suppress the Stokes pathway, so the counter-rotating term can become comparable to the anti-Stokes contribution and substantially alter the interference condition. The resulting evolution from a transparency-like dip to a peak-like enhancement is therefore a direct signature of the non-RWA dynamics expected in the USR.

The light-gray dashed lines in the experimental maps indicate the measured drift of the mechanical resonance frequency. The black dashed lines show the optical-spring shift calculated from the standard linearized backaction formula \cite{aspelmeyerCavityOptomechanics2014}, using the fitted parameters of the corresponding OMIT map. They therefore provide an independent reference demonstrating that the observed displacement of the dips and peaks is quantitatively consistent with optomechanical backaction. This backaction-induced shift is reproduced only by the full model and is absent in the expression using the RWA (see SI).

\begin{figure*}[!t]
	\includegraphics[width=1\linewidth]{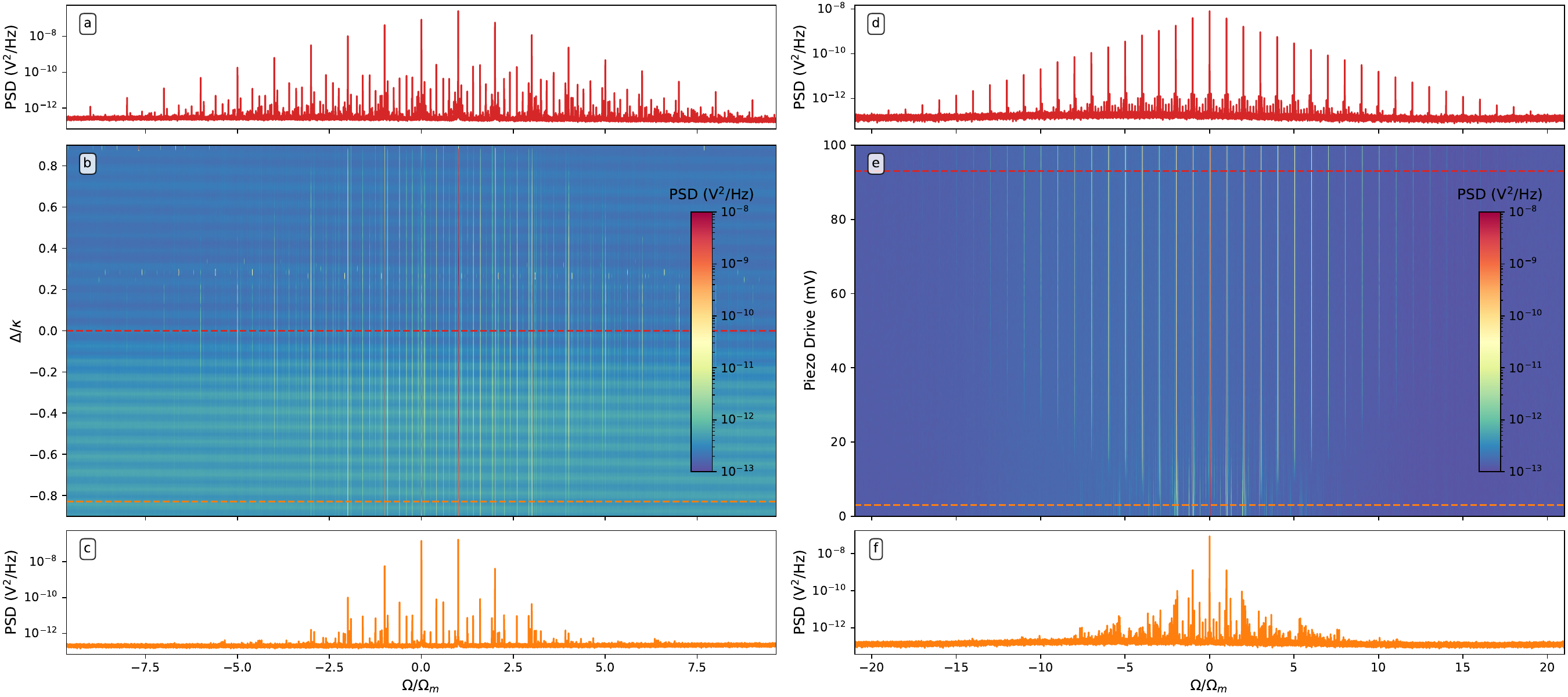}
	\caption{Optomechanical frequency-comb generation. (a--c) All-optical measurement with two coherent pump tones separated by the mechanical resonance frequency $\Omega_m$. The central panel shows the reflected power spectral density versus cavity detuning; the spectra above and below correspond to the detunings marked by the dashed lines. (d--f) Mechanically driven comb measurement with a single optical pump tone and a coherent piezo drive at $\Omega_m$. The central panel shows the power spectral density versus piezo-drive amplitude; the spectra above and below correspond to the marked amplitudes.}
	\label{fig:Fig6_Freq_Comb}
\end{figure*}

\textit{Detuning maps at fixed power---}
To explore the OMIT response over a broader parameter space, we perform two-dimensional measurements in which the probe offset $\Omega$ is swept around $\Omega_m$ while the pump detuning $\Delta$ is scanned across the cavity resonance at fixed input power. \Cref{fig:Fig5_OMIT_vs_Detuning} shows three representative maps acquired at different powers. In each case, the top panel displays the measured normalized response $R(\Omega,\Delta)$ and the bottom panel shows the corresponding response from the full USR model; representative line cuts from the center map are shown in the left column.

At the lowest power, the response is dominated by a narrow OMIT dip near $\Omega/\Omega_m \simeq 1$ over most of the scanned detuning range. As the power is increased, a region with $R>1$ develops near cavity resonance and extends over a broader range of detunings. In our normalization convention, this again corresponds to a probe-power enhancement relative to the off-mechanical baseline of the same trace. At the same time, the position of the dip/peak shifts away from the bare mechanical resonance in a manner consistent with the optical spring induced by dynamical backaction.

Beyond the fixed-detuning scans, the detuning maps show that the same USR phenomenology persists over a continuous range of cavity detunings. In particular, they show that the crossover from a transparency-like dip to a peak-like enhancement is not restricted to a narrow detuning window, but occupies an extended region of the experimentally accessible $(\Omega,\Delta)$ phase space.

The black dashed lines in \cref{fig:Fig5_OMIT_vs_Detuning} are computed as in the fixed-detuning maps and included only as a reference, showing that the observed displacement of the dips and peaks is quantitatively consistent with the optical spring. In the experimental panels they differ slightly from those in the fitted panels because the measured thermal drift is added to the optical-spring contribution.

Taken together, the fixed-detuning power sweeps and fixed-power detuning maps establish the key phenomenology of USR OMIT and provide a dynamical characterization of the optomechanical interaction in our hBN MIM system. The transparency window evolves into a probe-gain feature as power is increased, while the feature position follows the backaction-induced optical spring. The quantitative agreement with the full model, together with the failure of the RWA description, serves as a clear optical demonstrations of the predicted dip-to-peak phenomenology, alongside related microwave implementations \cite{kumarOptomechanicallyInducedTransparency2024}.

\subparagraph{Frequency Comb}
Equally spaced frequency combs are broadly important for precision frequency synthesis, spectroscopy, and signal generation. In optomechanical systems, the repetition rate is naturally set by the mechanical resonance, while the span is controlled by the cavity-frequency excursion produced by coherent motion, and therefore scales directly with the pull parameter $G$. This makes the hBN fiber-cavity platform interesting beyond OMIT: its large frequency-pull parameter and single-photon coupling place it in a favorable regime for low-repetition-rate optomechanical combs. Scaling estimates in the End Matter and SI show that nanometer-scale coherent motion at the maximum measured coupling would correspond to an ideal modulation index above $10^3$, with the detected span mainly limited by cavity transduction and linewidth.

A representative example is shown in \cref{fig:Fig6_Freq_Comb}(a--c). The cavity is driven by two coherent optical pump tones whose frequency separation is matched to the independently measured mechanical resonance frequency $\Omega_m$ of the selected hBN drum mode. This configuration is closely related to the OMIT measurement discussed above, with the weak pump--probe configuration replaced by a two-pump drive of comparable strength. As the cavity detuning is scanned near resonance, the reflected spectrum develops narrow equally spaced lines separated by $\Omega_m$; farther away, fewer tones remain visible, consistent with the reduced intracavity photon number. This behavior is qualitatively consistent with the resonantly enhanced two-tone mechanism proposed for USR systems in \cite{guOpticalFrequencyComb2024}; in this regime, comb formation can be more robust against chaos \cite{miriOptomechanicalFrequencyCombs2018}, although our accessible span is limited by cavity-lock and operating-point stability.

\Cref{fig:Fig6_Freq_Comb}(d--f) provides a direct control measurement. Here a single optical pump is combined with a coherent piezoelectric drive at $\Omega_m$. As the drive amplitude increases, the spectrum evolves from weak thermally driven sidebands to a broad comb with many harmonics spaced by $\Omega_m$. The correspondence between the all-optical and mechanically driven datasets shows that the comb regime is mediated by large coherent mechanical motion and can be accessed either optically or by direct resonant actuation. Together with previous optomechanical-comb demonstrations in microcavities and integrated resonators \cite{miriOptomechanicalFrequencyCombs2018,huGenerationOpticalFrequency2021,mercadeMicrowaveOscillatorFrequency2020,gouChipScaleOptomechanicalFrequency2025}, these results identify hBN fiber cavities as a reconfigurable MIM-like platform for optomechanical comb generation. Because the measured coupling is large while the cavity linewidth and drum geometry remain design parameters, our estimates indicate a realistic path toward one-to-two orders-of-magnitude broader detectable spans, or comparable spans at orders-of-magnitude lower optical power than several existing platforms.

\subparagraph{Conclusion}
We have demonstrated optomechanically induced transparency in a fiber-based Fabry--Perot microcavity coupled in a MIM geometry to a suspended hBN drum resonator, operating deep in the unresolved-sideband regime. After static and dynamical benchmarking, we resolved the predicted power-driven transition of the OMIT response from a transparency-like dip to a probe-gain feature. The measured power and detuning maps are quantitatively reproduced by the full linearized optomechanical response, while the RWA fails to capture the observed dip-to-peak evolution and backaction-induced feature displacement. This establishes the system as a direct optical realization of USR OMIT phenomenology and shows that counter-rotating scattering processes can play a central role in shaping the probe response in this regime \cite{yanOptomechanicallyInducedTransparency2020,yanOptomechanicallyInducedOptical2021,ojanenGroundstateCoolingMechanical2014}.

More broadly, the present system combines three ingredients that are rarely brought together in a single MIM device: a low-mass 2D mechanical resonator, a compact and widely tunable open-access fiber cavity, and strong single-photon coupling \cite{ruelleTunableFiberFabry2022,sanchezarribasRadiationPressureBackaction2023,jaegerMechanicalModeImaging2023}. It therefore identifies hBN fiber cavities as a promising platform in which small mechanical mass and optical mode volume can be used to enhance single-photon coupling while preserving the experimental flexibility of an open-cavity architecture \cite{foglianoMappingCavityOptomechanical2021a,vigneauUltrastrongCouplingElectron2022a,samantaNonlinearNanomechanicalResonators2023a,gislerEnhancingMembranebasedScanning2024}. The frequency-comb measurements further show that the same platform can access nonlinear optomechanical dynamics in the USR, with a realistic route toward broader detectable spans or lower-power comb generation \cite{miriOptomechanicalFrequencyCombs2018,guOpticalFrequencyComb2024}.

Several clear directions emerge for future work. On the mechanical side, further improvement of the hBN resonator quality factor, for example through phononic or acoustic engineering of the suspended 2D structure, could significantly increase cooperativity and mechanical coherence \cite{kirchhofTunableGraphenePhononic2021,yuTunableGraphenePhononic2023}. On the optical side, the fiber-cavity architecture offers a degree of flexibility that is difficult to achieve in fully integrated optomechanical crystals \cite{hungerFiberFabryPerot2010,ruelleTunableFiberFabry2022}. In particular, the cavity linewidth is not irreversibly fixed by fabrication: besides changing the cavity geometry and mirror set, the effective finesse can in practice also be modified by operating at different wavelengths within the finite stopband of the dielectric Bragg mirrors, where the mirror reflectivity decreases toward the band edge. Together with smaller drums or higher-frequency mechanical modes, this should enable systematic comparisons between deep unresolved-sideband operation, as studied here, and regimes approaching sideband resolution of order unity \cite{sanchezarribasRadiationPressureBackaction2023}. Finally, because hBN can combine nanomechanics with optically active crystal defects \cite{wangCavityQuantumElectrodynamics2021,qianEmitterOptomechanicalInteraction2025}, the present high-coupling platform may provide a useful starting point for future hybrid emitter-photon-phonon experiments, where the unresolved-sideband regime could be enriched by additional quantum and nonlinear degrees of freedom \cite{restrepoFullyCoupledHybrid2017,cernotikInterferenceEffectsHybrid2019,pirkkalainenCavityOptomechanicsMediated2015,carlonzambonEnhancedCavityOptomechanics2022a,raniwalaSpinoptomechanicalCavityInterfaces2025,joePurcellenhancedSpinPhonon2026}.

\section{Acknowledgments}
We thank Sascha Martin and his team in the machine shop of the Physics Department at the University of Basel for help in building the apparatus. We acknowledge the support of the Canton Aargau and the Swiss National Science Foundation (SNSF) under the Project grants no. 200020-178863 and no. 200020-207933.

\section{End Matter}
\subparagraph{Static-coupling calibration and coordinate convention}
The cavity-length axis is first calibrated from broadband white-light spectroscopy and then refined by imposing the free-spectral-range constraint. Static cavity scans through a bare hole of the Si$_3$N$_4$ support membrane provide a reference with negligible dispersive modulation, which is used to remove residual drift from the hBN datasets. Because the cavity length is scanned experimentally by moving only one fiber, the measured coordinates do not directly correspond to the standard symmetric membrane-in-the-middle configuration. We therefore transform the data to the equivalent symmetric frame using
\begin{equation}
	\Delta L_{\mathrm{cav}}^{\mathrm{sym}}=\Delta \tilde{L}_{\mathrm{cav}},
	\qquad
	\Delta z^{\mathrm{sym}}=\Delta \tilde{z}-\frac{\Delta \tilde{L}_{\mathrm{cav}}}{2},
\end{equation}
after which the cavity-frequency shift yields the local dispersive pull $G(z)$ and the corresponding single-photon coupling $g_0(z)$. The same standing-wave periodicity is used to calibrate the sample axis and compensate residual piezo nonlinearity. These static maps also show that the cavity linewidth rises strongly away from the nodes because of increased scattering and mode mixing. The backaction and OMIT measurements are therefore performed closer to a node, where stable cavity locking is maintained and excess linewidth broadening is reduced. The SI further contains the full static characterization of the bare-hole, Si$_3$N$_4$, and hBN scans, including the extracted $G$, $G_\kappa$, $\kappa$, and related fit parameters, with the Si$_3$N$_4$ dataset serving as a useful comparison to more standard membrane-in-the-middle systems.

\subparagraph{Backaction calibration and linewidth extraction}
For the backaction measurements, the displacement power spectral density is calibrated using both an optical reference tone and a mechanically injected reference tone. At each detuning, the thermomechanical spectrum is fitted with a single Lorentzian or, when a split peak is resolved, with a sum of two Lorentzians. The optical-spring curve is fitted first to extract the local coupling and effective cavity linewidth; the damping curve shown in the main text is then computed from the same parameters and compared to the apparent linewidth. Near cavity resonance, the apparent linewidth is more fragile because lock instability and the Brownian motion of the drum both fluctuate the instantaneous cavity detuning. Since the cavity responds instantaneously on the mechanical timescale in the USR, the measured spectrum can average over nearby detuning configurations rather than representing a single effective mechanical linewidth, similar to the large-fluctuation backaction distortions discussed in \cite{leijssenNonlinearCavityOptomechanics2017}. This explains the extra broadening and occasional two-peak structures observed near resonance. Additional stronger-coupling data and further fitting details are given in the SI.

\subparagraph{Experimental tones, normalized OMIT response, and measurement caveats}
All optical tones are generated from a single laser by electro-optic phase modulation. One sideband is used for cavity stabilization, a second sideband provides the OMIT pump tone at frequency $\omega_{L_1}$, and a weak probe tone at frequency $\omega_p$ is swept with a vector network analyzer. In the main text, the pump frequency is denoted by $\omega_l$. All RF sources share a common \qty{10}{\mega\hertz} reference. In the present configuration, the cavity is not addressed directly with the laser carrier, but with a GHz-shifted EOM sideband: the laser carrier acts as a local oscillator, while the OMIT pump and probe are high frequency sidebands. We define the pump detuning from cavity resonance as $\Delta=\omega_{L_1}-\omega_{\mathrm{cav}}$ and the pump--probe offset as $\Omega=\omega_p-\omega_{L_1}$.

The experimentally measured OMIT signal is the demodulated reflected response $Z(\Omega)$, which is proportional to the intracavity probe-sideband amplitude $A^-(\Omega)$. In the main text we plot the normalized scalar response
\begin{equation}
	R(\Omega)=\frac{|Z(\Omega)|}{|Z_{\mathrm{off}}|},
\end{equation}
with $|Z| = |a_{L_0}| \sqrt{\kappa_{\mathrm{ext}}}\,|A^-|$, where $Z_{\mathrm{off}}$ is the off-mechanical background of the same trace. With this convention, the off-mechanical response is normalized to unity for each trace, and the OMIT feature appears either as a dip ($R<1$) or as a probe-power enhancement feature ($R>1$).

This measurement geometry is important for interpreting the data. In other OMIT implementations, the measured response can contain additional nearby sidebands or image probes that must either be filtered by the cavity or included explicitly in the model \cite{karuzaOptomechanicallyInducedTransparency2013,butersStraightforwardMethodMeasure2017a,bodiyaSubhertzOptomechanicallyInduced2019,sbarraMultiphysicsModelUltrahigh2021}. In our case, because the cavity is addressed on a selected GHz-shifted sideband branch, the opposite PDH sideband and other image tones are far outside the relevant cavity band and their beat terms are rejected by the chosen demodulation and low-pass filtering. As a result, the measured complex response is directly proportional to $\sqrt{\kappa_{\mathrm{ext}}}A^-$, which is why the main-text OMIT data can be compared directly to the full theoretical expression for $A^-(\Omega)$. See the SI for the full derivation of the demodulated response and for additional measurement-scheme comparisons.

\subparagraph{OMIT fit conventions and backaction overlays}
The OMIT maps in the main text are compared to the full non-RWA expression for the intracavity probe amplitude $A^-(\Omega)$. For the fixed-detuning power sweeps of \cref{fig:Fig4_OMIT_vs_Power}, the cavity linewidth is fixed to $\kappa/2\pi=\qty{1.0}{\giga\hertz}$, while the effective local coupling $G$ and detuning $\Delta$ are treated as global fit parameters shared across the full dataset; the mechanical resonance frequency and linewidth are allowed to vary from trace to trace to account phenomenologically for slow temperature drift and lock-induced broadening.

For the fixed-power detuning maps of \cref{fig:Fig5_OMIT_vs_Detuning}, the input power is fixed by the measurement, while the effective local coupling and effective cavity linewidth are treated as global fit parameters for each map. In this case, the fitted linewidth should be interpreted as an effective linewidth that includes the thermal broadening contribution $\delta\kappa$. In both figures, the black dashed lines in the fitted panels indicate only the optical-spring shift predicted by the dynamical-backaction model. In the experimental panels, the black dashed lines show the sum of this model optical-spring shift and the measured slow drift of the mechanical resonance frequency, the latter being indicated separately by the light-gray dashed lines.

\subparagraph{Cavity-induced heating}
Absorption-driven photothermal and thermo-optic corrections can modify the modulated optomechanical response, as discussed for semiconductor resonators in \cite{sbarraMultiphysicsModelUltrahigh2021}. For the present hBN fiber-cavity system, we performed qualitative auxiliary checks of possible light-induced heating using the independent $\qty{780}{\nano\meter}$ readout of the drum motion. While this diagnostic was limited by the technical noise of the laser and was not accurate enough for a quantitative calibration, it did not indicate significant cavity-light-induced heating over the power range used here. This is consistent with the large bandgap of hBN and with the fact that the measured backaction and OMIT data are already well reproduced without introducing an additional photothermal degree of freedom.

\subparagraph{Frequency-comb scaling and supporting data}
For a coherent mechanical displacement $x(t)=x_0\cos(\Omega_m t)$, with peak amplitude $x_0$, the ideal modulation index is
\begin{equation}
	\mathcal{M}_{\mathrm{OM}} = \frac{Gx_0}{\Omega_m} = \frac{g_0}{\Omega_m}\frac{x_0}{x_{\mathrm{zpf}}}.
\end{equation}
It measures the cavity-frequency excursion in units of $\Omega_m$ and gives the ideal sideband-order scale. In the unresolved-sideband limit, the detected span is also limited by cavity transduction. For near-zero effective detuning, the sideband-power envelope is approximately geometric, $|a_n|^2\propto\rho^{2|n|}$, with
\begin{equation}
	\rho=\frac{|Gx_0|}{\sqrt{(\kappa/2)^2+(Gx_0)^2}+\kappa/2},
	\qquad
	N_\rho \simeq \frac{D_{\mathrm{dB}}}{-10\log_{10}(\rho^2)} .
\end{equation}
Here $|a_n|^2$ is the power in the $n$-th sideband, $D_{\mathrm{dB}}$ is the power dynamic range relative to the local noise floor, and $N_\rho$ estimates the largest visible positive comb order in this adiabatic approximation. Full derivation in SI.

The envelope in \cref{fig:Fig6_Freq_Comb}(d--f) is consistent with $G/2\pi\simeq\qty{2.75}{\giga\hertz\per\nano\meter}$, thermal motion of $x_0\simeq\qty{38}{\pico\meter}$ at zero drive, and $x_0\simeq\qty{0.6}{\nano\meter}$ at maximum drive, giving $\mathcal{M}_{\mathrm{OM}}\sim8\times10^2$ and about twenty visible positive-order lines. At the maximum measured static coupling, $G/2\pi\simeq\qty{11}{\giga\hertz\per\nano\meter}$ and $g_0/2\pi\simeq\qty{180}{\kilo\hertz}$, \qty{1}{\nano\meter} motion would give $\mathcal{M}_{\mathrm{OM}}\simeq5.5\times10^3$. With the present $\kappa/2\pi\simeq\qty{1}{\giga\hertz}$ this corresponds to order $10^2$ visible lines; reducing $\kappa/2\pi$ to $\sim\qty{100}{\mega\hertz}$, as in related fiber cavities \cite{rochauDynamicalBackactionUltrahighFinesse2021,sanchezarribasRadiationPressureBackaction2023}, would increase this to order $10^3$ while remaining in the USR.

For two resonant optical tones with intracavity photon numbers $n_1$ and $n_2$, the linear estimate
\begin{equation}
	\mathcal{M}_{\mathrm{OM}} \simeq \frac{4g_0^2\sqrt{n_1n_2}}{\Gamma_m\Omega_m}
\end{equation}
shows why large $g_0$ is valuable for low-power comb generation, although it neglects saturation, detuning errors, optical-spring shifts, heating, and lock stability. The SI comparison shows that long MIM cavities can have very small $\kappa$ but small $G$ \cite{planzMembraneinthemiddleOptomechanicsSoftclamped2023}, while photonic-crystal systems can have large $g_0$ or $Q_m$ but often broad effective linewidths or less reconfigurability
\cite{leijssenNonlinearCavityOptomechanics2017,guoIntegratedOpticalreadoutHighQ2022a}. By combining open-cavity tunability, large $G$, comparatively large $g_0$, and MHz mechanical frequencies, the present platform can reach comparable comb metrics with substantially lower estimated optical power than several existing optomechanical-comb implementations \cite{huGenerationOpticalFrequency2021,mercadeMicrowaveOscillatorFrequency2020,gouChipScaleOptomechanicalFrequency2025}.

The two-tone optical comb in \cref{fig:Fig6_Freq_Comb} contains fewer lines than the strongest piezo-actuated comb and shows weaker harmonics between dominant tones, resembling mechanically driven spectra at intermediate drive or slightly off resonance. We cannot distinguish unambiguously between weak effective drive and small detuning from the optimal condition, although limited lock stability suggests the former is likely. The SI shows a stronger two-tone transition to a cleaner comb, suggestive of entry into a more strongly synchronized or locking-enhanced regime, a piezo-frequency scan confirming that the comb weakens over several mechanical linewidths, and mechanically driven combs on SiN and higher-order hBN modes. 

An additional attractive perspective is that, in upcoming work, we show that the mechanical resonance frequency of hBN drum resonators can be tuned with temperature over an exceptionally large range, up to about 300\% of its original value. Combined with the present fiber-cavity architecture, this suggests a route toward widely tunable optomechanical frequency combs, with a comb spacing directly controlled through the mechanical resonance.


\bibliographystyle{apsrev4-2}
\bibliography{OMIT_Main}

\end{document}



\title{Unresolved-Sideband Optomechanics with Hexagonal Boron Nitride:\\ Induced Transparency, Gain, and Frequency Combs  --  Supplementary Information}


\author{Francesco Fogliano}
\affiliation{Physics Department, University of Basel}
\author{Thibaud Ruelle}
\affiliation{Physics Department, University of Basel}
\author{David Jaeger}
\affiliation{Physics Department, University of Basel}
\author{Martino Poggio}
\email[]{martino.poggio@unibas.ch}
\affiliation{Physics Department, University of Basel}


\date{\today}

%

\keywords{hBN, OMIT, fiber-cavity, frequency comb}

\maketitle



\section{Experimental Setup}

\subsection{Optics scheme}
The experimental platform is based on a tunable fiber-based Fabry--Perot microcavity operated in a membrane-in-the-middle geometry, following the general architecture developed in \cite{ruelleTunableFiberFabry2022}. The cavity is formed by two concave mirrors laser-machined on the end facets of optical fibers and coated with dielectric Bragg mirrors centered around \qty{940}{\nano\meter}. In the configuration used here, the input mirror is realized on a single-mode fiber, while the transmission mirror is realized on a multimode fiber. The two fibers are mounted on
independent positioning stages that provide coarse alignment as well as fine cavity-length control through shear piezo actuators, allowing the cavity to be tuned over many free spectral ranges.
\begin{figure}[!b]
	\includegraphics[width=0.52\linewidth]{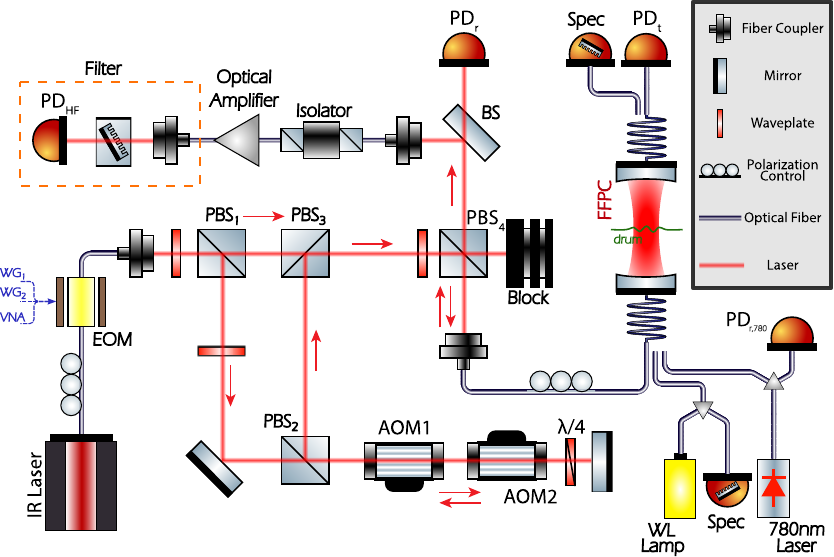}
	\hfill
	\raisebox{0.2\height}{\includegraphics[width=0.45\linewidth]{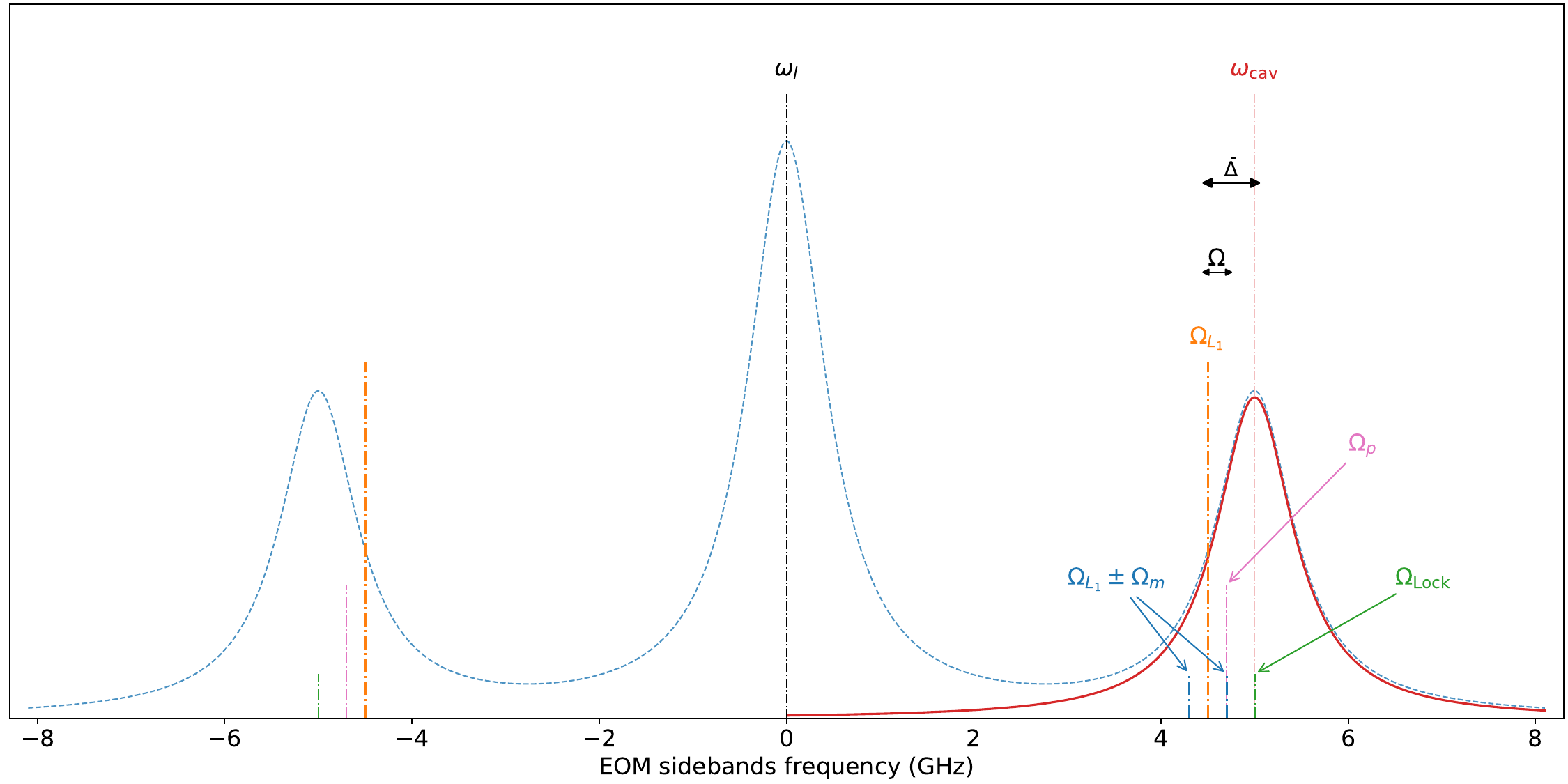}}
	\caption{Optical setup and RF-tone layout. Left: schematic of the fiber-based Fabry--Perot membrane-in-the-middle setup. The IR laser is phase modulated with an EOM, sent to the cavity through polarization optics, and monitored in reflection and transmission; the auxiliary \qty{780}{\nano\meter} beam and white-light path are used for imaging, alignment, and distance calibration. Right: optical sidebands generated by the RF tones used for cavity locking, pumping, and probing.}
	\label{fig:FigS1_Optics_scheme}
\end{figure}

The optical setup is designed to prepare the laser field before it is sent to the cavity and to collect both reflected and transmitted signals for cavity locking, cavity characterization, and optomechanical measurements. A schematic of the optical layout is shown in \cref{fig:FigS1_Optics_scheme}. The same narrow-linewidth laser provides the optical carrier for the cavity measurements and acts as a common local oscillator for the demodulated signals. The resonant optical path is combined with an auxiliary off-resonant beam at \qty{780}{\nano\meter}, which is used for sample imaging and alignment without perturbing the cavity resonance. The reflected optical signal from the cavity is separated from the input light, amplified, filtered and detected on a fast photodiode, while the transmitted signal is detected independently.

The mechanical resonator is introduced into the cavity field using an independent sample-positioning block, such that the cavity and the sample can be aligned separately. The sample stage provides transverse motion in the membrane plane and axial motion along the cavity standing-wave direction, which is essential both for bringing the device into the cavity mode and for selecting the local optomechanical working point used in the measurements reported in the main text. The experiment is operated in high vacuum and is compatible with both room-temperature and cryogenic operation; all data presented in the present work were acquired at \qty{300}{\kelvin}. Static OM measurements were performed also at \qty{4}{\kelvin}, but due to high losses at the splicing point occurred during the cool-down, it was not possible to lock the cavity and perform measurement of the dynamical effects.

\subsection{White light spectroscopy}
\label{sec:SI_white_light}
During cavity and sample alignment, we require a fast and practical method to estimate distances along the cavity axis over a broad range. The nanopositioners allow the fibers and the sample to be separated by several millimeters. Visual inspection allows us to reduce this distance to a few hundred micrometers, comparable to the fiber diameter and the thickness of the sample support frame. For shorter distances, we employ a white-light spectroscopy technique, which can operate from roughly \qty{200}{\micro\meter} (limited by the resolution of our spectrometer) down to fiber contact.

The method can be implemented both in reflection and in transmission. In practice, reflection is used during the first alignment steps, when the sample is positioned with respect to the input fiber while the second fiber remains far away, such that the measured signal is dominated by the interference between the input fiber and the sample. Once the sample is aligned and positioned in XY so that the beam passes through a hole in the SiN and the system behaves effectively as an empty cavity, the transmission-fiber mirror is approached and the white-light spectrum is recorded in transmission, which typically provides a better signal-to-noise ratio for the cavity-length calibration.
\begin{figure}[!b]
	\includegraphics[width=1\linewidth]{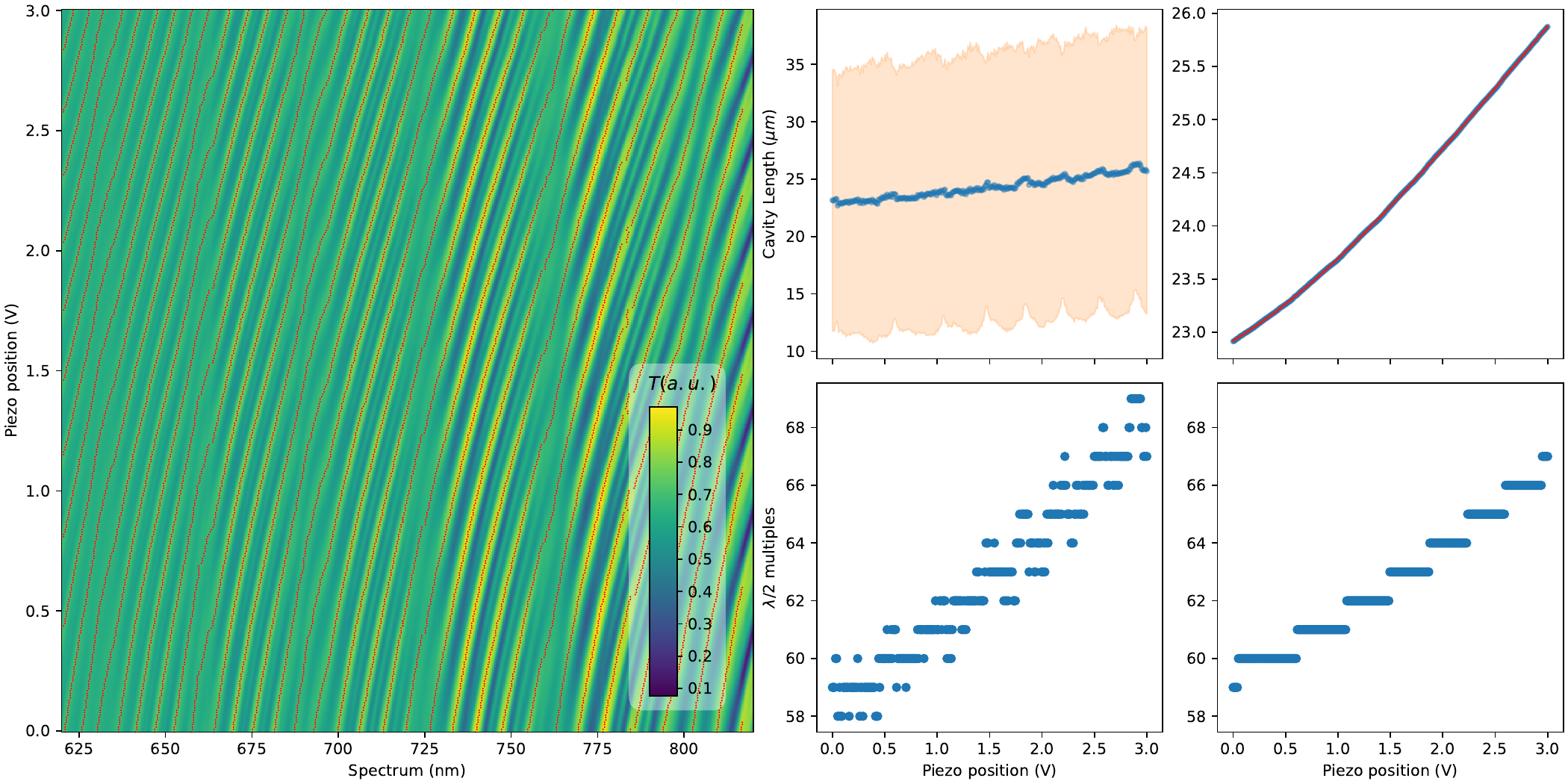}
	\caption{White-light spectroscopy used for coarse cavity-length estimation and calibration of the fiber-piezo axis. Left: normalized transmission spectrum as a function of wavelength and fiber-piezo position; red markers indicate the detected transmission maxima. Top center and bottom center: cavity length and longitudinal mode number obtained from the pairwise fit of adjacent maxima. Top right and bottom right: refined cavity length and longitudinal mode number obtained from the global fit using all detected maxima of each spectrum. The red curve in the top-right panel is a smooth fit used to convert fiber-piezo voltage into cavity length.}
	\label{fig:FigS2_white_light}
\end{figure}

A broadband spectrum map is acquired by scanning the fiber (or sample) piezo over a certain length range while recording the transmitted spectrum on the spectrometer. The acquisition time for a full map is of the order of one minute. A representative dataset is shown in \cref{fig:FigS2_white_light}. The left panel displays the normalized white-light transmission map as a function of wavelength and fiber-piezo position, with the extracted transmission maxima indicated by red markers. The remaining panels summarize the two-step fitting procedure used to estimate the cavity length and the associated longitudinal mode number for each piezo position.

The analysis is based on the cavity resonance condition including the wavelength-dependent reflection phase of the DBR mirrors. The reflectivity and phase of the coating are computed independently using the characteristic transfer-matrix method, starting from the nominal coating stack and matched to the calibration data provided by the coating manufacturer. As discussed in \cite{koksMicrocavityResonanceCondition2021a}, different penetration-depth concepts are relevant for different cavity observables. For the white-light calibration considered here, the relevant quantity is the wavelength-dependent phase penetration, which enters through the DBR reflection phase $\phi(\lambda)$. By contrast, the frequency penetration depth $L_\tau$ is required later when converting cavity length into frequency-space quantities in the
static optomechanical analysis.

In a first step the cavity length is estimated independently from each pair of adjacent transmission maxima $\lambda_q$ and $\lambda_{q+1}$ using
\begin{equation}
	L_{\mathrm{cav}} = \frac{\lambda_q \lambda_{q+1}}{2\left(\lambda_q-\lambda_{q+1}\right)}
				\left(1 + \frac{\phi(\lambda_q)}{\pi}-\frac{\phi(\lambda_{q+1})}{\pi} \right).
	\label{eq:SI_WL_dummy}
\end{equation}
Here the phase values $\phi(\lambda)$ are taken directly from the simulated DBR phase profile. For each piezo position, this procedure produces several pairwise estimates of the cavity length; after removing outliers, their mean and standard deviation provide a first estimate of the cavity length and its uncertainty. The corresponding longitudinal mode number is then assigned from the same set of maxima using the resonance condition, taking the mode at the highest wavelength within the selected spectral window as the reference. In \cref{fig:FigS2_white_light}, the top-center and bottom-center panels show the cavity length and longitudinal mode number obtained from this first step.

In a second step all detected maxima of a given spectrum are fitted simultaneously. Starting from the cavity-length estimate and longitudinal mode assignment obtained from the previous step, the algorithm searches over a discrete range of candidate longitudinal mode numbers and, for each candidate, minimizes the error function
\begin{equation}
	\sum_q
		\left|
			\lambda_q - \frac{2 L_{\mathrm{cav}} \pi}{q\pi-\phi(\lambda_q)}
		\right|^2.
	\label{eq:SI_WL_clever}
\end{equation}
In this way, the cavity length is treated as a continuous fit parameter, while the longitudinal mode index remains discrete. The resulting refined cavity-length estimate is shown in the top-right panel of \cref{fig:FigS2_white_light}, together with a smooth polynomial fit used to convert the fiber-piezo voltage into cavity length. The bottom-right panel shows the corresponding refined longitudinal mode number.

The main purpose of this procedure is to provide a rapid and robust estimate of the cavity length during alignment and to calibrate the fiber-piezo displacement axis for later measurements. The smooth polynomial fit of cavity length versus piezo voltage is used subsequently to express cavity scans on a physically meaningful length axis, while the absolute distance estimate is essential to avoid sample or fiber contact and to reproducibly set the desired cavity length.

\subsection{RF tones}
All optical tones used for cavity locking and OMIT measurements are generated from a single narrow-linewidth laser by phase modulation with an electro-optic modulator (EOM). The optical carrier therefore acts as a common local oscillator for all measurements, while the EOM produces the sidebands used for locking, pumping, and probing the cavity. In the experiments reported here, three phase-coherent RF sources drive the modulation chain: a first RF synthesizer WG1 generates the lock tone, a second RF synthesizer WG2 generates the OMIT pump tone, and a vector network analyzer (VNA) provides the swept probe tone. All RF instruments are referenced to a common \qty{10}{\mega\hertz} clock, ensuring stable relative phase between the different tones throughout the measurement.

The cavity is not locked to the carrier resonance itself, but to one of the PDH sidebands generated by the EOM. Following a procedure similar to \cite{shkarinQuantumOptomechanicsLiquid2019a}, the lock uses the phase (or angle) of the demodulated PDH signal as an error signal, rather than the standard PDH slope around the carrier resonance. This allows the system to lock directly to the upper or lower sideband and avoids the restriction of a conventional on-resonance PDH lock, which would only permit sweeps on one side of the locking point. In the present configuration, this is essential because it allows the probe tone to be swept around the locking point toward both lower and higher frequencies, and therefore to access both red- and blue-detuned OMIT configurations within the same general locking scheme.

The reflected optical signal is detected on a fast photodiode and split into two independent electrical paths: one path is used for cavity locking, while the second is routed to the OMIT readout chain. The RF generation and detection architecture follows a single-sideband upconversion and downconversion scheme. In practice, the lock sidebands are intentionally kept weaker than the main measurement tones so that they provide a robust error signal while minimally perturbing the cavity response, and the OMIT probe is the weakest tone in the modulation chain.

A graphical representation of the optical tones and sidebands used in the experiment is shown in \cref{fig:FigS1_Optics_scheme}. In the notation used throughout the paper, the OMIT pump is the optical tone at frequency $\omega_{L_1}$ and the probe is the weaker tone at frequency $\omega_p$. We define the pump detuning from the cavity resonance as
\begin{equation}
	\Delta = \omega_{L_1} - \omega_{\mathrm{cav}},
\end{equation}
and the pump--probe offset as
\begin{equation}
	\Omega = \omega_p - \omega_{L_1}.
\end{equation}
Experimentally, the detuning $\Delta$ is set by the frequency difference between the pump RF tone and the lock RF tone, while the VNA sweeps the probe offset $\Omega$ in a narrow range around the mechanical resonance $\Omega_m$. The derivation of the corresponding demodulated OMIT signal and its relation to the intracavity probe sideband amplitude are given in the \textit{"OMIT in the USR"} section below.

\subsection{Fiber-cavity parameters}
\begin{figure}[!htb]
	\includegraphics[width=1\linewidth]{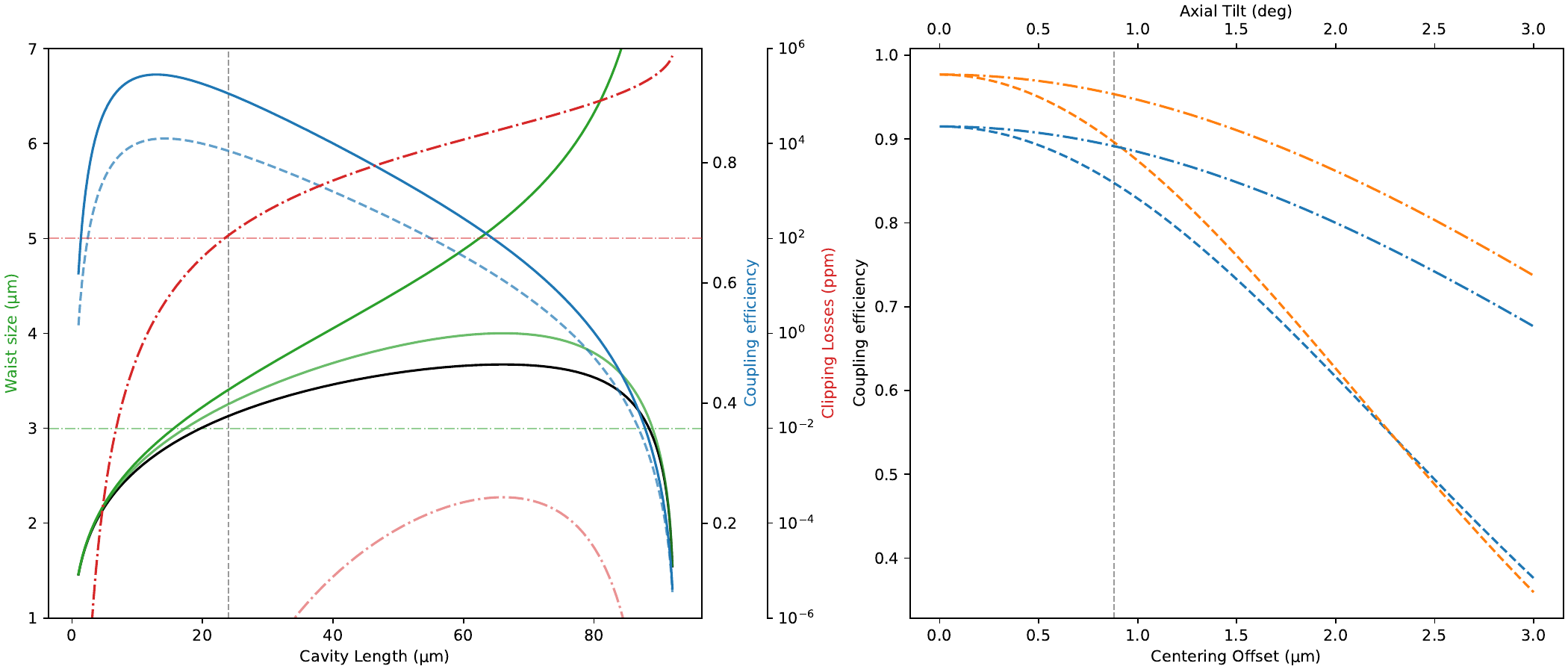}
	\caption{Estimated fiber-cavity geometric parameters and alignment tolerances. Left: cavity waist $w_0$, cavity mode sizes on the two mirrors $w_1$ and $w_2$, ideal input-fiber-to-cavity mode-matching efficiency $\epsilon_{\mathrm{cav}}$, reduced coupling efficiency including the measured off-centering and a finite tilt, and clipping losses on the two mirrors, all plotted as a function of cavity length. The vertical dashed line marks the nominal operating length used in the main text. Right: sensitivity of the coupling efficiency at the nominal cavity length to lateral centering offset and axial tilt, shown for both the cavity-coupled beam (blue) and the directly reflected beam (orange).}
	\label{fig:FigS4_FC_properties}
\end{figure}

The compact fiber cavity used in this work is designed to combine a small optical mode volume with good mode matching to the input single-mode fiber and sufficiently low clipping losses on both fiber mirrors. In this section we summarize the Gaussian-beam model used to estimate the main cavity parameters shown in \cref{fig:FigS4_FC_properties}. The calculation is based on the measured radii of curvature and diameters of the laser-machined mirrors, together with the nominal fiber mode-field diameter and refractive index.

For a cavity of length $L$ and mirror radii of curvature $R_1$ and $R_2$, we define the standard resonator parameters, cavity waist and corresponding Rayleigh range as 
\begin{equation}
	g_i = 1-\frac{L}{R_i}, \qquad
	w_0^2 = \frac{L\lambda}{\pi} \sqrt{\frac{g_1 g_2 (1-g_1 g_2)}{(g_1+g_2-2g_1 g_2)^2}}, \qquad
	z_R = \frac{\pi w_0^2}{\lambda}.
\end{equation}
The cavity mode size on the two mirrors is given by
\begin{equation}
	w_1^2 = \frac{L\lambda}{\pi}\sqrt{\frac{g_2}{g_1(1-g_1 g_2)}},
	\qquad
	w_2^2 = \frac{L\lambda}{\pi}\sqrt{\frac{g_1}{g_2(1-g_1 g_2)}},
\end{equation}
while the waist position relative to the two mirrors is
\begin{equation}
	z_1 = L\frac{g_2(1-g_1)}{g_1+g_2-2g_1 g_2},
	\qquad
	z_2 = L\frac{g_1(1-g_2)}{g_1+g_2-2g_1 g_2}.
\end{equation}

To estimate the coupling between the input single-mode fiber and the cavity mode, we model the fiber mode as a Gaussian beam modified by the lensing action of the curved fiber facet. Its effective radius of curvature is
\begin{equation}
	\mathcal{R}^f = \frac{\mathcal{R}_{\mathrm{mirror}}}{n_f-1},
\end{equation}
from which the effective fiber waist, waist position, and Rayleigh range are
\begin{equation}
	w_0^f = \frac{w^f}{\sqrt{1+\left(\frac{\pi (w^f)^2}{\lambda \mathcal{R}^f}\right)^2}},
	\qquad
	z_0^f = \frac{\mathcal{R}^f}{1+\left(\frac{\lambda \mathcal{R}^f}{\pi (w^f)^2}\right)^2},
	\qquad
	z_R^f = \frac{\pi (w_0^f)^2}{\lambda},
\end{equation}
with $w^f=\mathrm{MFD}/2$. The ideal cavity-coupling efficiency is then written as
\begin{equation}
	\epsilon_{\mathrm{cav}} =
	\frac{4}{
		\left(\frac{w_0^f}{w_0}+\frac{w_0}{w_0^f}\right)^2
		+
		\frac{(z_0-z_0^f)^2}{z_R z_R^f}
	},
\end{equation}
where $z_0$ denotes the cavity-waist position with respect to the input mirror.

In addition to this ideal mode-matching efficiency, we estimate the sensitivity to lateral offset $d$ and axial tilt $\theta$ via
\begin{equation}
	\epsilon_d = \epsilon_{\mathrm{cav}} \exp\!\left[-\left(\frac{d}{d_e}\right)^2\right],
	\qquad
	\epsilon_\theta = \epsilon_{\mathrm{cav}} \exp\!\left[-\left(\frac{\theta}{\theta_e}\right)^2\right],
\end{equation}
with characteristic tolerances
\begin{equation}
	d_e^2 = \frac{2}{\epsilon_{\mathrm{cav}}\left(\frac{1}{(w_0^f)^2}+\frac{1}{w_0^2}\right)},
	\qquad
	\theta_e^2 = \frac{2}{\epsilon_{\mathrm{cav}}\left(\frac{\pi}{\lambda}\right)^2\left[(w^f)^2+w^2\right]},
\end{equation}
where $w$ denotes the relevant cavity-mode size at the coupling plane.

Finally, clipping losses on the mirrors are estimated as
\begin{equation}
	\mathcal{L}_{\mathrm{cl},i} =
	\exp\!\left[-2\left(\frac{D_i/2}{w_i}\right)^2\right],
\end{equation}
with $D_i$ the effective mirror diameter and $w_i$ the cavity mode size on the corresponding mirror.

\begin{table}[t]
	\caption{Parameters used for the estimates shown in \cref{fig:FigS4_FC_properties}.}
	\label{tab:SI_FC_parameters}
	\centering
	\begin{minipage}{0.7\linewidth}
		\begin{ruledtabular}
			\begin{tabular}{lcr}
				Parameter 								& Symbol 						& Value \\
				\hline
				Laser wavelength 						& $\lambda_0$ 					& \qty{919}{\nano\meter} \\
				Input mirror radius of curvature 		& $R_{d1}$ 						& \qty{92.2}{\micro\meter} \\
				Output mirror radius of curvature		& $R_{d2}$ 						& \qty{126.5}{\micro\meter} \\
				Input mirror diameter 					& $D_{\mathrm{sph},1}$ 			& \qty{14.5}{\micro\meter} \\
				Output mirror diameter 					& $D_{\mathrm{sph},2}$ 			& \qty{26.4}{\micro\meter} \\
				Input-fiber lateral offset				& $d_{1,\mathrm{off}}$ 			& \qty{0.88}{\micro\meter} \\
				Output-fiber lateral offset 			& $d_{2,\mathrm{off}}$ 			& \qty{0.99}{\micro\meter} \\
				Nominal cavity length 					& $L_{\mathrm{cav}}$ 			& \qty{24.0}{\micro\meter} \\
				Input-fiber mode-field diameter 		& MFD 							& \qty{6.0}{\micro\meter} \\
				Output-fiber numerical aperture 		& $\mathrm{NA}_{\mathrm{MM}}$ 	& 0.22 \\
				Fiber refractive index 					& $n_f$ 						& 1.4597 \\
			\end{tabular}
		\end{ruledtabular}
	\end{minipage}
\end{table}

The resulting estimates are summarized in \cref{fig:FigS4_FC_properties}. The left panel shows the cavity waist $w_0$ together with the cavity mode sizes on the two mirrors, $w_1$ and $w_2$, as a function of cavity length. On the same plot we also show the ideal input-fiber-to-cavity mode-matching efficiency $\epsilon_{\mathrm{cav}}$, a reduced coupling efficiency including the measured off-centering and a finite angular mismatch, and the expected clipping losses on both mirrors. The vertical dashed line marks the nominal operating length used in the main text. This representation makes clear the tradeoff between reducing the cavity length to obtain a smaller optical mode volume and maintaining acceptable clipping losses and alignment tolerance.

The right panel of \cref{fig:FigS4_FC_properties} shows the sensitivity of the coupling efficiency to lateral centering offset and axial tilt at the nominal cavity length. For comparison, we plot both the cavity-coupled case and the directly reflected beam. The latter is less sensitive to imperfect matching to the cavity mode, which is relevant during alignment because a strong directly reflected signal does not by itself guarantee optimal cavity coupling. The same analysis also shows that the output multimode fiber is not expected to be the dominant limitation in the present geometry, since the cavity divergence remains well within its numerical aperture over the cavity-length range of interest.

\subsection{hBN sample -- Fabrication and navigation}
The mechanical element used in this work is an hBN drum resonator fabricated on the hybrid hBN/Si$_3$N$_4$ platform reported in Ref.~\cite{jaegerMechanicalModeImaging2023}; detailed fabrication and mechanical-characterization procedures can be found there. In brief, exfoliated hBN flakes are selected on a Si/SiO$_2$ substrate, transferred by a PMMA-assisted wet-transfer process onto a prepatterned high-stress Si$_3$N$_4$ membrane, and cleaned by solvent and UV/ozone treatment. The Si$_3$N$_4$ membrane is supported by a silicon frame with through-etched windows, providing optical access from both sides of the sample. This geometry is particularly important for the present experiment, because it allows one cavity fiber to approach the sample from the front and the other from the back through the silicon window, enabling membrane-in-the-middle cavities with very short length while maintaining independent sample alignment.

The same sample architecture can support both nearly isolated hBN drum modes and hybridized hBN/Si$_3$N$_4$ modes. In the measurements reported in the main text, we focus on the fundamental hBN drum mode, which is sufficiently well isolated from the surrounding Si$_3$N$_4$ modes to retain a comparatively small effective mass and therefore comparatively strong radiation-pressure coupling. The detailed mechanical characterization of these devices, including mode imaging and the effect of hybridization, is reported in Ref.~\cite{jaegerMechanicalModeImaging2023}.
\begin{figure}[!htb]
	\includegraphics[width=0.8\linewidth]{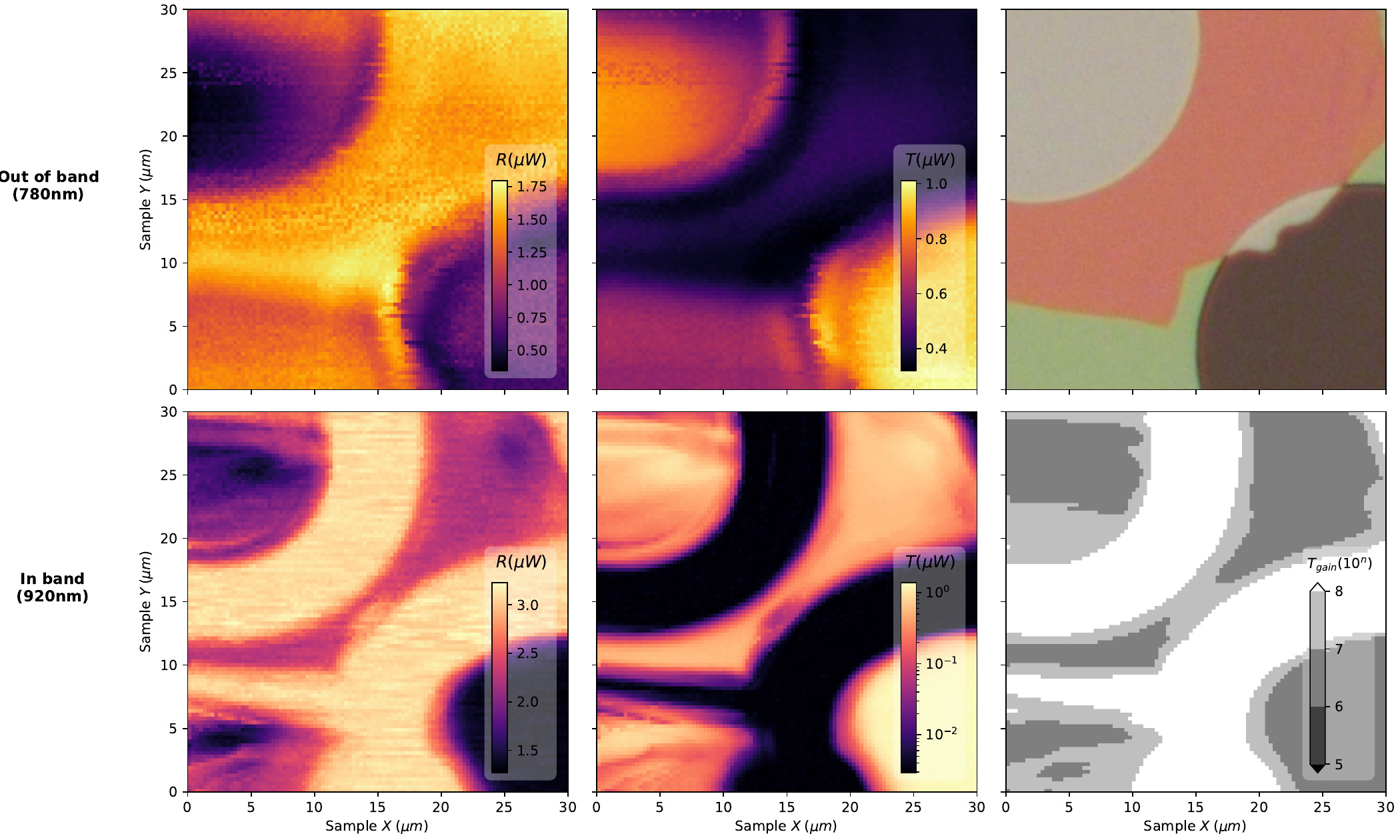}
	\caption{Sample imaging and navigation. Top row: optical microscope image and reflected/transmitted scans obtained with the auxiliary \qty{780}{\nano\meter} beam for coarse identification of empty and hBN-covered holes. Bottom row: maps obtained with the cavity laser around \qty{920}{\nano\meter}, using the maximum transmission and minimum reflection extracted from a cavity scan at each position. The latter highlight low-scattering regions and vignetting around the membrane apertures.}
	\label{fig:FigS5_sample_imaging}
\end{figure}

A representative sample image and navigation sequence are shown in \cref{fig:FigS5_sample_imaging}. The first row of images is obtained with the auxiliary \qty{780}{\nano\meter} beam and is used for coarse sample navigation. The reflected and transmitted signals allow empty holes, hBN-covered holes, and the surrounding Si$_3$N$_4$ membrane to be distinguished while scanning the sample in the transverse plane. An optical microscope image of the same region is shown for comparison.

The second row of \cref{fig:FigS5_sample_imaging} is obtained using the cavity laser around \qty{920}{\nano\meter}. For each point of the scan, a cavity-length sweep is performed and the maximum transmission together with the minimum reflection are recorded. The resulting maps provide a direct indication of the cavity regions with the lowest scattering losses due to sample tilt and fabrication imperfection. In particular, they clearly reveal the vignetting associated with the finite cavity waist around each hole. These scans are therefore used not only to identify the target hBN drum but also to select the regions of the sample that are best suited for stable cavity operation and low-loss optomechanical measurements.

Once the desired region is brought into the scanning window, the sample is approached along the cavity axis and its distance from the input fiber is monitored using the white-light technique described above. This procedure provides a robust route to navigate the sample, align the cavity through the selected aperture, and finally position the hBN drum at the desired location within the cavity mode.

\section{OM Coupling}
\subsection{Bare-cavity reference scan through an empty hole}
\label{sec:SI_OMcoupling_hole}

The start of the static optomechanical analysis is a reference measurement performed through an empty hole of the Si$_3$N$_4$ support membrane, such that the cavity behaves effectively as a bare fiber Fabry--Perot cavity. The purpose of this measurement is threefold. First, it provides a reference map of the cavity reflection and transmission in the absence of dispersive and dissipative optomechanical coupling. Second, it is used to validate the cavity-axis calibration and to track several longitudinal and transverse cavity modes across the sample scan. Third, the small residual drift of the cavity resonances with sample position, caused by experimental imperfections such as residual contraction/dilatation or electrostatic effects during the scan, is extracted from this dataset and later used as a correction baseline for the SiN and hBN scans shown in \cref{fig:FigS7_OMcoupling_SiN,fig:FigS8_OMcoupling_hBN}.

\begin{figure}[!t]
	\includegraphics[width=1\linewidth]{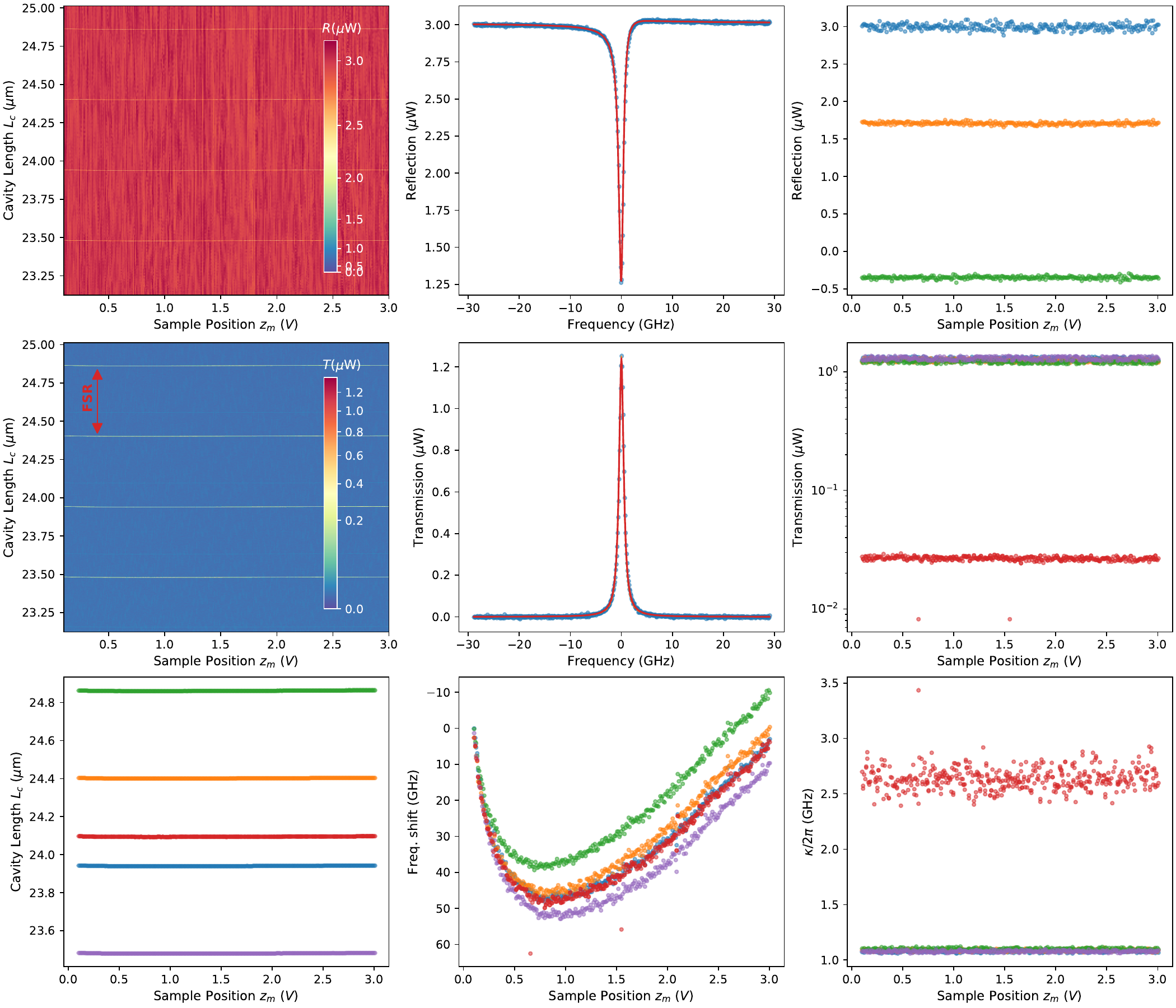}
	\caption{Bare-cavity reference scan through an empty hole of the Si$_3$N$_4$ support membrane. Left: reflected and transmitted cavity maps and extracted peak positions of four longitudinal modes and one transverse mode. Center: representative reflection and transmission fits, and residual mode-frequency drift versus sample position. Right: fitted reflection parameters $A$, $B$, and $C$, transmission amplitude $E$, and cavity linewidth $\kappa$ for the tracked modes.}
	\label{fig:FigS6_OMcoupling_hole}
\end{figure}

The cavity-length axis is calibrated in two steps. A first estimate is obtained from the white-light calibration described above, which converts the fiber-piezo voltage into cavity length. This axis is then refined by imposing the physical constraint that adjacent longitudinal modes must be separated by one free spectral range, corresponding to
\begin{equation}
	\Delta L_{\mathrm{FSR}} = \frac{\lambda}{2}.
\end{equation}
In practice, the measured peak-to-peak spacing across the cavity scan is used to correct residual piezo nonlinearity and flatten the cavity-length axis. The calibrated cavity-length axis can then be converted into frequency space by expanding the cavity resonance condition around the selected longitudinal mode at length $L_{\mathrm{cav}}$ using
\begin{equation}
	\Delta \nu_{\mathrm{cav}} \simeq -\frac{c}{\lambda}\frac{\Delta L_{\mathrm{cav}}}{L_{\mathrm{cav}}+ 2 L_{\tau}}.
	\label{eq:SI_Lcav2Nucav}
\end{equation}
Here $L_\tau$ is the frequency penetration depth of the DBR mirrors, defined from the wavelength-dependent reflection phase through
\begin{equation}
	L_\tau(\lambda)=\frac{c}{2}\frac{\partial \phi(\omega)}{\partial \omega},
\end{equation}
and evaluated numerically from the simulated coating phase following \cite{koksMicrocavityResonanceCondition2021a}. By contrast, the sample-position axis in \cref{fig:FigS6_OMcoupling_hole} is kept in piezo voltage, since for an effectively empty cavity there is no internal standing-wave reference that would allow a reliable conversion of the sample displacement into absolute micrometers.

To fit the reflection and transmission resonances, we use the cavity-response expressions following the mode-matching analysis of \cite{gallegoHighfinesseFiberFabry2016a}. For a single cavity resonance centered at detuning $\Delta=0$, the reflected and transmitted powers can be written as
\begin{equation}
	C_R(\Delta) =
	A
	- B \frac{1}{1+\left(\frac{\Delta}{\kappa/2}\right)^2}
	- C \frac{\frac{\Delta}{\kappa/2}}{1+\left(\frac{\Delta}{\kappa/2}\right)^2}
	+ D \frac{\left(\frac{\Delta}{\kappa/2}\right)^2}{1+\left(\frac{\Delta}{\kappa/2}\right)^2},
	\label{eq:SI_CR_full}
\end{equation}
\begin{equation}
	C_T(\Delta) =
	E \frac{1}{1+\left(\frac{\Delta}{\kappa/2}\right)^2}.
	\label{eq:SI_CT_full}
\end{equation}

Here $A$, $B$, $C$, $D$, and $E$ are fit parameters related to the spatial mode matching and interference of the directly reflected and cavity-coupled fields, while $\kappa$ is the cavity linewidth. The overlap parameters are defined such that
\begin{equation}
	|\alpha|^2=\epsilon_1,
	\qquad
	|\beta|^2=\epsilon_{\mathrm{DR}},
	\qquad
	|o|^2=\epsilon_2,
\end{equation}
where $\epsilon_1$ is the mode matching between the input-fiber mode and the cavity mode, $\epsilon_{\mathrm{DR}}$ is the mode matching between the input-fiber mode and the directly reflected beam, and $\epsilon_2$ is the mode matching between the cavity mode and the output-fiber mode. The coefficients entering \cref{eq:SI_CR_full} and \cref{eq:SI_CT_full} can then be written as
\begin{equation}
	A = |\beta|^2 r_1^2 \simeq |\beta|^2,
\end{equation}
\begin{equation}
	B =	|\alpha|^2 |\beta| \frac{\kappa_{\mathrm{ex}}}{\kappa/2}
			\left[
					2\cos(\theta)\sqrt{r_1 r_2} - \frac{|\alpha|^2}{|\beta|} \frac{r_2}{r_1} \frac{\kappa_{\mathrm{ex}}}{\kappa/2}
			\right]
		\simeq |\alpha|^2 |\beta|
			\left[
					2\cos(\theta)-\frac{|\alpha|^2}{|\beta|}
			\right],
\end{equation}
\begin{equation}
	C = 2 |\alpha|^2 |\beta| \sin(\theta) \frac{\kappa_{\mathrm{ex}}}{\kappa/2}
		\simeq 2 |\alpha|^2 |\beta| \sin(\theta),
\end{equation}
\begin{equation}
	D = 2 |\alpha|^2 |\beta| \cos(\theta) \kappa_{\mathrm{ex}} \frac{L}{c}	=
			|\alpha|^2 |\beta| \cos(\theta)\frac{t_1^2}{2} \simeq 0,
\end{equation}
\begin{equation}
	E = |\alpha|^2 |o|^2.
\end{equation}
The approximation is taken in the high-reflectivity limit $r_1^2 \simeq r_2^2 \rightarrow 1$, $t_1^2 \rightarrow 0$ and $\kappa_{\mathrm{ex}}\simeq \kappa/2$.

The results are summarized in \cref{fig:FigS6_OMcoupling_hole}. The left column shows the raw reflected and transmitted cavity maps recorded while scanning the sample position and cavity length, together with the extracted peak positions of four longitudinal modes and one transverse mode (red). The central column shows representative fits of a single reflection peak and a single transmission peak, demonstrating that the above expressions reproduce the measured line shapes well. The bottom panel of the central column shows the residual frequency drift of the tracked modes as a function of sample position, after setting the initial value of each mode to zero. In an ideal empty cavity these curves would be perfectly flat; experimentally, however, small systematic drifts remain, which are attributed to residual mechanical or electrostatic effects occurring during the scan. These drift curves are therefore retained as a reference correction and later subtracted from the SiN and hBN datasets.

The right column of \cref{fig:FigS6_OMcoupling_hole} reports the fitted parameters extracted from the scan. The top panel shows the reflection parameters $A$, $B$, and $C$ for the longitudinal mode centered around $L_c \sim \qty{24}{\micro\meter}$. The middle panel shows the transmission amplitude parameter $E$ for the four longitudinal modes and one transverse mode, while the bottom panel shows the fitted cavity linewidth $\kappa$ for the same set of modes. Together, these data provide a compact reference characterization of the bare cavity and establish the baseline against which the static dispersive and dissipative effects of the SiN and hBN resonators are quantified.

\subsection{Static dispersive and dissipative coupling in Si$_3$N$_4$ and hBN}
\begin{figure}[!t]
	\includegraphics[width=1\linewidth]{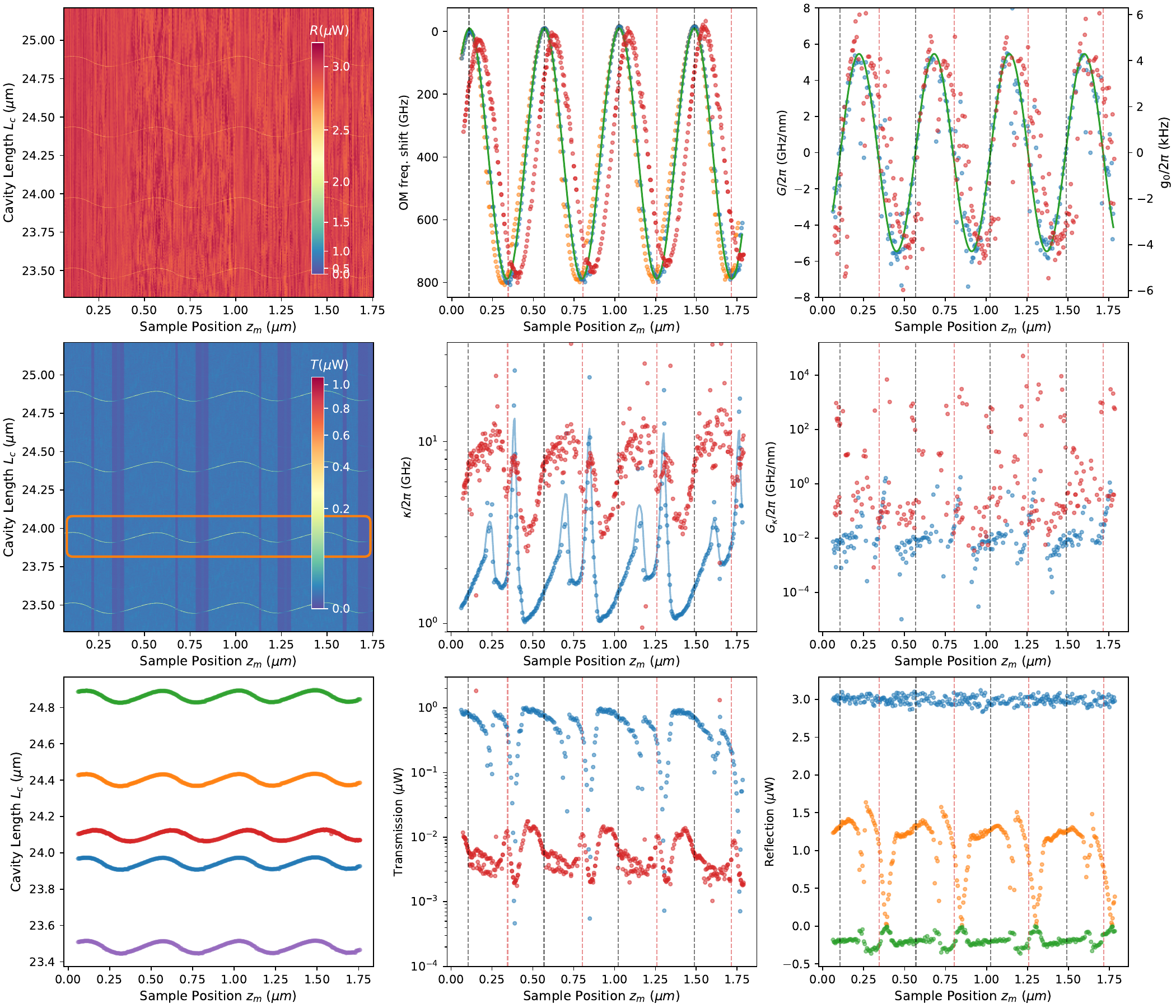}
	\caption{Static optomechanical coupling measured on a bare Si$_3$N$_4$ region. Left column: reflected and transmitted cavity maps together with the extracted peak positions of the tracked longitudinal and transverse modes. Center column: cavity resonance frequency shift, linewidth $\kappa$, and transmission amplitude parameter $E$ as a function of sample position; orange points show the raw longitudinal-mode data in the experimental frame, while blue and red points show the corrected longitudinal and transverse-mode data in the symmetric reference frame. The green curve in the top panel is a fit to the corrected longitudinal-mode frequency shift. Right column: corresponding dispersive coupling $G$, dissipative coupling $G_\kappa$, and reflection fit parameters $A$, $B$, and $C$. The green curve in the top-right panel is the derivative of the fitted frequency-shift curve shown in the top-center panel.}
	\label{fig:FigS7_OMcoupling_SiN}
\end{figure}

\begin{figure}[!t]
	\includegraphics[width=1\linewidth]{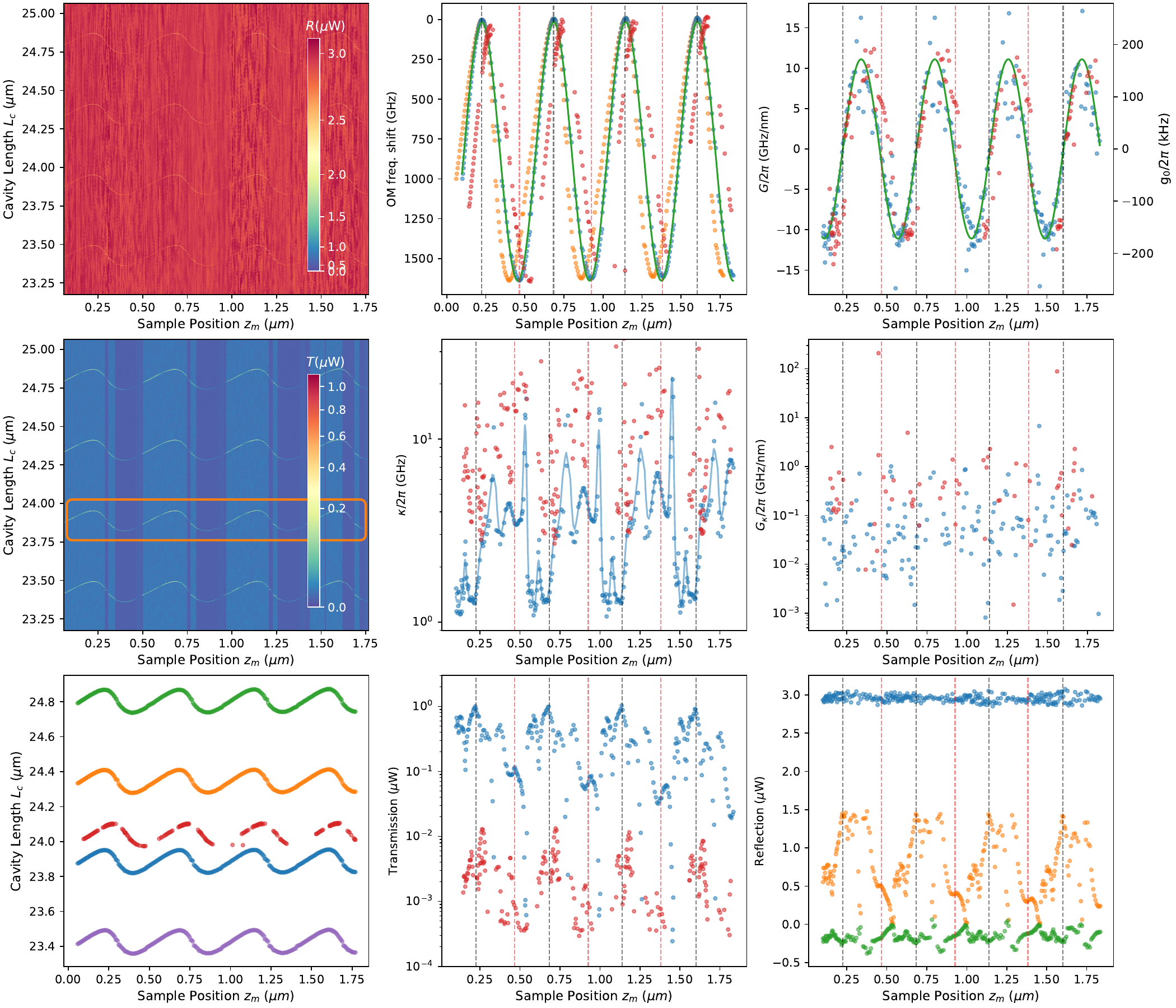}
	\caption{Static optomechanical coupling measured on the suspended hBN drum. Left column: reflected and transmitted cavity maps together with the extracted peak positions of the tracked longitudinal and transverse modes. Center column: cavity resonance frequency shift, linewidth $\kappa$, and transmission amplitude parameter $E$ as a function of sample position; orange points show the raw longitudinal-mode data in the experimental frame, while blue and red points show the corrected longitudinal and transverse-mode data in the symmetric reference frame. The green curve in the top panel is a fit to the corrected longitudinal-mode frequency shift. Right column: corresponding dispersive coupling $G$, dissipative coupling $G_\kappa$, and reflection fit parameters $A$, $B$, and $C$. The green curve in the top-right panel is the derivative of the fitted frequency-shift curve shown in the top-center panel.}
	\label{fig:FigS8_OMcoupling_hBN}
\end{figure}

After establishing the bare-cavity reference through the empty-hole scan of \cref{fig:FigS6_OMcoupling_hole}, we perform the same cavity-length and sample-position scans on the bare Si$_3$N$_4$ region and on the suspended hBN drum. The corresponding datasets are shown in \cref{fig:FigS7_OMcoupling_SiN} and \cref{fig:FigS8_OMcoupling_hBN}. The analysis follows the same fitting procedure introduced above for the bare-cavity scan: for each sample position, the reflection and transmission resonances are fitted with the cavity-response expressions of \cref{eq:SI_CR_full,eq:SI_CT_full}, yielding the resonance frequency, linewidth, and amplitude parameters of the tracked cavity modes.

The left column of \cref{fig:FigS7_OMcoupling_SiN,fig:FigS8_OMcoupling_hBN} shows the raw reflected and transmitted cavity maps as a function of sample position and cavity length. As in the bare-cavity reference case, several optical modes can be tracked across the scan. In the present analysis, we focus on the longitudinal mode around $L_{\mathrm{cav}} \sim \qty{24}{\micro\meter}$, together with one transverse mode that is sufficiently visible over the scan range to provide a useful comparison. The residual frequency drift measured in the bare-cavity scan is subtracted mode by mode from the measured resonance positions in order to isolate the sample-induced frequency shift.

A second important correction concerns the sample-position axis. Experimentally, the cavity length is changed by moving only one fiber, which means that the directly measured coordinates do not correspond to a symmetric displacement of the membrane with respect to the two cavity mirrors. In order to compare the data to the standard membrane-in-the-middle picture, the measured coordinates are transformed into a symmetric reference frame, in which the two mirrors are displaced symmetrically about the membrane position. This can be achieved using the simple transformation
\begin{equation}
	\begin{cases}
		\Delta L_{\mathrm{cav}}^{\mathrm{sym}} &= \Delta \tilde{L}_{\mathrm{cav}} \\
		\Delta z_m^{\mathrm{sym}} &= \Delta \tilde{z}_m - \frac{\Delta \tilde{L}_{\mathrm{cav}}}{2}
	\end{cases}
\end{equation}
In this frame, the periodic cavity response is directly related to the standing-wave structure of the intracavity field, and nodes and antinodes are separated by $\lambda/2$. This periodicity is also used to properly calibrate the sample-axis position and compensate for piezo non-linearities. The corrected data shown in the center and right columns of \cref{fig:FigS7_OMcoupling_SiN,fig:FigS8_OMcoupling_hBN} are expressed in the symmetric reference frame.

The center column of \cref{fig:FigS7_OMcoupling_SiN,fig:FigS8_OMcoupling_hBN} summarizes the main fitted quantities. The top panels show the cavity resonance frequency shift as a function of sample position. The orange points correspond to the raw data in the experimental frame for the longitudinal mode around $L_{\mathrm{cav}} \sim \qty{24}{\micro\meter}$, while the blue points show the same longitudinal-mode data after correction into the symmetric reference frame. The green curve is a fit to the corrected longitudinal-mode frequency shift, and the red points show the corrected data for the selected transverse mode. The middle panels show the fitted linewidth $\kappa$ for the same longitudinal and transverse modes, while the bottom panels show the fitted transmission amplitude parameter $E$.

From the corrected resonance-frequency shift, we extract the dispersive frequency-pull parameter
\begin{equation}
	G = \frac{\partial \omega_c}{\partial z_m},
\end{equation}
shown in the top-right panels of \cref{fig:FigS7_OMcoupling_SiN,fig:FigS8_OMcoupling_hBN}. The blue and red points again correspond to the corrected longitudinal and transverse modes, while the green curve is not an independent fit to $G$, but simply the derivative of the fitted longitudinal-mode frequency-shift curve shown in the top-center panel. The same derivative procedure is applied to the linewidth variation, yielding the dissipative coupling parameter
\begin{equation}
	G_\kappa = \frac{\partial \kappa}{\partial z_m},
\end{equation}
shown in the middle-right panels. Finally, the bottom-right panels report the reflection fit parameters $A$, $B$, and $C$ for the tracked longitudinal mode.

The single-photon optomechanical coupling rate can then be estimated from
\begin{equation}
	g_0 = G x_{\mathrm{zpf}},
	\qquad
	x_{\mathrm{zpf}}=\sqrt{\frac{\hbar}{2M_{\mathrm{eff}}\Omega_m}}.
\end{equation}
For the Si$_3$N$_4$ device of \cref{fig:FigS7_OMcoupling_SiN}, the conversion shown on the secondary vertical axis uses $M_{\mathrm{eff}}=\qty{11}{\nano\gram}$ and $\Omega_m/2\pi=\qty{1.23}{\mega\hertz}$. For the hBN device of \cref{fig:FigS8_OMcoupling_hBN}, the corresponding values are $M_{\mathrm{eff}}=\qty{17}{\pico\gram}$ and $\Omega_m/2\pi=\qty{2.05}{\mega\hertz}$. The large difference in effective mass between the two systems strongly amplifies the difference in single-photon coupling rate, such that the hBN device reaches values of $g_0$ that are much larger than for the Si$_3$N$_4$ membrane, well beyond the increase expected solely from the higher reflectivity of the hBN flake itself.

Several additional features are evident in these data. First, the transverse-mode response is shifted with respect to the longitudinal-mode response, indicating a different effective node position for the two optical modes. Second, the dissipative coupling becomes particularly strong away from the nodes, where scattering and clipping losses are enhanced. In addition, some regions of especially large linewidth variation are consistent with local hybridization or mixing with higher-order optical modes, which modifies the cavity loss landscape beyond a simple single-mode standing-wave picture. These effects are visible in both the Si$_3$N$_4$ and hBN scans, but are most striking in the hBN device, where the stronger overall optomechanical interaction makes the contrast between node-centered and off-node operation especially clear. This strong increase of $\kappa$ away from the nodes is also the main reason why the dynamical measurements reported in the main text were performed close to a node of the standing-wave field. Far from these operating points, the increased scattering losses strongly reduce the cavity signal and make stable cavity locking difficult or impossible.

Taken together, \cref{fig:FigS7_OMcoupling_SiN,fig:FigS8_OMcoupling_hBN} provide the static optomechanical characterization of the two resonators. The Si$_3$N$_4$ scan serves as a useful intermediate benchmark to compare to other work in the literature, while the hBN scan demonstrates the large dispersive coupling and correspondingly large single-photon coupling rate that enable the unresolved-sideband OMIT and dynamical backaction measurements discussed in the main text.

\subsection{Theoretical estimate for hBN drum optomechanics}
To complement the static optomechanical measurements of \cref{fig:FigS7_OMcoupling_SiN,fig:FigS8_OMcoupling_hBN}, we estimate the expected mechanical, optical, and optomechanical properties of suspended hBN drums over a broad range of geometrical parameters. The purpose of this calculation is twofold. First, it provides a compact overview of the design space accessible to hBN membrane-in-the-middle devices and helps identify the combinations of thickness and drum diameter that are most favorable for strong optomechanical coupling. Second, it serves as a consistency check for the experimental results reported in the main text, by comparing the measured single-photon coupling rates to the theoretical maximum expected for the present cavity geometry.

\begin{figure}[!t]
	\includegraphics[width=1\linewidth]{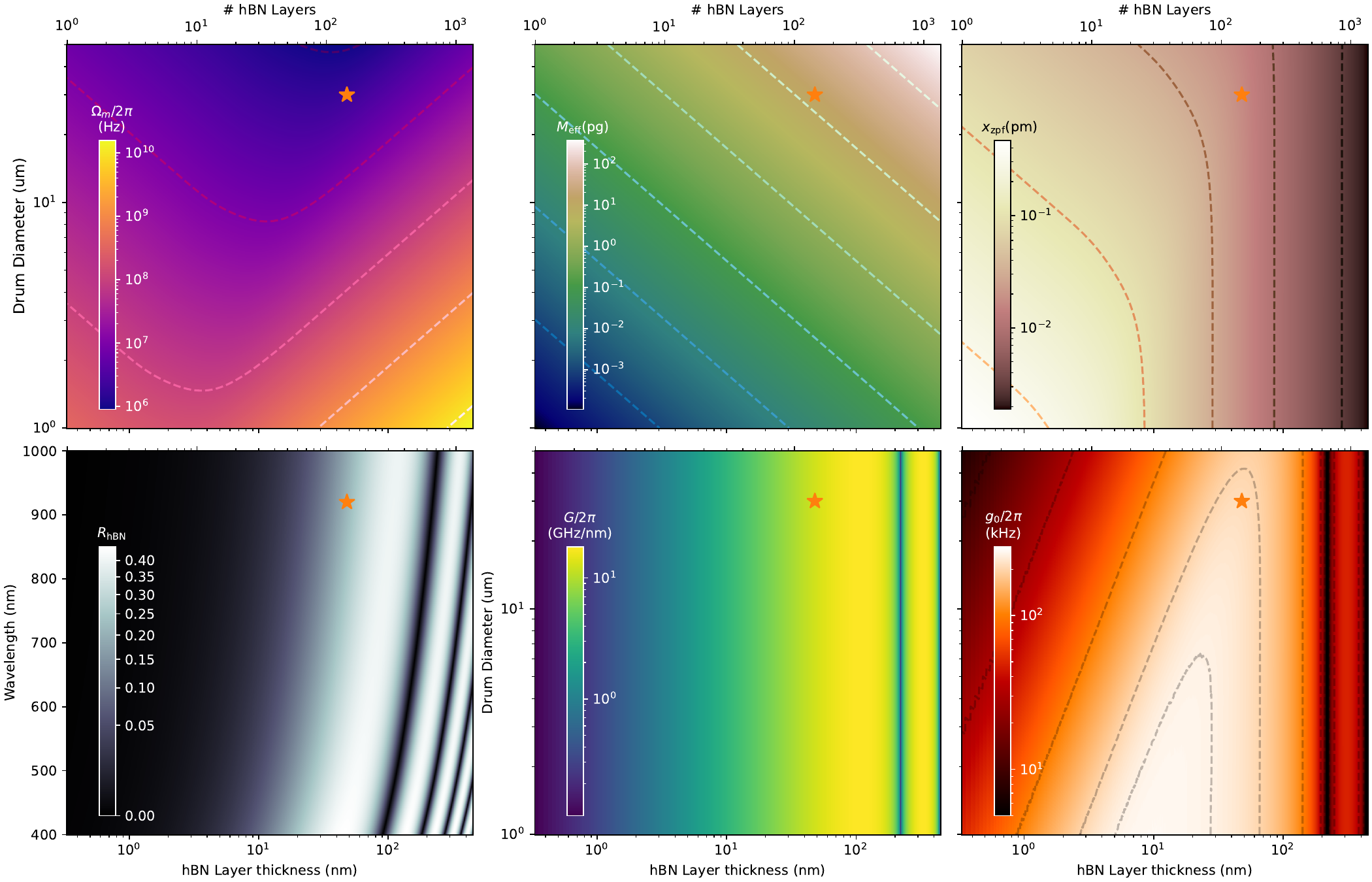}
	\caption{Theoretical estimate for hBN drum optomechanics. Top row: mechanical resonance frequency $\Omega_m/2\pi$, effective mass $M_{\mathrm{eff}}$, and zero-point fluctuation $x_{\mathrm{zpf}}$ of the fundamental hBN drum mode as a function of drum diameter and hBN thickness. Bottom row: hBN reflectivity $R_{\mathrm{hBN}}$ versus thickness and wavelength, dispersive frequency-pull parameter $G/2\pi$, and resulting single-photon coupling rate $g_0/2\pi$ expected for the membrane-in-the-middle cavity. The orange marker indicates the approximate geometry of the device used in the experiment.}
	\label{fig:FigS9_OMcoupling_Theo}
\end{figure}

The results are summarized in \cref{fig:FigS9_OMcoupling_Theo}, where the horizontal axis is the hBN thickness and the vertical axis is the drum diameter. The orange marker indicates the approximate geometry of the device used in the experiment. The top row of the figure summarizes the mechanical properties of the fundamental hBN drum mode. The bottom row shows the optical reflectivity of the hBN flake and the corresponding optomechanical coupling expected in the membrane-in-the-middle cavity.

The top-left panel shows the expected mechanical resonance frequency of the fundamental drum mode. To estimate this quantity, we use the standard description of a circular drum resonator in the crossover between the membrane and plate limits, as discussed in \cite{jaegerMechanicalModeImaging2023}. Writing the drum radius as $r=D/2$, the mode frequency can be expressed as
\begin{equation}
	\omega_m =
	k_{mn}\sqrt{\frac{D_{\mathrm{p}}}{\rho t\, r^4}\left(k_{mn}^2+\frac{T_{\mathrm{p}} r^2}{D_{\mathrm{p}}}\right)},
	\label{eq:SI_hBN_drum_freq}
\end{equation}
where $t$ is the hBN thickness, $\rho$ its mass density, $T_{\mathrm{p}}$ the pretension, and
\begin{equation}
	D_{\mathrm{p}}=\frac{E_Y t^3}{12(1-\nu^2)}
\end{equation}
is the bending rigidity, with $E_Y$ the Young's modulus and $\nu$ the Poisson ratio. The quantity $k_{mn}$ is the dimensionless eigenvalue of the corresponding drum mode and interpolates between the membrane and plate limits. As expected, large and relatively thick drums can reach low mechanical frequencies by operating close to the crossover between tension-dominated and bending-dominated behavior.

The top-center panel shows the corresponding effective mass of the fundamental drum mode. For the circular geometry considered here, we use the standard estimate
\begin{equation}
	M_{\mathrm{eff}} \simeq \frac{\rho \pi r^2 t}{5},
	\label{eq:SI_hBN_meff}
\end{equation}
which is appropriate for the fundamental mode of an ideal drum. The top-right panel then shows the corresponding zero-point fluctuation amplitude $x_{\mathrm{zpf}}$, calculated from \cref{eq:SI_hBN_meff} together with the resonance frequency of \cref{eq:SI_hBN_drum_freq}. As expected, $x_{\mathrm{zpf}}$ is largest for light and low-frequency drums, i.e. for small effective mass and correspondingly soft mechanical motion.

The bottom-left panel shows the optical reflectivity of a suspended hBN slab as a function of thickness and wavelength. This quantity is computed using a standard characteristic transfer-matrix model for a single hBN layer embedded in air. Around the cavity operating wavelength near \qty{920}{\nano\meter}, the reflectivity increases with flake thickness, which directly enhances the ability of the membrane to perturb the cavity resonance.

The bottom-center panel shows the corresponding dispersive frequency-pull parameter $G/2\pi$ expected in the membrane-in-the-middle geometry. To estimate this quantity, we simulate the full fiber-cavity--membrane system with the same characteristic transfer-matrix approach, now including the two cavity mirrors and the hBN membrane. The cavity resonance is evaluated for two membrane positions separated by $\lambda/4$, corresponding approximately to a node and an antinode of the standing-wave field, so as to extract the maximum resonance shift induced by the membrane. This shift is then converted into the corresponding dispersive frequency-pull parameter $G=\partial \omega_c/\partial z_m$. 

Finally, the bottom-right panel shows the resulting single-photon coupling rate $g_0/2\pi$, obtained from \cref{eq:SI_hBN_meff} and the dispersive coupling estimate of the previous panel. The calculation shows that the present device geometry lies in a region of already very strong coupling, with a theoretical maximum value of $g_0/2\pi$ around \qty{210}{\kilo\hertz}, in good agreement with the experimentally observed maximum value of about \qty{180}{\kilo\hertz}. The same map further indicates that, for the present cavity length, smaller drum diameters could push the maximum coupling toward \qty{260}{\kilo\hertz}. More generally, keeping realistic cavity lengths of order \qty{10}{\micro\meter} and moving to shorter optical wavelengths would make it possible to approach single-photon coupling rates in the MHz range, comparable to the mechanical resonance frequency itself.

Overall, \cref{fig:FigS9_OMcoupling_Theo} provides a useful design-space overview for hBN membrane-in-the-middle devices. It shows explicitly how the optomechanical coupling arises from the interplay between mechanical frequency, effective mass, and membrane reflectivity, and it confirms that the large coupling achieved in the present experiment is close to the theoretical optimum expected for the chosen drum geometry.

\section{OM Backaction}
Radiation pressure modifies the effective mechanical susceptibility of the resonator, giving rise to the optical spring effect and to optomechanically induced damping or amplification. In the main text, these effects are used as an independent dynamical benchmark of the static optomechanical coupling extracted in the previous section. Here we provide additional details on the measurements and show a second dataset acquired at stronger coupling, where the backaction is larger but the spectra become correspondingly less stable and more difficult to interpret quantitatively.
\begin{figure}[!b]
	\includegraphics[width=1\linewidth]{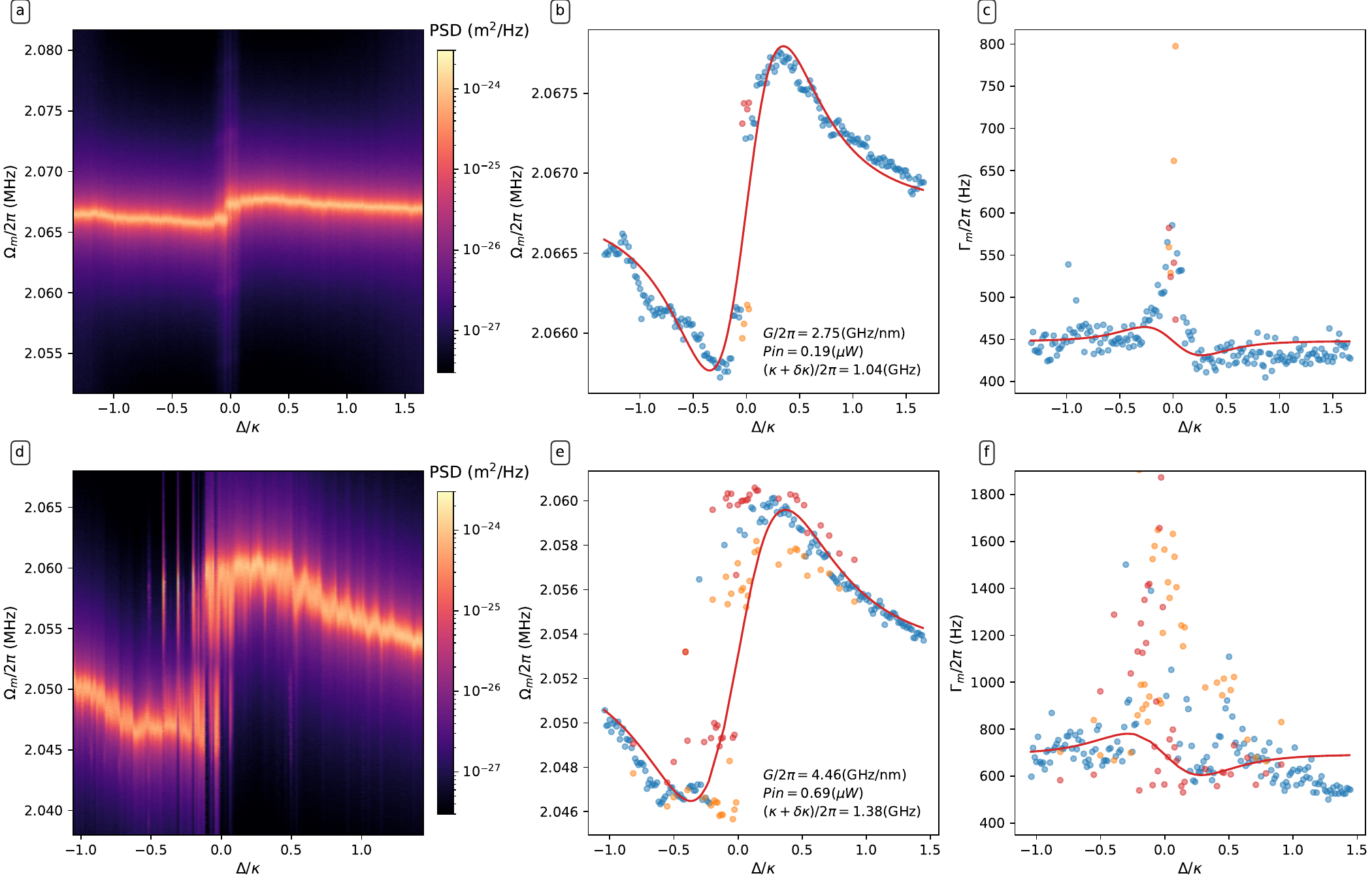}
	\caption{Radiation-pressure dynamical backaction of the selected hBN mechanical mode. Top row: same dataset discussed in the main text. Bottom row: measurement performed at higher optical power and slightly stronger local coupling. Left column: calibrated displacement power spectral density versus normalized detuning $\Delta/\kappa$ and mechanical frequency. Center column: extracted mechanical resonance frequency $\Omega_m/2\pi$ versus detuning; blue points correspond to single-Lorentzian fits, while orange and red points indicate the lower and upper branches obtained when a two-Lorentzian fit is required. The solid curve is the optical-spring fit. Right column: extracted apparent mechanical linewidth $\Gamma_m/2\pi$ together with the damping curve computed from the parameters obtained in the frequency fit.}
	\label{fig:FigS10_Backaction}
\end{figure}

The measurements are performed by sweeping the pump detuning $\Delta$ across the cavity resonance at fixed input power while recording the thermomechanical spectrum of the selected hBN mode. The displacement power spectral density is calibrated using two reference tones. An optical reference tone provides a stable calibration marker for the cavity response and detuning axis, while a second reference tone is injected mechanically through the piezo actuator used elsewhere in the experiment for coherent membrane excitation.

For each detuning, the thermomechanical peak is fitted either with a single Lorentzian or, when the spectrum develops a clearly resolved double structure, with a sum of two Lorentzians. In this way, the effective mechanical resonance frequency and linewidth are extracted across the full detuning range. The optical spring is then modeled using the standard linearized backaction expression $\Omega_{\mathrm{eff}}(\Delta)$, while the damping is compared to the corresponding $\Gamma_{\mathrm{eff}}(\Delta)$ computed using the same parameters obtained from the spring fit. In practice, the frequency shift provides the more robust and reliable observable for extracting the local coupling strength, while the linewidth is more strongly affected by technical instability close to resonance.

The first row of \cref{fig:FigS10_Backaction} reproduces the same dataset discussed in the main text. The left panel shows the calibrated displacement power spectral density as a function of normalized detuning $\Delta/\kappa$ and mechanical frequency. The thermomechanical peak shifts across the map due to the optical spring effect. The central panel shows the extracted mechanical resonance frequency. Blue points correspond to spectra that are well described by a single Lorentzian, while orange and red points indicate the lower and upper branches obtained when a two-Lorentzian description is required. The solid curve is the best fit to the optical-spring model, from which the local coupling strength and effective cavity linewidth are inferred. The optical power is obtained from an independent measurement of the cavity transmission. The right panel shows the corresponding apparent mechanical linewidth together with the damping curve computed from the same fitted parameters.

The second row of \cref{fig:FigS10_Backaction} shows a similar measurement performed at higher optical power and slightly farther from the node of the standing-wave field. This increases the local dispersive coupling and therefore enhances the backaction effect. At the same time, however, it also increases the instability of the system near cavity resonance, where the transmitted signal becomes more sensitive to fluctuations of the instantaneous detuning. As a result, the spectra more frequently develop broadened and split double-peak structures, and the extracted linewidth becomes more strongly distorted by averaging over fluctuating detuning configurations. This behavior is visible in the lower-row frequency and linewidth panels, where the qualitative spring and damping trends remain evident, but the quantitative agreement with the minimal model is less clean than in the weaker-coupling dataset.

These additional measurements support the interpretation adopted in the main text. The optical spring provides a robust dynamical confirmation of substantial radiation-pressure coupling in the device and yields values of the local coupling that are consistent with the static characterization. By contrast, very close to cavity resonance, especially at stronger coupling, the apparent linewidth is significantly affected by lock instability, thermally induced detuning fluctuations, and mode splitting, and should therefore be interpreted more cautiously.

\newpage
\section{OMIT in the USR}
\label{sec:SI_OMIT_USR}
In this section we summarize the theoretical framework used to interpret the optomechanically induced transparency (OMIT) measurements of the main text for an optomechanical system operating in the unresolved-sideband regime (USR), $\kappa \gg \Omega_m$. Starting from the standard radiation-pressure Hamiltonian and adopting the conventions of \cite{aspelmeyerCavityOptomechanics2014,bowenQuantumOptomechanics2015a}, we derive the linearized probe response while retaining both Stokes and anti-Stokes sidebands, following the general strategy of \cite{weisOptomechanicallyInducedTransparency2010}. We then compare our formulation to the commonly used expressions of \cite{weisOptomechanicallyInducedTransparency2010,agarwalElectromagneticallyInducedTransparency2010,yanOptomechanicallyInducedOptical2021}, discuss the role of the additional sideband rotating-wave approximation and the effect of thermal broadening across different sideband-resolution regimes, and finally connect the intracavity solution to the experimentally measured demodulated response through input-output theory and the specific measurement geometry used here.

\subsection{Model derivation}
We derive here the OMIT response relevant to our experiment starting from the standard optomechanical Hamiltonian in a frame rotating at the laser frequency $\omega_l$. After fixing notation and conventions, we linearize the coupled optical and mechanical equations of motion around a strong pump field and a weak probe, and solve them in frequency space while retaining both Stokes and anti-Stokes sidebands. This yields the full intracavity probe response appropriate to the USR, where the additional sideband rotating-wave approximation is not generally valid.
\subsubsection{Notation and Conventions}
\begin{itemize}
	\item  Operators and frames:
	\begin{itemize}
		\item Optical mode: $\hat{a}$, $\hat{a}^{\dagger}$ are the annihilation/creation operators of the driven cavity mode, written in a frame rotating at the laser frequency $\omega_l$.
		\item Mechanical mode: $\hat{b}$, $\hat{b}^{\dagger}$ are the phonon operators; displacement and momentum operators are $\hat{q} = x_{\mathrm{zpf}}(\hat{b} + \hat{b}^{\dagger})$ and $\hat{p}=ip_{\mathrm{zpf}}(\hat{b}^{\dagger}-\hat{b})$, with $x_{\mathrm{zpf}} = \sqrt{\frac{\hbar}{2m \Omega_m}}$ and $p_{\mathrm{zpf}} = \sqrt{\frac{\hbar m \Omega_m}{2}}$. A useful relation is: $\Omega_m x_{\mathrm{zpf}} = p_{\mathrm{zpf}} / m$. 
	\end{itemize}
	\item Frequencies and detuning:
	\begin{itemize}
		\item Bare cavity resonance: $\omega_{\mathrm{cav}}$; laser: $\omega_l$; mechanical resonance: $\Omega_m$.
		\item Detuning: $\Delta = \omega_l - \omega_{\mathrm{cav}}$
		\item Corrected detuning: $\bar{\Delta} = \Delta + G \bar{q}$
	\end{itemize}
	\item  Optical driving field:
	\begin{itemize}
		\item Coherent input field:  $a_{\mathrm{in}}(t)$ (photon-flux normalized, units $\sqrt{\mathrm{s}^{-1}}$)
		\item Pump + probe tones: $a_{\mathrm{in}}(t) = \bar{a}_{\mathrm{in}} + a_p e^{-i\Omega t}$, with $\Omega=\omega_p-\omega_l$
		\item Full input operator: $\hat{a}_{\mathrm{in}}(t) = a_{\mathrm{in}}(t) + \hat{a}_{\mathrm{vac}}(t)$
	\end{itemize}
	\item Decay rates:
	\begin{itemize}
		\item Cavity linewidth: $\kappa$
		\item External coupling: $\kappa_{\mathrm{ext}}$ (in our case of a symmetric fiber cavity: $\kappa_{\mathrm{ext}} = \kappa/2$) 
		\item Mechanical damping: $\Gamma_m$; mechanical quality factor: $Q = \Omega_m / \Gamma_m$.
	\end{itemize}
	\item Couplings:
	\begin{itemize}
		\item Dispersive frequency pull: $G = - \partial \omega_{\mathrm{cav}} / \partial q$
		\item Single-photon optomechanical coupling: $g_0 = G x_{\mathrm{zpf}}$
		\item Field-enhanced coupling: $g = g_0 \sqrt{\bar{n}_{\mathrm{cav}}}$, with $ \bar{n}_{\mathrm{cav}} = |\!\left<\hat{a}\right>\!|^2 = |\bar{a}|^2 = \frac{\kappa_{\mathrm{ext}}}{\left(\frac{\kappa}{2}\right)^2 + \bar{\Delta}^2} \frac{P_{\mathrm{in}}}{\hbar \omega_l} $ the intracavity photon number
	\end{itemize}
	
\end{itemize}

These conventions are used throughout the derivation below. In the later measurement section, where we discuss the experimentally generated EOM sidebands and the demodulated reflected signal, it is convenient to switch to a notation centered on the pump sideband branch; that change of variables will be introduced explicitly there.

\subsubsection{The Optomechanical Hamiltonian}
We model our system as a single optical cavity mode dispersively coupled to a single mechanical mode via radiation pressure. The total Hamiltonian that describes the system, in a frame rotating at $\omega_l$ and with $\Delta = \omega_l - \omega_{\mathrm{cav}}$, reads:
\begin{equation}
	\label{eq:H_RWA}
	\hat{H}_{\mathrm{RWA}} =	- \hbar \Delta \hat{a}^{\dagger}\hat{a}
	+ \hbar \Omega_m \hat{b}^{\dagger}\hat{b}
	- \hbar g_0 \hat{a}^{\dagger}\hat{a} \left( \hat{b}^{\dagger}+\hat{b} \right)
	+ i\hbar \sqrt{\kappa_{\mathrm{ext}}} \left( a_{\mathrm{in}}(t) \hat{a}^{\dagger} - a_{\mathrm{in}}^*(t) \hat{a} \right)
\end{equation}
The first term is the optical field. The second describes the mechanical resonator:
\begin{equation}
	\hat{H}_{\mathrm{mech}} = \hbar \Omega_m \hat{b}^{\dagger}\hat{b} = \frac{\hat{p}^2}{2m} + \frac{1}{2} m \Omega_m^2 \hat{q}^2
\end{equation}
The third term describes the dispersive optomechanical interaction, which originates from the position dependence of the cavity frequency, $\omega_{\mathrm{cav}}(q) \approx \omega_{\mathrm{cav}} - G q$, with $G = - \partial \omega_{\mathrm{cav}} / \partial q$. To leading order this gives the standard radiation-pressure coupling
\begin{equation}
	\hat{H}_{\mathrm{int}} = -\hbar G \hat{q} \hat{a}^\dagger \hat{a} = -\hbar g_0 \hat{a}^\dagger \hat{a} (\hat{b}+\hat{b}^\dagger),
\end{equation}
where $g_0 = G x_{\mathrm{zpf}}$ is the single-photon optomechanical coupling rate and $x_{\mathrm{zpf}} = \sqrt{\frac{\hbar}{2m \Omega_m}}$ is the zero-point displacement.
Finally, the optical mode is driven coherently through the external port at rate $\kappa_{\mathrm{ext}}$ by an input field $a_{\mathrm{in}}(t)$ normalized to photon flux,
\begin{equation}
	\hat{H}_{\mathrm{drive}} = i\hbar \sqrt{\kappa_{\mathrm{ext}}} \left( a_{\mathrm{in}}(t) \hat{a}^{\dagger} - a_{\mathrm{in}}^*(t) \hat{a} \right)
\end{equation}
Note that we decompose the input operator as $\hat{a}_{\mathrm{in}}(t) = a_{\mathrm{in}}(t) + \hat{a}_{\mathrm{vac}}(t)$; $a_{\mathrm{in}}(t)$ is the coherent drive (pump + probe), and $\hat{a}_{\mathrm{vac}}$ represents quantum noise entering through the optical ports.

\subsubsection{Quantum Markovian Langevin equation and Linearization}
To obtain equations of motion that include both coherent dynamics and damping, we start from the rotating‑frame Hamiltonian \eqref{eq:H_RWA} and use the Heisenberg–Langevin formalism \cite{bowenQuantumOptomechanics2015a}. Coupling the cavity and the mechanics to their environments leads to loss terms with rates $\kappa$ and $\Gamma_m$, and to corresponding input noise operators that continuously replenish fluctuations ($\hat{a}_{\mathrm{vac}}(t)$ and $\hat{F}(t)$). Assuming the standard Markov approximation for the bath coupling, we can derive the quantum Markovian Langevin equation for a general observable $\hat{O}$.
\paragraph{Equation of motion}
We will be using the "rotating wave" Langevin equation for the optical field:
\begin{equation}
	\dot{\hat{O}} =	  \frac{1}{i\hbar} \left[ \hat{O}, \hat{H}_{\mathrm{RWA}} \right]
	- \left[ \hat{O}, \hat{a}^{\dagger} \right] \left( \frac{\kappa}{2}\hat{a} -\sqrt{\kappa} \hat{a}_{\mathrm{vac}}(t) \right)
	+ \left( \frac{\kappa}{2}\hat{a}^{\dagger} -\sqrt{\kappa} \hat{a}^{\dagger}_{\mathrm{vac}}(t) \right) \left[ \hat{O}, \hat{a} \right] ,
\end{equation}
while we will use the general Markov Langevin equation for the mechanical oscillator:
\begin{equation}
	\dot{\hat{O}} =	  \frac{1}{i\hbar} \left[ \hat{O}, \hat{H}_{\mathrm{RWA}} \right]
	- \frac{1}{i\hbar} \left[ \hat{O}, x_{\mathrm{zpf}} \left( \hat{b}^{\dagger} + \hat{b} \right) \right] \hat{F}(t)
	+ \frac{m}{2i\hbar} \left\{ \left[ \hat{O}, x_{\mathrm{zpf}} \left( \hat{b}^{\dagger} + \hat{b} \right) \right],
	\Gamma_m x_{\mathrm{zpf}} \left( \dot{\hat{b}}^{\dagger} + \dot{\hat{b}} \right) \right\}_+ .
\end{equation}
Solving these equations yields the equations for the open-system dynamics of our optomechanical system. For the optical field (photon) operators we obtain:
\begin{equation}
	\label{eq:phot_eq_motion}
	\begin{cases}
		\dot{\hat{a}} 			&\kern-0.5em =	\; + i\Delta\hat{a}
		+ i g_0 \hat{a} \left( \hat{b}^{\dagger} + \hat{b} \right) 
		+ \sqrt{\kappa_{\mathrm{ext}}} a_{\mathrm{in}} (t) - \frac{\kappa}{2} \hat{a} + \sqrt{\kappa} \hat{a}_{\mathrm{vac}} \\
		
		\dot{\hat{a}}^{\dagger}	&\kern-0.5em =	\; - i\Delta\hat{a}^{\dagger} 
		- i g_0 \hat{a}^{\dagger} \left( \hat{b}^{\dagger} + \hat{b} \right)
		+ \sqrt{\kappa_{\mathrm{ext}}} a_{\mathrm{in}}^* (t) - \frac{\kappa}{2} \hat{a}^{\dagger} + \sqrt{\kappa} \hat{a}^{\dagger}_{\mathrm{vac}} .\\
	\end{cases}
\end{equation}
For the phonon operators we get:
\begin{equation}
	\label{eq:phon_eq_motion}
	\begin{cases}
		\dot{\hat{b}} 			&\kern-0.5em =	\, - i\Omega_m \hat{b}
		+ i g_0 \hat{a}^{\dagger}\hat{a}
		+ \dfrac{i x_{\mathrm{zpf}}}{\hbar} \hat{F}(t)
		+ \dfrac{m \Gamma_m x^2_{\mathrm{zpf}}}{2i \hbar} 2 \left( \dot{\hat{b}}^{\dagger} + \dot{\hat{b}} \right) \\
		
		\dot{\hat{b}}^{\dagger}	&\kern-0.5em \, =	+ i\Omega_m \hat{b}^{\dagger}
		- i g_0 \hat{a}^{\dagger}\hat{a}
		- \dfrac{i x_{\mathrm{zpf}}}{\hbar} \hat{F}(t)
		- \dfrac{\left( \dot{\hat{b}}^{\dagger} + \dot{\hat{b}} \right)}{2iQ} \\
	\end{cases}
\end{equation}
The latter can also be expressed in terms of the displacement and momentum operators:
\begin{equation}
	\begin{cases}
		\dot{\hat{q}}	&\kern-0.5em =	\; x_{\mathrm{zpf}}  \left( \dot{\hat{b}}^{\dagger} + \dot{\hat{b}} \right) =
		i \Omega_m x_{\mathrm{zpf}} \left( \hat{b}^{\dagger} - \hat{b} \right) = \hat{p} / m \\
		\dot{\hat{p}}	&\kern-0.5em = \; i p_{\mathrm{zpf}}  \left( \dot{\hat{b}}^{\dagger} - \dot{\hat{b}} \right) =
		i p_{\mathrm{zpf}} \left[ i \Omega_m \left( \hat{b}^{\dagger} + \hat{b} \right) -2ig_0\hat{a}^{\dagger}\hat{a}
		- \dfrac{2i x_{\mathrm{zpf}}}{\hbar} \hat{F}(t) - \Gamma_m  \left( \hat{b}^{\dagger} - \hat{b} \right) \right] = \\
		&\kern-0.5em = \; - m \Omega^2_m \hat{q} + \hbar G \hat{a}^{\dagger}\hat{a} + \hat{F}(t) - \Gamma_m \hat{p} .
	\end{cases}
\end{equation}
\paragraph{Linearization}
Because we operate in a regime in which we can consider our coherent driving to be suitably "strong", the system dynamics can be well approximated by a linearized description. This means that we can consider only the small fluctuations around the "displaced" semiclassical steady states of both the intracavity field and the mechanical position. Mathematically this is equivalent to performing the following substitution into our equation of motion:
\begin{equation}
	\quad \hat{a} \rightarrow \bar{a} + \delta\hat{a}, 
	\quad \hat{a}^{\dagger} \rightarrow \bar{a}^* + \delta\hat{a}^{\dagger},
	\quad \hat{b} \rightarrow \bar{b} + \delta\hat{b},
	\quad \hat{b}^{\dagger} \rightarrow \bar{b}^* + \delta\hat{b}^{\dagger}
	\quad \left( \text{or } \hat{q} \rightarrow \bar{q} + \delta\hat{q} \right).
\end{equation}
We can assume without loss of generality for both $\bar{a}$ and $\bar{b}$ to be real, which is equivalent to provide a phase reference for the input and output fields.
In a similar way we can also rewrite the coherent driving field, in a way that will allow us to more easily separate the pump and probe fields in the following:
\begin{equation}
	a_{\mathrm{in}}(t) \rightarrow \bar{a}_{\mathrm{in}} + \delta a_{\mathrm{in}}(t)
\end{equation}
Note that while the expansion of $\hat{b}$ will translate into a static displacement that we will later incorporate into the cavity detuning (see the \textit{"Solving coupled OM equations"} section below), the expansion of $\hat{a}$ can be used to express the Hamiltonian in its standard linearized form. Often this is the starting point before solving the EOM. Expressing the Hamiltonian this way allows to take the full linearized interaction ($\propto (\delta\hat{a}+\delta\hat{a}^{\dagger})(\delta\hat{b}+\delta\hat{b}^{\dagger})$) and to clearly separate both beam-splitter-like and two-mode-squeezing like terms, thus both upper and lower optical sidebands. In our derivation we keep all terms automatically. Any further rotating-wave approximation (e.g. useful in the resolved-sideband limit) will be discussed explicitly later (see the \textit{"Rotating wave approximation"} section).

Starting from the annihilation operator in \cref{eq:phot_eq_motion} we obtain:
\begin{equation}
	\begin{aligned}
		\dot{\bar{a}} + \delta\dot{\hat{a}} &	\begin{multlined}[t]
			= i\Delta (\bar{a} + \delta\hat{a}) 
			+ ig_0 (\bar{a} + \delta\hat{a}) (\bar{b} + \bar{b}^* + \delta\hat{b} + \delta\hat{b}^{\dagger}) \\
			\shoveleft[0.32\linewidth] + \sqrt{\kappa_{\mathrm{ext}}} (\bar{a}_{\mathrm{in}} + \delta a_{\mathrm{in}}(t))
			- \frac{\kappa}{2} (\bar{a} + \delta\hat{a})
			+ \sqrt{\kappa} \hat{a}_{\mathrm{vac}}(t) 
		\end{multlined}\\
		& 
		\begin{multlined}[t]
			= i\Delta \bar{a} + i\Delta \delta\hat{a}
			+ ig_0 \bar{a}(\bar{b} + \bar{b}^*) + ig_0 \delta\hat{a}(\bar{b} + \bar{b}^*) 
			+ ig_0 \bar{a}(\delta\hat{b} + \delta\hat{b}^{\dagger}) + ig_0 \delta\hat{a}(\delta\hat{b} + \delta\hat{b}^{\dagger}) \\
			\shoveleft[0.32\linewidth] + \sqrt{\kappa_{\mathrm{ext}}} \left( \bar{a}_{\mathrm{in}} + \delta a_{\mathrm{in}}(t) \right)
			- \frac{\kappa}{2} \left( \bar{a} + \delta\hat{a} \right) + \sqrt{\kappa}\hat{a}_{\mathrm{vac}}(t) 
		\end{multlined}\\
	\end{aligned}
\end{equation}
Separating the steady state from the fluctuations terms we can extract:	
\begin{equation}
	\begin{aligned}
		\dot{\bar{a}}	= 0 = i\Delta \bar{a} &+ ig_0 \bar{a}(\bar{b} + \bar{b}^*) 
		+ \sqrt{\kappa_{\mathrm{ext}}} \bar{a}_{\mathrm{in}} - \frac{\kappa}{2}\bar{a} \\
		&	\quad \Rightarrow \quad 
		\bar{a} =	\frac{\sqrt{\kappa_{\mathrm{ext}}} \bar{a}_{\mathrm{in}}}{\frac{\kappa}{2}
			- i \left[ \Delta + g_0(\bar{b} + \bar{b}^*) \right]} 
		=	\frac{\sqrt{\kappa_{\mathrm{ext}}} \bar{a}_{\mathrm{in}}}{\frac{\kappa}{2}
			- i \left( \Delta + G \bar{q} \right)}\\
	\end{aligned}
\end{equation}
The fluctuation term is instead (ignoring second order terms):
\begin{equation}
	\label{eq:a_eq_motion}
	\begin{aligned}
		\delta\dot{\hat{a}}	&=	i\Delta \delta\hat{a} + ig_0 \delta\hat{a}(\bar{b} + \bar{b}^*) + ig_0 \bar{a}(\delta\hat{b} + \delta\hat{b}^{\dagger})
		+ \sqrt{\kappa_{\mathrm{ext}}} \delta a_{\mathrm{in}}(t) - \frac{\kappa}{2}\delta\hat{a} + \sqrt{\kappa}\hat{a}_{\mathrm{vac}}(t) \\
		&= \left[ i \left( \Delta + g_0(\bar{b} + \bar{b}^*)\right) - \frac{\kappa}{2} \right] \delta\hat{a} 
		+ ig_0 \bar{a}(\delta\hat{b} + \delta\hat{b}^{\dagger}) + \sqrt{\kappa_{\mathrm{ext}}} \delta a_{\mathrm{in}}(t)
		+ \sqrt{\kappa}\hat{a}_{\mathrm{vac}}(t) \\
		&= \left[ i \left( \Delta + G \bar{q}\right) - \frac{\kappa}{2} \right] \delta\hat{a} + i G \bar{a}\delta\hat{q}
		+ \sqrt{\kappa_{\mathrm{ext}}} \delta a_{\mathrm{in}}(t)	+ \sqrt{\kappa}\hat{a}_{\mathrm{vac}}(t) \\
	\end{aligned}
\end{equation}
If we do the same for the creation operator in \cref{eq:phot_eq_motion} we obtain:
\begin{equation}
	\label{eq:a_dag_eq_motion}
	\begin{aligned}
		\delta\dot{\hat{a}}^{\dagger}	&=	-\left[i \left(\Delta + g_0(\bar{b} + \bar{b}^*) \right) + \frac{\kappa}{2}\right]\delta \hat{a}^{\dagger}
		- ig_0 \bar{a}(\delta \hat{b}^{\dagger} + \delta \hat{b}) + \sqrt{\kappa_{\mathrm{ext}}} \delta a_{\mathrm{in}}^*(t)
		+ \sqrt{\kappa}\hat{a}_{\mathrm{vac}}(t) \\
		&=	-\left[ i \left( \Delta + G \bar{q}\right) + \frac{\kappa}{2} \right] \delta\hat{a}^{\dagger}
		- i G \bar{a}\delta\hat{q} + \sqrt{\kappa_{\mathrm{ext}}} \delta a_{\mathrm{in}}^*(t) 
		+ \sqrt{\kappa}\hat{a}^{\dagger}_{\mathrm{vac}}(t) \\
	\end{aligned}
\end{equation}
If we now move to the phonon annihilation operator in \cref{eq:phon_eq_motion} and perform the linearization we can write:
\begin{equation}
	\begin{aligned}
		\dot{\bar{b}} + \delta\dot{\hat{b}}	&=	-i\Omega_m (\bar{b} + \delta\hat{b}) + ig_0 (\bar{a} + \delta\hat{a}^{\dagger})(\bar{a} + \delta\hat{a})
		+ \frac{i x_{\mathrm{zpf}}}{\hbar} \hat{F}(t) + \frac{\Gamma_m}{2} (\delta\hat{b}^{\dagger} - \delta\hat{b}) \\
		&=	-i\Omega_m \bar{b} - i\Omega_m \delta\hat{b} + ig_0 |\bar{a}|^2 
		+ ig_0 \bar{a} ( \delta\hat{a}^{\dagger} + \delta\hat{a} )+ ig_0 \delta\hat{a}^{\dagger}\delta\hat{a}
		+ \frac{i x_{\mathrm{zpf}}}{\hbar} \hat{F}(t) + \frac{\Gamma_m}{2} (\delta\hat{b}^{\dagger} - \delta\hat{b}) \\
	\end{aligned}
\end{equation}
Again, separating the steady state from the fluctuations terms we can extract:
\begin{equation}
	\begin{aligned}
		\dot{\bar{b}}	=	0 = -i\Omega_m \bar{b} + ig_0 |\bar{a}|^2 
		&	\quad \Rightarrow \quad
		\bar{b} = \frac{g_0 |\bar{a}|^2}{\Omega_m} \\
		&	\quad \Rightarrow \quad
		x_{\mathrm{zpf}} (\bar{b} + \bar{b}^*) = \bar{q} = \frac{2 x_{\mathrm{zpf}}^2 G |\bar{a}|^2 }{\Omega_m} =
		\frac{\hbar G |\bar{a}|^2 }{m \Omega_m^2}
	\end{aligned}
\end{equation}
while for the fluctuations we have:
\begin{equation}
	\label{eq:b_eq_motion}
	\begin{cases}
		\delta\dot{\hat{b}}	&\kern-0.5em =	\;	-i\Omega_m \delta\hat{b} + ig_0 \bar{a} ( \delta\hat{a}^{\dagger} + \delta\hat{a}) 
		+ \dfrac{i x_{\mathrm{zpf}}}{\hbar} \hat{F}(t) + \dfrac{\Gamma_m}{2} (\delta\hat{b}^{\dagger} - \delta\hat{b}) \\
		\delta\dot{\hat{b}}^{\dagger}	&\kern-0.5em =	\;	+i\Omega_m \delta\hat{b}^{\dagger} - ig_0 \bar{a} ( \delta\hat{a}^{\dagger} + \delta\hat{a}) 
		- \dfrac{i x_{\mathrm{zpf}}}{\hbar} \hat{F}(t) - \dfrac{\Gamma_m}{2} (\delta\hat{b}^{\dagger} - \delta\hat{b}) \\
	\end{cases}
\end{equation}
Rewriting the latter equations in terms of the displacement and momentum operator:
\begin{equation}
	\begin{cases}
		\delta\dot{\hat{q}}	&\kern-0.5em =	\;	\frac{\delta\hat{p}}{m} \\
		\delta\dot{\hat{p}}	&\kern-0.5em =	\; 	i p_{\mathrm{zpf}} (\delta\dot{\hat{b}}^{\dagger} - \delta\dot{\hat{b}}) =
		-m \Omega_m^2 \delta\hat{q} + \hbar G \bar{a} (\delta\hat{a}^{\dagger} + \delta\hat{a})
		+ \hat{F}(t) - \Gamma_m \delta \hat{p}
	\end{cases}
\end{equation}
Finally, these can be combined into one single equation:
\begin{equation}
	\label{eq:q_eq_motion}
	\delta\ddot{\hat{q}} + \Gamma_m \delta\dot{\hat{q}} + \Omega^2_m \delta\hat{q} =
	\frac{\hbar G}{m} \bar{a} (\delta\hat{a}^{\dagger} + \delta\hat{a}) + \hat{F}(t)
\end{equation}
With the quantum Langevin equations established and linearized about the steady states, we can now move to solve them in the frequency domain.

\subsubsection{Solving the coupled OM equations in frequency space (phonon "b" frame)}
\label{subsec:solving_EOM}
We can now solve our coupled equations in the frequency domain for a strong pump control laser and a weak probe tone. We will do so by applying the following:
\begin{itemize}
	\item we identify the operators with their expectation values: $y(t) = \left<\hat{y}(t)\right>$ ;
	\item we ignore the vacuum and thermal fluctuations ($\hat{a}_{\mathrm{vac}}(t)$, $\hat{F}(t)$); 
	\item we express the driving (probe) term as: $\delta a_{\mathrm{in}}(t) = a_p e^{-i\Omega t}$ with $\Omega = \omega_p - \omega_l$ ;
	\item we make use of the "corrected" detuning: $\bar{\Delta} = \Delta + G\bar{q}$ ;
	\item we make use of the following ANSATZ:
\end{itemize}
\begin{equation}
	\begin{cases}
		\delta a(t)		&=	A^- e^{-i\Omega t} + A^+ e^{+i\Omega t} \\
		\delta a^*(t)	&=	(A^+)^* e^{-i\Omega t} + (A^-)^* e^{+i\Omega t} \\
		\delta q(t)		&=	X e^{-i\Omega t} + X^* e^{+i\Omega t} \\
		\delta b(t)		&=	B^- e^{-i\Omega t} + B^+ e^{+i\Omega t} \\
	\end{cases}
\end{equation}
In the laboratory frame the term $A^- e^{-i\Omega t}$ corresponds to an intracavity field at $\omega_l+\Omega$ (upper, anti‑Stokes sideband), while $A^+ e^{+i\Omega t}$ corresponds to $\omega_l - \Omega$ (lower, Stokes sideband). For a probe at $\omega_p = \omega_l + \Omega$, $A^-$ is the intracavity amplitude at the probe frequency and $A^+$ is the ‘counter‑rotating’ lower sideband. In the resolved‑sideband regime and for red detuning, we will later show that $A^-$ and $A^+$ can be mapped directly to the anti‑Stokes and Stokes sideband amplitudes, and that $A^+$ is strongly suppressed and can be neglected. In contrast, in the USR both sidebands lie well within the cavity linewidth, so $A^+$ is no longer negligible and the coupling between the two sidebands via the mechanics (the ‘counter‑rotating’ contribution, encoded through $A^+$ and the associated terms) can no longer be ignored.

We start from the equation of motion \cref{eq:a_eq_motion,eq:a_dag_eq_motion,eq:b_eq_motion}:
\begin{subequations}
	\begin{empheq}[left=\empheqlbrace]{align}
		&\delta\dot{\hat{a}} =				\left[i(\Delta + 2 g_0\bar{b}) - \frac{\kappa}{2}\right]\delta \hat{a} + ig_0 \bar{a}(\delta \hat{b}^{\dagger} + \delta \hat{b})
		+ \sqrt{\kappa_{\mathrm{ext}}} \delta a_{\mathrm{in}}(t) + \sqrt{\kappa}\hat{a}_{\mathrm{vac}}(t)		\label{eq:a_eq_mot}\\
		&\delta\dot{\hat{a}}^{\dagger} =	-\left[i(\Delta + 2 g_0\bar{b}) + \frac{\kappa}{2}\right]\delta \hat{a}^{\dagger}
		- ig_0 \bar{a}(\delta \hat{b}^{\dagger} + \delta \hat{b}) + \sqrt{\kappa_{\mathrm{ext}}} \delta a_{\mathrm{in}}^*(t)
		+ \sqrt{\kappa}\hat{a}_{\mathrm{vac}}(t)														\label{eq:a_dag_eq_mot}
	\end{empheq}
\end{subequations}
\begin{subequations}
	\begin{empheq}[left=\empheqlbrace]{align}
		&\delta\dot{\hat{b}} =				-i\Omega_m \delta \hat{b} + ig_0 \bar{a}(\delta \hat{a}^{\dagger} + \delta \hat{a}) + i \frac{x_{\mathrm{zpf}}}{\hbar}\hat{F}(t)
		+ \frac{\Gamma_m}{2}(\delta \hat{b}^{\dagger} - \delta \hat{b}) 								\label{eq:b_eq_mot}\\
		&\delta\dot{\hat{b}}^{\dagger} =	+i\Omega_m \delta \hat{b}^{\dagger} - ig_0 \bar{a}(\delta \hat{a}^{\dagger} + \delta \hat{a}) - i \frac{x_{\mathrm{zpf}}}{\hbar}\hat{F}(t)
		- \frac{\Gamma_m}{2}(\delta \hat{b}^{\dagger} - \delta \hat{b})									\label{eq:b_dag_eq_mot}
	\end{empheq}
\end{subequations}
Applying the operations described above to \cref{eq:a_eq_mot} we get:
\begin{multline}
	-i\Omega A^- e^{-i\Omega t} + i\Omega A^+ e^{i\Omega t} = 
	\left(i\bar{\Delta} - \frac{\kappa}{2}\right)\left(A^- e^{-i\Omega t} + A^+ e^{i\Omega t}\right) \\
	+ ig_0 \bar{a}\left(B^- e^{-i\Omega t} + B^+ e^{i\Omega t} + (B^+)^* e^{-i\Omega t}	+ (B^-)^* e^{i\Omega t}\right) 
	+ \sqrt{\kappa_{\mathrm{ext}}} a_p e^{-i\Omega t} 
\end{multline}
We can then separate the contributions of the $\pm \Omega$ terms:
\begin{subequations}
	\begin{empheq}[left=\empheqlbrace]{align}
		&\left[\frac{\kappa}{2} - i(\bar{\Delta} + \Omega)\right]A^- = ig_0 \bar{a} \left( B^- + (B^+)^* \right) + \sqrt{\kappa_{\mathrm{ext}}} a_p		&; \quad e^{-i\Omega t} \label{eq:Am_eq_mot}\\
		&\left[\frac{\kappa}{2} - i(\bar{\Delta} - \Omega)\right]A^+ = ig_0 \bar{a} \left( B^+ + (B^-)^* \right)										&; \quad e^{+i\Omega t} \label{eq:Ap_eq_mot}
	\end{empheq}
\end{subequations}
We can do the same for \cref{eq:a_dag_eq_mot}:
\begin{multline}
	-i\Omega (A^+)^* e^{-i\Omega t} + i\Omega (A^-)^* e^{i\Omega t} =
	-\left(\frac{\kappa}{2} + i\bar{\Delta}\right)\left((A^+)^* e^{-i\Omega t} + (A^-)^* e^{i\Omega t}\right) \\
	- ig_0 \bar{a}\left(B^- e^{-i\Omega t} + B^+ e^{i\Omega t} + (B^+)^* e^{-i\Omega t} + (B^-)^* e^{i\Omega t}\right)
	+ \sqrt{\kappa_{\mathrm{ext}}} a_p^* e^{i\Omega t}
\end{multline}
Again, separating the contributions of the $\pm \Omega$ terms:
\begin{subequations}
	\begin{empheq}[left=\empheqlbrace]{align}
		&\left[\frac{\kappa}{2} + i(\bar{\Delta} - \Omega)\right](A^+)^* = -ig_0 \bar{a}\left(B^- + (B^+)^*\right) 										&; \quad e^{-i\Omega t} \label{eq:Apcc_eq_mot}\\
		&\left[\frac{\kappa}{2} + i(\bar{\Delta} + \Omega)\right](A^-)^* = -ig_0 \bar{a}\left(B^+ + (B^-)^*\right) + \sqrt{\kappa_{\mathrm{ext}}} a_p^*	&; \quad e^{+i\Omega t} \label{eq:Amcc_eq_mot}
	\end{empheq}
\end{subequations}
Now we repeat the same calculation for the phonon annihilation operator \cref{eq:b_eq_mot}:
\begin{multline}
	-i\Omega B^- e^{-i\Omega t} + i\Omega B^+ e^{i\Omega t} =
	-i\Omega_m\left(B^- e^{-i\Omega t} + B^+ e^{i\Omega t}\right) \\
	+ ig_0 \bar{a}\left(A^- e^{-i\Omega t} + A^+ e^{i\Omega t} + (A^+)^* e^{-i\Omega t} + (A^-)^* e^{i\Omega t}\right) \\
	+ \frac{\Gamma_m}{2}\left[ - B^- e^{-i\Omega t} - B^+ e^{i\Omega t} + (B^+)^* e^{-i\Omega t} + (B^-)^* e^{i\Omega t} \right]
\end{multline}
\begin{subequations}
	\begin{empheq}[left=\empheqlbrace]{align}
		&\left[i(\Omega_m - \Omega) + \frac{\Gamma_m}{2}\right]B^- - \frac{\Gamma_m}{2}(B^+)^* = ig_0 \bar{a}\left[A^- + (A^+)^*\right]	&; \quad e^{-i\Omega t} \label{eq:Bm_eq_mot}\\
		&\left[i(\Omega_m + \Omega) + \frac{\Gamma_m}{2}\right]B^+ - \frac{\Gamma_m}{2}(B^-)^* = ig_0 \bar{a}\left[A^+ + (A^-)^*\right]	&; \quad e^{+i\Omega t} \label{eq:Bp_eq_mot}
	\end{empheq}
\end{subequations}
and for the  phonon creation operator \cref{eq:b_dag_eq_mot}:
\begin{multline}
	-i\Omega (B^+)^* e^{-i\Omega t} + i\Omega (B^-)^* e^{i\Omega t} = +i\Omega_m\left[(B^+)^* e^{-i\Omega t} + (B^-)^* e^{i\Omega t}\right] \\
	- ig_0 \bar{a}\left[A^- e^{-i\Omega t} + A^+ e^{i\Omega t} + (A^+)^* e^{-i\Omega t} + (A^-)^* e^{i\Omega t}\right] \\
	- \frac{\Gamma_m}{2}\left[- B^- e^{-i\Omega t} - B^+ e^{i\Omega t} + (B^+)^* e^{-i\Omega t} + (B^-)^* e^{i\Omega t} \right]
\end{multline}
\begin{subequations}
	\begin{empheq}[left=\empheqlbrace]{align}
		&\left[-i(\Omega_m + \Omega) + \frac{\Gamma_m}{2}\right](B^+)^* - \frac{\Gamma_m}{2}B^- = -ig_0 \bar{a}\left[A^- + (A^+)^*\right]	&; \quad e^{-i\Omega t} \label{eq:Bpcc_eq_mot}\\
		&\left[-i(\Omega_m - \Omega) + \frac{\Gamma_m}{2}\right](B^-)^* - \frac{\Gamma_m}{2}B^+ = -ig_0 \bar{a}\left[A^+ + (A^-)^*\right]	&; \quad e^{+i\Omega t} \label{eq:Bmcc_eq_mot}
	\end{empheq}
\end{subequations}		
We can now use previous equations to derive an expression for $A^-$. We start by summing \cref{eq:Bm_eq_mot,eq:Bpcc_eq_mot} and we get an expression for $(B^+)^*$:
\begin{equation}
	i(\Omega_m - \Omega) B^- - i(\Omega_m + \Omega)(B^+)^* = 0
	\qquad \Rightarrow \qquad
	(B^+)^* = \frac{\Omega_m - \Omega}{\Omega_m + \Omega} B^-
\end{equation}
Summing \cref{eq:Am_eq_mot,eq:Apcc_eq_mot} instead, we obtain for $(A^+)^*$:
\begin{equation}
	\left[\frac{\kappa}{2} - i(\bar{\Delta} + \Omega)\right]A^- + \left[\frac{\kappa}{2} + i(\bar{\Delta} - \Omega)\right](A^+)^* = \sqrt{\kappa_{\mathrm{ext}}} a_p
	\quad \Rightarrow \quad
	(A^+)^* = \frac{\sqrt{\kappa_{\mathrm{ext}}} a_p - \left[\frac{\kappa}{2} - i(\bar{\Delta} + \Omega)\right]A^-}{\left[\frac{\kappa}{2} + i(\bar{\Delta} - \Omega)\right]}
\end{equation}
Substituting these last two equations into \cref{eq:Apcc_eq_mot} we can write for $B^-$:
\begin{equation}
	\left[\frac{\kappa}{2} + i(\bar{\Delta} - \Omega)\right](A^+)^* = -ig_0 \bar{a}\left(1 + \frac{\Omega_m - \Omega}{\Omega_m + \Omega} \right) B^-
	\quad \Rightarrow \quad
	B^- = \frac{\sqrt{\kappa_{\mathrm{ext}}} a_p - \left[\frac{\kappa}{2} - i(\bar{\Delta} + \Omega)\right]A^-}{-ig_0 \bar{a}\left(1 + \frac{\Omega_m - \Omega}{\Omega_m + \Omega} \right)}
\end{equation}
Finally we can plug the expressions for $B^-$, $(B^+)^*$ and $(A^+)^*$ into \cref{eq:Bm_eq_mot}:
\begin{equation}
	\begin{aligned}
		&\left[ i(\Omega_m - \Omega) +\frac{\Gamma_m}{2} -\frac{\Gamma_m}{2} \left(\frac{\Omega_m - \Omega}{\Omega_m + \Omega}\right) \right]
		\frac{\sqrt{\kappa_{\mathrm{ext}}} a_p - \left[\frac{\kappa}{2} - i(\bar{\Delta} + \Omega)\right]A^-}{-ig_0 \bar{a}\left(1 + \frac{\Omega_m - \Omega}{\Omega_m + \Omega} \right)} = 
		i g_0 \bar{a} \left[ \frac{\sqrt{\kappa_{\mathrm{ext}}} a_p + 2i\bar{\Delta}A^-}{\frac{\kappa}{2}+i(\bar{\Delta}-\Omega)} \right] \\
		\Rightarrow
		&\begin{multlined}[t]
			\left[i(\Omega_m - \Omega) + \frac{\Gamma_m}{2}\frac{2\Omega}{\Omega_m + \Omega}\right] \left[ \frac{\kappa}{2} +i(\bar{\Delta}- \Omega) \right]
			\left\{ \sqrt{\kappa_{\mathrm{ext}}} a_p - \left[\frac{\kappa}{2} - i(\bar{\Delta} + \Omega)\right]A^- \right\} = \\
			\shoveleft[0.6\linewidth]	g_0^2\bar{a}^2\left( \frac{2\Omega_m}{\Omega_m + \Omega}\right) \left[\sqrt{\kappa_{\mathrm{ext}}} a_p + 2i\bar{\Delta}A^-\right]
		\end{multlined} \\
		\Rightarrow
		&\begin{multlined}[t]
			\left\{
			\left[i(\Omega_m - \Omega) + \frac{\Gamma_m}{2}\frac{2\Omega}{\Omega_m + \Omega}\right]
			\left[\frac{\kappa}{2} + i(\bar{\Delta} - \Omega)\right]
			- g_0^2\bar{a}^2 \frac{2\Omega_m}{\Omega_m + \Omega}
			\right\} \sqrt{\kappa_{\mathrm{ext}}} a_p = \\
			\shoveleft[0.08\linewidth]	\left\{
			\left[i(\Omega_m - \Omega) + \frac{\Gamma_m}{2}\frac{2\Omega}{\Omega_m + \Omega}\right]
			\left[\frac{\kappa}{2} + i(\bar{\Delta} - \Omega)\right] \left[\frac{\kappa}{2} - i(\bar{\Delta} + \Omega)\right]
			+ 2ig_0^2\bar{a}^2\bar{\Delta} \frac{2\Omega_m}{\Omega_m + \Omega}
			\right\} A^-
		\end{multlined} \\
	\end{aligned}
\end{equation}
If now we multiply both the left and right hand term by $-i(\Omega_m + \Omega)$ we obtain:
\begin{multline}
	\left\{
	\left[ \Omega_m^2 - \Omega^2 -i\Gamma_m \Omega \right] \left[\frac{\kappa}{2} + i(\bar{\Delta} - \Omega)\right] + i g_0^2\bar{a}^2 2\Omega_m
	\right\} \sqrt{\kappa_{\mathrm{ext}}} a_p = \\
	\left\{
	\left[ \Omega_m^2 - \Omega^2 -i\Gamma_m \Omega \right] \left[\frac{\kappa}{2} + i(\bar{\Delta} - \Omega)\right] \left[\frac{\kappa}{2} - i(\bar{\Delta} + \Omega)\right]
	+ 2 g_0^2 \bar{a}^2 \bar{\Delta} 2\Omega_m
	\right\} A^-
\end{multline}
Using the following definitions:
\begin{equation}
	\chi_m (\Omega) = \frac{1}{m \left( \Omega_m^2 - \Omega^2 - i\Gamma_m \Omega \right)}
	\qquad
	g_0^2 = G^2 x_{\mathrm{zpf}}^2 = G^2 \frac{\hbar}{2m\Omega_m}
	\qquad
	\alpha = g_0^2 \bar{a}^2 2 m \Omega_m = \hbar G^2 \bar{a}^2
\end{equation}
where $\chi_m$ is the mechanical susceptibility, and multiplying both side by $m$, we can finally arrive to the following expression for $A^-$:
\begin{equation}
	\label{eq:Aminus}
	\boxed{	A^- =	\frac{\frac{1}{\chi_m} \left[ \frac{\kappa}{2} + i(\bar{\Delta}-\Omega) \right] + i\alpha}
		{\frac{1}{\chi_m} \left[ \left(\frac{\kappa}{2} - i\Omega \right)^2 + \bar{\Delta}^2 \right] + 2\bar{\Delta}\alpha}
		\sqrt{\kappa_{\mathrm{ext}}} a_p }
\end{equation}

\subsubsection{Solving the coupled OM equations in freq. space (displacement "q" frame)}
\label{subsec:q_frame}
We can perform the same derivation of $A^-$ using our OM equations expressed in the displacement "q" frame instead of the phonon "b" frame. This serves to check the consistency of our results and to more easily compare to other derivations in literature. This time we start from \cref{eq:a_eq_motion,eq:a_dag_eq_motion,eq:q_eq_motion}
\begin{subequations}
	\begin{empheq}[left=\empheqlbrace]{align}
		&\delta\dot{\hat{a}} =				\left[ i \left( \Delta + G \bar{q}\right) - \frac{\kappa}{2} \right] \delta\hat{a} + i G \bar{a}\delta\hat{q}
		+ \sqrt{\kappa_{\mathrm{ext}}} \delta a_{\mathrm{in}}(t)	+ \sqrt{\kappa}\hat{a}_{\mathrm{vac}}(t)		\label{eq:a_eq_mot2}\\
		&\delta\dot{\hat{a}}^{\dagger} =	-\left[ i \left( \Delta + G \bar{q}\right) + \frac{\kappa}{2} \right] \delta\hat{a}^{\dagger}
		- i G \bar{a}\delta\hat{q} + \sqrt{\kappa_{\mathrm{ext}}} \delta a_{\mathrm{in}}^*(t) 
		+ \sqrt{\kappa}\hat{a}^{\dagger}_{\mathrm{vac}}(t)												\label{eq:a_dag_eq_mot2}\\
		&\delta\ddot{\hat{q}} + \Gamma_m \delta\dot{\hat{q}} + \Omega^2_m \delta\hat{q} =
		\frac{\hbar G}{m} \bar{a} (\delta\hat{a}^{\dagger} + \delta\hat{a}) + \hat{F}(t)				\label{eq:q_dag_eq_mot}
	\end{empheq}
\end{subequations}
and using the same ANSATZ and approximation as before, we get for the $\delta\hat{a}$ operator:
\begin{multline}
	-i\Omega A^- e^{-i\Omega t} + i\Omega A^+ e^{i\Omega t} = \\
	\left(i\bar{\Delta} - \frac{\kappa}{2}\right)\left(A^- e^{-i\Omega t} + A^+	e^{i\Omega t}\right) 
	+ iG\bar{a}\left(X e^{-i\Omega t} + X^* e^{i\Omega t}\right) + \sqrt{\kappa_{\mathrm{ext}}} a_p e^{-i\Omega t}
\end{multline}
\begin{subequations}
	\begin{empheq}[left=\empheqlbrace]{align}
		&\left[\frac{\kappa}{2} - i(\bar{\Delta} + \Omega)\right]A^- = iG\bar{a}X + \sqrt{\kappa_{\mathrm{ext}}} a_p		&; \quad e^{-i\Omega t} \label{eq:Am_eq_mot2}\\
		&\left[\frac{\kappa}{2} - i(\bar{\Delta} - \Omega)\right]A^+ = iG\bar{a}X^*  										&; \quad e^{+i\Omega t} \label{eq:Ap_eq_mot2}
	\end{empheq}
\end{subequations}
similarly for the $\delta\hat{a}^{\dagger}$ operator:
\begin{multline}
	-i\Omega (A^+)^* e^{-i\Omega t} + i\Omega (A^-)^* e^{i\Omega t} = \\
	-\left(\frac{\kappa}{2} + i\bar{\Delta}\right) \left[ (A^+)^* e^{-i\Omega t} + (A^-)^* e^{i\Omega t} \right]
	- iG\bar{a}\left(X e^{-i\Omega t} + X^* e^{i\Omega t}\right) + \sqrt{\kappa_{\mathrm{ext}}} a_p^* e^{i\Omega t}
\end{multline}
\begin{subequations}
	\begin{empheq}[left=\empheqlbrace]{align}
		&\left[\frac{\kappa}{2} + i(\bar{\Delta} - \Omega)\right](A^+)^* = -iG\bar{a}X											&; \quad e^{-i\Omega t} \label{eq:Apcc_eq_mot2}\\
		&\left[\frac{\kappa}{2} + i(\bar{\Delta} + \Omega)\right](A^-)^* = -iG\bar{a}X^* + \sqrt{\kappa_{\mathrm{ext}}} a_p^*	&; \quad e^{+i\Omega t} \label{eq:Amcc_eq_mot2}
	\end{empheq}
\end{subequations}
and for the $\delta \hat{q}$ operator:
\begin{multline}
	-\Omega^2 X e^{-i\Omega t} - \Omega^2 X^* e^{i\Omega t} + \Gamma_m \left(-i\Omega X e^{-i\Omega t} + i\Omega X^* e^{i\Omega t}\right)
	+ \Omega_m^2\left(X e^{-i\Omega t} + X^* e^{i\Omega t}\right) = \\
	\frac{\hbar G}{m}\bar{a}\left[ \left( A^- + (A^+)^* \right) e^{-i\Omega t} + \left( A^+ + (A^-)^* \right) e^{i\Omega t} \right]
\end{multline}
\begin{subequations}
	\begin{empheq}[left=\empheqlbrace]{align}
		&m \left( \Omega_m^2 - \Omega^2 - i\Gamma_m \Omega \right) X = \hbar G \bar{a} \left[ A^- + (A^+)^* \right]			&; \quad e^{-i\Omega t} \label{eq:X_eq_mot}\\
		&m \left( \Omega_m^2 - \Omega^2 + i\Gamma_m \Omega \right) X^* = \hbar G \bar{a} \left[ A^+ + (A^-)^* \right]		&; \quad e^{+i\Omega t} \label{eq:Xcc_eq_mot}
	\end{empheq}
\end{subequations}
From \cref{eq:Apcc_eq_mot2} we can now derive
\begin{equation}
	(A^+)^* = -\frac{iG\bar{a}X}{\left[\frac{\kappa}{2} + i(\bar{\Delta} - \Omega)\right]}
\end{equation}
and then from \cref{eq:X_eq_mot}, using again the mechanical susceptibility $\chi_m$, we get:
\begin{equation}
	\label{eq:x_chi_m}
	\frac{X}{\chi_m} = \hbar G \bar{a} A^- - \frac{i \hbar G^2 \bar{a}^2 X}{\left[\frac{\kappa}{2} + i(\bar{\Delta} - \Omega)\right]}
\end{equation}
Rewriting this equation as a function of $A^-$ we have:
\begin{equation}
	\left[ \frac{\kappa}{2} + i(\bar{\Delta}-\Omega) \right] \frac{X}{\chi_m} + i\hbar G^2 \bar{a}^2 X =
	\left[ \frac{\kappa}{2} + i(\bar{\Delta}-\Omega) \right] \hbar G \bar{a} A^-
	\quad \Rightarrow \quad
	X = \frac{\left[ \frac{\kappa}{2} + i(\bar{\Delta}-\Omega) \right] \hbar G \bar{a}}{\frac{1}{\chi_m} \left[ \frac{\kappa}{2} + i(\bar{\Delta}-\Omega) \right] + i\hbar G^2 \bar{a}^2} A^-
\end{equation}
We can then substitute this expression in \cref{eq:Am_eq_mot2} to obtain:
\begin{equation}
	\begin{aligned}
		&\begin{multlined}[t]
			\frac{1}{\chi_m} \left[ \frac{\kappa}{2} + i(\bar{\Delta}-\Omega) \right] \left[\frac{\kappa}{2} - i(\bar{\Delta} + \Omega)\right] A^-
			+ \left[\frac{\kappa}{2} - i(\bar{\Delta} + \Omega)\right] i\hbar G^2 \bar{a}^2 A^- = \\
			\shoveleft[0.2\linewidth] \left[ \frac{\kappa}{2} + i(\bar{\Delta}-\Omega) \right] i \hbar G^2 \bar{a}^2 A^-
			+ \sqrt{\kappa_{\mathrm{ext}}} a_p \left\{ \frac{1}{\chi_m} \left[ \frac{\kappa}{2} + i(\bar{\Delta}-\Omega) \right] + i\hbar G^2 \bar{a}^2 \right\}
		\end{multlined} \\
		\Rightarrow
		&\begin{multlined}[t]
			\frac{1}{\chi_m} \left[ \frac{\kappa^2}{4} -i\frac{\kappa}{2}\bar{\Delta} -i\frac{\kappa}{2}\Omega 
			+i\frac{\kappa}{2}\bar{\Delta} -i\frac{\kappa}{2}\Omega + \left( \bar{\Delta}^2 - \Omega^2 \right) \right] A^- \\
			\shoveleft[0.2\linewidth] + \left[ \frac{\kappa}{2} -i\bar{\Delta} -i\Omega -\frac{\kappa}{2} -i\bar{\Delta} + i\Omega \right] i \hbar G^2 \bar{a}^2 A^- =\\
			\shoveleft[0.4\linewidth] \sqrt{\kappa_{\mathrm{ext}}} a_p \left\{ \frac{1}{\chi_m} \left[ \frac{\kappa}{2} + i(\bar{\Delta}-\Omega) \right] + i\hbar G^2 \bar{a}^2 \right\}
		\end{multlined} \\
		\Rightarrow
		&\frac{1}{\chi_m} \left[ \frac{\kappa^2}{4} -i\kappa\Omega + \left( \bar{\Delta}^2 - \Omega^2 \right) \right] A^- + 2\bar{\Delta}\hbar G^2 \bar{a}^2 A^- =
		\sqrt{\kappa_{\mathrm{ext}}} a_p \left\{ \frac{1}{\chi_m} \left[ \frac{\kappa}{2} + i(\bar{\Delta}-\Omega) \right] + i\hbar G^2 \bar{a}^2 \right\} \\
		\Rightarrow
		&\frac{1}{\chi_m} \left[ \left( \frac{\kappa}{2} - i\Omega \right)^2 + \bar{\Delta}^2 \right] A^- + 2 \bar{\Delta} \hbar G^2 \bar{a}^2 A^- = 
		\sqrt{\kappa_{\mathrm{ext}}} a_p \left\{ \frac{1}{\chi_m} \left[ \frac{\kappa}{2} + i(\bar{\Delta}-\Omega) \right] + i\hbar G^2 \bar{a}^2 \right\}
	\end{aligned}
\end{equation}
Finally using the same definition $\alpha = \hbar G^2 \bar{a}^2$ as before, we get to the same expression for $A^-$ as in \cref{eq:Aminus}:
\begin{equation}
	\label{eq:Aminus2}
	\boxed{	A^- =	\frac{\frac{1}{\chi_m} \left[ \frac{\kappa}{2} + i(\bar{\Delta}-\Omega) \right] + i\alpha}
		{\frac{1}{\chi_m} \left[ \left(\frac{\kappa}{2} - i\Omega \right)^2 + \bar{\Delta}^2 \right] + 2\bar{\Delta}\alpha}
		\sqrt{\kappa_{\mathrm{ext}}} a_p }
\end{equation}

Following an analogous procedure we can also derive the expression for $A^+$. From \cref{eq:Amcc_eq_mot2} we have
\begin{equation}
	(A^-)^* = \frac{-iG\bar{a}X^* + \sqrt{\kappa_{\mathrm{ext}}} a_p^*}{\left[\frac{\kappa}{2} + i(\bar{\Delta} + \Omega)\right]}
\end{equation}
Combining this with \cref{eq:Xcc_eq_mot} it follows
\begin{equation}
	\label{eq:x_chi}
	\frac{X^*}{\chi^*_m}	= \hbar G \bar{a} \left[ A^+ + (A^-)^* \right]
	= \hbar G \bar{a} A^+ - \frac{i\hbar G^2 \bar{a}^2 X^*}{\left[\frac{\kappa}{2} + i(\bar{\Delta} + \Omega)\right]} 
	+ \frac{\hbar G \bar{a} \sqrt{\kappa_{\mathrm{ext}}} a_p^*}{\left[\frac{\kappa}{2} + i(\bar{\Delta} + \Omega)\right]}
\end{equation}
where we define $\chi_m^*$ as:
\begin{equation}
	\chi_m^* (\Omega) = \frac{1}{m \left( \Omega_m^2 - \Omega^2 + i\Gamma_m \Omega \right)}
\end{equation}
Rewriting this equation as a function of $A^+$ we have:
\begin{multline}
	\left\{ \left[\frac{\kappa}{2} + i(\bar{\Delta} + \Omega)\right] + i\hbar G^2 \bar{a}^2 \chi^*_m  \right\} X^*
	= \chi^*_m \hbar G \bar{a} \left[\frac{\kappa}{2} + i(\bar{\Delta} + \Omega)\right] A^+ + \chi^*_m \hbar G \bar{a} \sqrt{\kappa_{\mathrm{ext}}} a_p^* \\
	\Rightarrow \quad X^* = \frac{\chi^*_m \hbar G \bar{a} \left[\frac{\kappa}{2} + i(\bar{\Delta} + \Omega)\right]}
	{\left[\frac{\kappa}{2} + i(\bar{\Delta} + \Omega)\right] + i\hbar G^2 \bar{a}^2 \chi^*_m} A^+
	+ \frac{\chi^*_m \hbar G \bar{a} \sqrt{\kappa_{\mathrm{ext}}} a_p^*}
	{\left[\frac{\kappa}{2} + i(\bar{\Delta} + \Omega)\right] + i\hbar G^2 \bar{a}^2 \chi^*_m}
\end{multline}
Lastly, substituting this expression in \cref{eq:Ap_eq_mot2} returns
\begin{multline}
	\left[\frac{\kappa}{2} - i(\bar{\Delta} - \Omega)\right]A^+ = 
	\frac{i \chi^*_m \hbar G^2 \bar{a}^2 \left[\frac{\kappa}{2} + i(\bar{\Delta} + \Omega)\right]}
	{\left[\frac{\kappa}{2} + i(\bar{\Delta} + \Omega)\right] + i\hbar G^2 \bar{a}^2 \chi^*_m} A^+
	+ \frac{i \chi^*_m \hbar G^2 \bar{a}^2 \sqrt{\kappa_{\mathrm{ext}}} a_p^*}
	{\left[\frac{\kappa}{2} + i(\bar{\Delta} + \Omega)\right] + i\hbar G^2 \bar{a}^2 \chi^*_m} \\
	\Rightarrow 	\left[\frac{\kappa}{2} - i(\bar{\Delta} - \Omega)\right] \left[\frac{\kappa}{2} + i(\bar{\Delta} + \Omega)\right] A^+ 
	+ i \chi^*_m \hbar G^2 \bar{a}^2 \left[\frac{\kappa}{2} - i(\bar{\Delta} - \Omega) -\frac{\kappa}{2} - i(\bar{\Delta} + \Omega) \right] A^+ =
	i \chi^*_m \hbar G^2 \bar{a}^2 \sqrt{\kappa_{\mathrm{ext}}} a_p^*
\end{multline}
Finally we get to the expression:
\begin{equation}
	A^+ =	\frac{i \alpha \sqrt{\kappa_{\mathrm{ext}}} a_p^*}
	{\frac{1}{\chi^*_m} \left[\frac{\kappa}{2} - i(\bar{\Delta} - \Omega)\right] \left[\frac{\kappa}{2} + i(\bar{\Delta} + \Omega)\right] + 2\bar{\Delta}\alpha} =
	\boxed{ \frac{i \alpha \sqrt{\kappa_{\mathrm{ext}}} a_p^*}
		{\frac{1}{\chi^*_m} \left[ \left(\frac{\kappa}{2} +i\Omega \right)^2 + \bar{\Delta}^2 \right] + 2\bar{\Delta}\alpha} }
\end{equation}

\subsection{Comparison with other derivations in literature}
\label{sec:comparison}
Derivations and notation for OMIT vary significantly across the literature. In this section we compare our formulation to several commonly used results in order to highlight both the equivalences and the key differences.

\subsubsection{\citeauthor{agarwalElectromagneticallyInducedTransparency2010} (\citeyear{agarwalElectromagneticallyInducedTransparency2010})} 
This work is among the first to theoretically describe the OMIT effect and to draw the parallel with the EIT effect in AMO physics. We can prove the equivalence of eq. (5) in the paper with our expression in \cref{eq:Aminus,eq:Aminus2}. The conversion of notations that we need to use is reported in \cref{tab:agarwal}.
\begin{table}[ht]
	\centering
	\begin{tabular}{l|c|c}
		\toprule
		Parameter			& Agarwal ('10)			& This work \\
		\midrule
		Mech. Freq. 		& $\omega_m$			& $\Omega_m$ \\
		Mech. Decay			& $\gamma_m$			& $\Gamma_m$ \\
		OM coupling			& $\chi_0$				& $\hbar G$ \\
		Cav. static field	& $\tilde{c}_0$			& $\bar{a}$ \\
		Strength parameter	& $\beta = \frac{\chi_0^2 |\tilde{c}_0|^2}{2m\hbar\omega_m}$	& $\frac{\alpha}{2m\Omega_m} = \frac{\hbar G^2 |\bar{a}|^2}{2m \Omega_m}$ \\
		Cavity Freq.		& $\omega_0$			& $\omega_{\mathrm{cav}}$ \\
		Pump/Control laser	& $\omega_c$			& $\omega_l$ \\
		Probe laser			& $\omega_p$			& $\omega_p$ \\
		Cavity Decay		& $\kappa$				& $\kappa/2$ \\
		Detuning			& $\omega_0 - \omega_c$	& $\Delta = \omega_l - \omega_{\mathrm{cav}}$ \\
		Mech. static displ. & $q_0 = \frac{\chi_0 |\tilde{c}_0|^2}{m \omega_m^2}$			& $\bar{q} = \frac{\hbar G |\bar{a}|^2}{m \Omega_m^2}$ \\
		Corrected detuning	& $\Delta = \omega_0 - \omega_c - \frac{\chi_0}{\hbar} q_0 = \omega_0 - \omega_c - \frac{2\beta}{\omega_m}$ 
		& $\bar{\Delta} = \Delta + G\bar{q} = \Delta + \frac{\alpha}{m \Omega_m^2}$ \\
		Pump-probe detuning	& $\delta = \omega_p - \omega_c$	& $\Omega = \omega_p - \omega_l$ \\
		\bottomrule
	\end{tabular}
	\caption{Notation convention to convert equations from \cite{agarwalElectromagneticallyInducedTransparency2010} to this work.}
	\label{tab:agarwal}
\end{table}

If we consider equation (5) in \cite{agarwalElectromagneticallyInducedTransparency2010} and use the equivalences in \cref{tab:agarwal} to convert it to our notations, we can express it as follows:
\begin{equation}
	\label{eq:equiv_agarwal}
	\begin{aligned}
		\varepsilon_T	&= \kappa_{\mathrm{ext}}
		\frac{\left(\Omega^2 - \Omega_m^2 + i\Gamma_m\Omega \right) \left[\frac{\kappa}{2} - i(-\bar{\Delta}+\Omega) \right] -2 i \Omega_m \frac{\alpha}{2m\Omega_m}}
		{\left( \Omega^2-\Omega_m^2 +i\Gamma_m \Omega \right) \left[ (\frac{\kappa}{2}-i\Omega)^2 + \bar{\Delta}^2 \right] -4\bar{\Delta} \Omega_m \frac{\alpha}{2m\Omega_m}} \\
		&= \kappa_{\mathrm{ext}} \frac{\frac{1}{\chi_m} \left[ \frac{\kappa}{2} + i(\bar{\Delta}-\Omega) \right] + i\alpha}
		{\frac{1}{\chi_m} \left[ \left(\frac{\kappa}{2} - i\Omega \right)^2 + \bar{\Delta}^2 \right] + 2\bar{\Delta}\alpha}
		= \kappa_{\mathrm{ext}} \frac{A^-}{\varepsilon_p} = \sqrt{\kappa_{\mathrm{ext}}} \frac{A^-}{a_p}
	\end{aligned}
\end{equation}
Note that $\varepsilon_T$ in Eq. (5) in \cite{agarwalElectromagneticallyInducedTransparency2010} is a dimensionless transmission coefficient (ratio of the output probe amplitude to the input $\varepsilon_p$), whereas our $A^-$ is the intracavity probe field amplitude and therefore carries units.  To compare the two, we used the relation $\varepsilon_p = \sqrt{\kappa_{\mathrm{ext}}} a_p$ between Agarwal’s intracavity probe-drive amplitude and our probe photon-flux amplitude. We also take into account the typo in Eq. (6) of \cite{agarwalElectromagneticallyInducedTransparency2010}, where an extra factor $\chi_0$ appears in the definition of $\Delta_{\mathrm{Agarwal}}$; our definitions in \cref{tab:agarwal} and \cref{eq:equiv_agarwal} already incorporate this correction. In addition, the detuning conventions differ by a sign, so that $\Delta_{\mathrm{Agarwal}} = -\bar{\Delta}$ in our notation. Finally, the factor $\kappa$ that multiplies $\varepsilon_T$ in Eq. (5) of \cite{agarwalElectromagneticallyInducedTransparency2010} corresponds to the coupling rate through the driven port and is therefore identified with our external coupling $\kappa_{\mathrm{ext}}$, while the total cavity linewidth $\kappa$ appearing in the rest of the expression remains unchanged. This is consistent with the input–output conventions that we will discuss later on. It is also useful to note that $A_-$ and $A_+$ in \cite{agarwalElectromagneticallyInducedTransparency2010} are not the same and are not related to our $A^-$ and $A^+$. We emphasize that while Eq. (5) of Agarwal still contains the combined effect of Stokes and anti‑Stokes channels (analogous to our full expression for $A^-$), from Eq. (7) onward Agarwal’s analysis focuses on the sideband‑resolved regime and employs further approximations tailored to that limit.

\subsubsection{\citeauthor{weisOptomechanicallyInducedTransparency2010} (\citeyear{weisOptomechanicallyInducedTransparency2010})}
To compare our results to the result derived in \cite{weisOptomechanicallyInducedTransparency2010}, we can start from our derivation in the previous section -- which is completely analogous -- and show that the expressions differ only in terms of notation. Starting from \cref{eq:x_chi_m}, if we define $f(\Omega)$ as: 
\begin{equation}
	f(\Omega)  = \frac{\hbar G^2 \bar{a}^2 \chi_m }{\left[ \frac{\kappa}{2} +i(\bar{\Delta} - \Omega) \right]}
\end{equation}

we can then obtain:
\begin{equation}
	X = \hbar G \bar{a} \chi_m A^- -i f X
	\qquad \Rightarrow \qquad
	X = \frac{\hbar G \bar{a} \chi_m A^-}{1 + if}
\end{equation}
Substituting this expression in \cref{eq:Am_eq_mot2} we get:
\begin{equation}
	\left[ \frac{\kappa}{2} - i(\bar{\Delta} + \Omega) \right] A^- = i \hbar G^2 \bar{a}^2 \frac{\chi_m}{1+if}A^- + \sqrt{\kappa_{\mathrm{ext}}} a_p
	\; \Rightarrow \;
	\sqrt{\kappa_{\mathrm{ext}}} a_p =
	\left\{ \frac{ \left[ \frac{\kappa}{2} - i(\bar{\Delta} + \Omega) \right] (1+if) - i\hbar G^2 \bar{a}^2 \chi_m }{1+if} \right\} A^-
\end{equation}
The numerator in the expression inside the curly bracket can be re-written as:
\begin{multline}
	\left[ \frac{\kappa}{2} - i(\bar{\Delta} + \Omega) \right]
	+ \left[ \frac{\kappa}{2} - i(\bar{\Delta} + \Omega) \right] \frac{i\hbar G^2 \bar{a}^2 \chi_m}{ \left[ \frac{\kappa}{2} + i(\bar{\Delta} - \Omega) \right] }
	- \frac{i\hbar G^2 \bar{a}^2 \chi_m \left[ \frac{\kappa}{2} + i(\bar{\Delta} - \Omega) \right] }{ \left[ \frac{\kappa}{2} + i(\bar{\Delta} - \Omega) \right] } = \\
	= \frac{\kappa}{2} - i(\bar{\Delta} + \Omega) + if(-2i\bar{\Delta})  = \frac{\kappa}{2} - i(\bar{\Delta} + \Omega) + 2\bar{\Delta}f
\end{multline}
Finally we obtain the expression for $A^-$:
\begin{equation}
	\label{eq:Aminus_Weis}
	\boxed{	A^- = \frac{ (1+if) \sqrt{\kappa_{\mathrm{ext}}} a_p }{ \frac{\kappa}{2} - i(\bar{\Delta} + \Omega) + 2\bar{\Delta}f } },
\end{equation}
with $\kappa_{\mathrm{ext}}=\eta_c\kappa$ in the notation of \cite{weisOptomechanicallyInducedTransparency2010}. Notation aside, we can see that \cref{eq:Aminus_Weis} is completely analogous to our expression in \cref{eq:Aminus,eq:Aminus2}. The derivation in \cite{weisOptomechanicallyInducedTransparency2010} then proceeds in the resolved‑sideband regime and neglects $A^+$ and the associated terms, which is equivalent to dropping the counter‑rotating contribution.

We can anyway derive an expression for $A^+$ using this notation. If we define $F(\Omega)$ as:
\begin{equation}
	F(\Omega)  = \frac{\hbar G^2 \bar{a}^2 \chi_m^* }{\left[ \frac{\kappa}{2} +i(\bar{\Delta} + \Omega) \right]}
\end{equation}
we can rewrite \cref{eq:x_chi} as
\begin{equation}
	X^* = \hbar G \bar{a} \chi_m^* A^+ - iF(\Omega) X^* + \frac{F(\Omega)\sqrt{\kappa_{\mathrm{ext}}} a_p^*}{G \bar{a}}
	\; \Rightarrow \;
	X^* = \frac{\hbar G \bar{a}\chi_m^*}{1+i F(\Omega)} A^+ + \frac{F(\Omega)\sqrt{\kappa_{\mathrm{ext}}} a_p^*}{G \bar{a} \left( 1+i F(\Omega) \right)}
\end{equation}
Using this expression inside \cref{eq:Ap_eq_mot2} returns:
\begin{equation}
	\left[ \frac{\kappa}{2} - i (\bar{\Delta}-\Omega) \right] A^+ =
	i\hbar G^2 \bar{a}^2 \frac{\chi_m^*}{1+i F(\Omega)} A^+ + \frac{i F(\Omega) \sqrt{\kappa_{\mathrm{ext}}} a_p^*}{1 + i F(\Omega)}
\end{equation}
The terms that multiply $A^+$ can be simplify as:
\begin{multline}
	\frac{ \left[ \frac{\kappa}{2} - i (\bar{\Delta}-\Omega) \right] - i\hbar G^2 \bar{a}^2 \chi_m^*}{1 + i F(\Omega)} = \\
	\left\{ \left[ \frac{\kappa}{2} - i (\bar{\Delta}-\Omega) \right] 
	+ \left[ \frac{\kappa}{2} - i (\bar{\Delta}-\Omega) \right] \frac{i\hbar G^2 \bar{a}^2 \chi_m^*}{\left[ \frac{\kappa}{2} + i (\bar{\Delta}+\Omega) \right]} 
	- \frac{i\hbar G^2 \bar{a}^2 \chi_m^* \left[ \frac{\kappa}{2} + i (\bar{\Delta}+\Omega) \right]}{\left[ \frac{\kappa}{2} + i (\bar{\Delta}+\Omega) \right]} \right\} 
	\frac{1}{1 + i F(\Omega)} = \\
	\left\{ \left[ \frac{\kappa}{2} - i (\bar{\Delta}-\Omega) \right] +i F(\Omega) (-2i \bar{\Delta}) \right\} \frac{1}{1 + i F(\Omega)} = 
	\frac{\left[ \frac{\kappa}{2} - i (\bar{\Delta}-\Omega) \right] + 2\bar{\Delta} F(\Omega)}{1 + i F(\Omega)}
\end{multline}
Finally:
\begin{equation}
	\boxed{ A^+ = \frac{i F(\Omega) \sqrt{\kappa_{\mathrm{ext}}} a_p^*}{\left[ \frac{\kappa}{2} - i (\bar{\Delta}-\Omega) \right] + 2\bar{\Delta} F(\Omega)} }
\end{equation}

\subsubsection{\citeauthor{yanOptomechanicallyInducedTransparency2020} (\citeyear{yanOptomechanicallyInducedTransparency2020,yanOptomechanicallyInducedOptical2021})}
In order to compare our results to the works of Yan \cite{yanOptomechanicallyInducedTransparency2020,yanOptomechanicallyInducedOptical2021}, we want to rewrite our expression for $A^-$ in a way that will be useful for our comparison between the unresolved-sideband regime (USR) and the more standard resolved case (RSR) (see next section). Starting from \cref{eq:Aminus}, we can write:
\begin{equation}
	\label{eq:Aminus3}
	\begin{aligned}
		A^- &=	\frac{\frac{1}{\chi_m} \left[ \frac{\kappa}{2} + i(\bar{\Delta}-\Omega) \right] + i\alpha}
		{\frac{1}{\chi_m} \left[ \left(\frac{\kappa}{2} - i\Omega \right)^2 + \bar{\Delta}^2 \right] + 2\bar{\Delta}\alpha}
		\sqrt{\kappa_{\mathrm{ext}}} a_p = \\
		&=	\frac{\left[ \frac{\kappa}{2} + i(\bar{\Delta}-\Omega) \right] + i\alpha\chi_m}
		{\left\{ \left[ \frac{\kappa}{2} + i(\bar{\Delta}-\Omega) \right] + i\alpha\chi_m \right\} 
			\left\{ \left[ \frac{\kappa}{2} - i(\bar{\Delta}+\Omega) \right] - i\alpha\chi_m \right\} -\alpha^2\chi_m^2}
		\sqrt{\kappa_{\mathrm{ext}}} a_p = \\
		&=	\frac{\sqrt{\kappa_{\mathrm{ext}}} a_p}
		{\left[ \frac{\kappa}{2} - i(\bar{\Delta}+\Omega) \right] - i\alpha\chi_m 
			- \frac{\alpha^2\chi_m^2}{\left[ \frac{\kappa}{2} + i(\bar{\Delta}-\Omega) \right] + i\alpha\chi_m}} = \\
		&=	\frac{\sqrt{\kappa_{\mathrm{ext}}} a_p}
		{\left[ \frac{\kappa}{2} - i(\bar{\Delta}+\Omega) \right] 
			- \frac{ i\alpha\chi_m \left[ \frac{\kappa}{2} + i(\bar{\Delta}-\Omega) \right] -\alpha^2\chi_m^2 +\alpha^2\chi_m^2}
			{\left[ \frac{\kappa}{2} + i(\bar{\Delta}-\Omega) \right] + i\alpha\chi_m}} = \\
		&=	\boxed{ \frac{\sqrt{\kappa_{\mathrm{ext}}} a_p}
			{\left[ \frac{\kappa}{2} - i(\bar{\Delta}+\Omega) \right] 
				+ \frac{\alpha}{\frac{i}{\chi_m} - \frac{\alpha}{\left[ \frac{\kappa}{2} + i(\bar{\Delta}-\Omega) \right]}} } }
	\end{aligned}
\end{equation}
Note that in the previous expression we made use of the following equivalence:
\begin{multline}
	\left\{ \left[ \frac{\kappa}{2} + i(\bar{\Delta}-\Omega) \right] + i\alpha\chi_m \right\} 
	\left\{ \left[ \frac{\kappa}{2} - i(\bar{\Delta}+\Omega) \right] - i\alpha\chi_m \right\} = \\
	\left[ \left(\frac{\kappa}{2} - i\Omega \right)^2 + \bar{\Delta}^2 \right] + 2\bar{\Delta}\alpha \chi_m + \alpha^2 \chi_m^2 = \\
	\left[ \frac{\kappa}{2} + i(\bar{\Delta}-\Omega) \right] \left[ \frac{\kappa}{2} - i(\bar{\Delta}+\Omega) \right]
	+ 2\bar{\Delta}\alpha \chi_m + \alpha^2 \chi_m^2
\end{multline}
Expressing $A^-$ in this form we can directly compare our results to eq. (A9) in \citeauthor{yanOptomechanicallyInducedTransparency2020} (\citeyear{yanOptomechanicallyInducedTransparency2020}) \cite{yanOptomechanicallyInducedTransparency2020}. Yan uses the same notation as Agarwal, so we can again employ the conversion from \cref{tab:agarwal} and it is easy to prove that the two equations are equivalent. A similar procedure can be used to prove the equivalence with eq. (8) in \citeauthor{yanOptomechanicallyInducedOptical2021} (\citeyear{yanOptomechanicallyInducedOptical2021}) \cite{yanOptomechanicallyInducedOptical2021}. The only difference in this case is that we have to replace $2\kappa$ (and not just $\kappa$) with $\kappa/2$ for the notations to match. It is also useful to remember that $\beta = \frac{\alpha}{2m\Omega_m}$ and that in our definition $\alpha = \hbar G^2 \bar{a}^2 = g_0^2 \bar{a}^2 2m\Omega_m$. Finally, it is important to note that in most of its analysis Yan usually employs the simplification $\Omega \sim \Omega_m \sim \Delta$. While the first equivalence will usually be true also in our experiment, for the most part we will work with  $\Delta \sim \kappa \gg \Omega_m$, so the specific near‑resonant simplifications used in Yan are not directly applicable to our USR parameter regime.

\subsection{Rotating wave approximation, thermal effect and comparison with the resolved-sideband regime}
\label{sec:RSB_vs_USR}
In the previous sections we have derived the full expression for the intracavity probe-sideband amplitude $A^-$ starting from the standard optomechanical Hamiltonian, without making any rotating-wave approximation on the linearized interaction. In particular, the frequency-domain solution obtained from the $q$-frame and $b$-frame equations retains both optical sidebands $A^-$ and $A^+$ in the ansatz, and the final result for $A^-$ (see \cref{eq:Aminus,eq:Aminus2} and its rearranged form \cref{eq:Aminus3}) reflects the influence of both the upper (anti-Stokes) and lower (Stokes) sidebands. This “full” expression is the one relevant for our experiment, which operates deeply in the unresolved-sideband regime (USR), $\kappa \gg \Omega_m$, where the cavity cannot filter out one of the sidebands.

It is useful to distinguish this “sideband” rotating-wave approximation from the much more basic RWA that is already implicit in our starting point. When we move to a frame rotating at the laser frequency $\omega_l$ and write the drive and cavity dynamics in terms of slowly varying operators (cf. \cite{aspelmeyerCavityOptomechanics2014}), we have already neglected terms oscillating at optical frequencies such as $2\omega_l$; this is the standard optical RWA and is always assumed throughout. The additional approximation discussed in this section is a second RWA performed at the \emph{mechanical} (sideband) scale: after linearization we selectively keep only those fluctuation terms whose residual time dependence is slow near a chosen detuning (e.g. $\bar{\Delta}\simeq -\Omega_m$), and drop terms oscillating at $\sim 2\Omega_m$. In the resolved-sideband regime these “fast” terms correspond precisely to the off-resonant Stokes (or, for blue-detuning, off-resonant anti-Stokes) sideband, and this is what leads to the simplified beam-splitter or two-mode-squeezing Hamiltonians (see \cite{aspelmeyerCavityOptomechanics2014}).

In contrast, most common OMIT derivations (including \cite{weisOptomechanicallyInducedTransparency2010,agarwalElectromagneticallyInducedTransparency2010}) assume the resolved-sideband regime (RSR), typically with a control laser red-detuned close to the lower motional sideband ($\bar{\Delta} \simeq -\Omega_m$) and with $\kappa \ll \Omega_m$. In this limit the cavity susceptibility strongly suppresses the lower (Stokes) sideband, so that the upper (anti-Stokes) sideband dominates the dynamics. At the level of the equations of motion this corresponds to neglecting the $+\Omega$ components in the ansatz we choose before, i.e. setting $A^+ \simeq 0$ together with the associated mechanical amplitudes, and hence dropping the counter-rotating Stokes channel. At the Hamiltonian level this is equivalent to starting from the linearized optomechanical Hamiltonian and keeping only the beam-splitter terms (red-detuned RWA), while discarding the two-mode-squeezing contribution.

The aim of this section is threefold. First, we derive explicitly the standard resolved-sideband/RWA expression for $A^-$ in our notation, by starting from the full coupled equations derived above and neglecting the $A^+$, $X^*$ and $B^+$ terms. \Cref{eq:Am_eq_mot2,eq:X_eq_mot} then reduce to:
\begin{subequations}
	\begin{empheq}[left=\empheqlbrace]{align}
		&\left[\frac{\kappa}{2} - i(\bar{\Delta} + \Omega)\right]A^- = iG\bar{a}X + \sqrt{\kappa_{\mathrm{ext}}} a_p		&; \quad e^{-i\Omega t}\\
		&m \left( \Omega_m^2 - \Omega^2 - i\Gamma_m \Omega \right) X = \hbar G \bar{a} A^- 									&; \quad e^{-i\Omega t}
	\end{empheq}
\end{subequations}
It is then straightforward to obtain
\begin{equation}
	\label{eq:Aminus_RWA}
	X = \hbar G \bar{a} \chi_m A^-
	\quad \Rightarrow \quad
	A^- = \frac{\sqrt{\kappa_{\mathrm{ext}}} a_p}{\left[\frac{\kappa}{2} - i(\bar{\Delta} + \Omega)\right] - i\hbar G^2 \bar{a}^2 \chi_m}
	= \boxed{ \frac{\sqrt{\kappa_{\mathrm{ext}}} a_p}{\left[\frac{\kappa}{2} - i(\bar{\Delta} + \Omega)\right] + \frac{\alpha}{i/\chi_m}} }
\end{equation}
This directly exhibits how the usual OMIT formula arises as an approximation to our general result. Comparing this expression to \cref{eq:Aminus3}, we can see that now the term $-\frac{\alpha}{\left[ \frac{\kappa}{2} + i(\bar{\Delta}-\Omega) \right]}$ is absent. This encodes the effect of the counter-rotating terms, and the advantage of using \cref{eq:Aminus3} is that we can easily put this term at zero if we want to compare the full expression with the approximated version for different parameter ranges.\\
We can apply the same procedure to \cref{eq:Am_eq_mot,eq:Bm_eq_mot}, which results in:
\begin{subequations}
	\begin{empheq}[left=\empheqlbrace]{align}
		&\left[\frac{\kappa}{2} - i(\bar{\Delta} + \Omega)\right]A^- = ig_0 \bar{a} B^- + \sqrt{\kappa_{\mathrm{ext}}} a_p		&; \quad e^{-i\Omega t} \\
		&\left[i(\Omega_m - \Omega) + \frac{\Gamma_m}{2}\right]B^- = ig_0 \bar{a}A^- 											&; \quad e^{-i\Omega t}
	\end{empheq}
\end{subequations}
From there
\begin{multline}
	B^- = \frac{i g_0 \bar{a}A^-}{\left[ i(\Omega_m-\Omega) + \frac{\Gamma_m}{2} \right]}
	\quad \Rightarrow \quad \\
	\quad \Rightarrow \quad
	\left\{ \left[\frac{\kappa}{2} - i(\bar{\Delta} + \Omega)\right] \left[ i(\Omega_m-\Omega) + \frac{\Gamma_m}{2} \right] + g_0^2 \bar{a}^2 \right\} A^-
	= \left[ i(\Omega_m-\Omega) + \frac{\Gamma_m}{2} \right] \sqrt{\kappa_{\mathrm{ext}}} a_p
\end{multline}
and finally for $A^-$ we get
\begin{equation}
	\begin{aligned}
		A^- &= \frac{\left[ i(\Omega_m-\Omega) + \frac{\Gamma_m}{2} \right] \sqrt{\kappa_{\mathrm{ext}}} a_p}
		{\left[\frac{\kappa}{2} - i(\bar{\Delta} + \Omega)\right] \left[ i(\Omega_m-\Omega) + \frac{\Gamma_m}{2} \right] + g_0^2 \bar{a}^2} =\\
		&= \frac{\sqrt{\kappa_{\mathrm{ext}}} a_p}
		{\left[\frac{\kappa}{2} - i(\bar{\Delta} + \Omega)\right] + \frac{g_0^2 \bar{a}^2}{\left[ i(\Omega_m-\Omega) + \frac{\Gamma_m}{2} \right]}
			\left( \frac{-i2m\Omega_m}{-i2m\Omega_m} \right)} = \\
		&= \frac{\sqrt{\kappa_{\mathrm{ext}}} a_p}{\left[\frac{\kappa}{2} - i(\bar{\Delta} + \Omega)\right] + \frac{\alpha}{i/\chi_m}}
	\end{aligned}
\end{equation}
where we use the approximation
\begin{multline}
	\left[ i(\Omega_m-\Omega) + \frac{\Gamma_m}{2} \right] \left( -i2m\Omega_m \right) =
	m \left[ 2\Omega_m (\Omega_m-\Omega) - i\Gamma_m\Omega_m \right] \simeq \\
	\simeq m \left[ (\Omega_m+\Omega)(\Omega_m-\Omega) -i\Gamma_m\Omega_m \right] =
	m \left[ \Omega_m^2-\Omega^2 - i\Gamma_m\Omega_m \right] = \frac{1}{\chi_m}
\end{multline}
which is valid for $\Omega_m \simeq \Omega$. As expected, once again we obtain the same approximate result for $A^-$.

We could re-derive the same approximate expression starting directly from the linearized Hamiltonian in the beam-splitter (or, alternatively, in the two-mode-squeezing) regime. This highlights how the “resonant” versus “non-resonant” terms in the Hamiltonian map onto the upper and lower sidebands in our frequency-domain treatment.

\subparagraph{Critical frequency and critical power}
One useful feature of the full expression \cref{eq:Aminus3} is that it admits parameter values for which the intracavity probe-sideband amplitude vanishes, $A^-(\Omega)=0$, corresponding to a perfect destructive-interference point (‘perfect OMIT’). It is indeed possible to find values for which
\begin{equation}
	\frac{i}{\chi_m} - \frac{\alpha}{\left[ \frac{\kappa}{2} + i(\bar{\Delta} - \Omega) \right]} = 0
\end{equation}
and consequently $A^- = 0$. We can rewrite this equation as
\begin{align}
	&i \left[\Omega_m^2 - \Omega^2 - i\Gamma_m\Omega \right] \left[ \frac{\kappa}{2} + i (\bar{\Delta}-\Omega) \right] = \frac{\alpha}{m} \\
	\Rightarrow \quad &i(\Omega_m^2-\Omega^2)\frac{\kappa}{2} + i\Gamma_m\Omega(\bar{\Delta}-\Omega) - (\Omega_m^2-\Omega^2)(\bar{\Delta}-\Omega) + \Gamma_m\Omega\frac{\kappa}{2}
	= \frac{\alpha}{m},
\end{align}
from which if we equate separately the real and imaginary part we arrive at:
\begin{subequations}
	\begin{empheq}[left=\empheqlbrace]{align}
		&(\Omega_m^2-\Omega^2)\frac{\kappa}{2} + \Gamma_m\Omega(\bar{\Delta}-\Omega) = 0 \\
		&\Gamma_m\Omega\frac{\kappa}{2} - (\Omega_m^2-\Omega^2)(\bar{\Delta}-\Omega) = \frac{\alpha}{m}
	\end{empheq}
\end{subequations}
The first equation allows to compute a "critical frequency" at which the OMIT window reaches zero, while the second tells us, for a given single-photon OM coupling, for what input power this is going to happen (remember that $\alpha = \hbar G^2 \bar{a}^2 = g_0^2 \bar{a}^2 2m\Omega_m$). Solving the first equation we get:
\begin{equation}
	\Omega_m^2 \frac{\kappa}{2} - \Omega^2 \frac{\kappa}{2} + \Gamma_m\bar{\Delta}\Omega - \Gamma_m\Omega^2 = 0
	\Rightarrow \left( \frac{\kappa}{2} + \Gamma_m \right)\Omega^2 - \Gamma_m\bar{\Delta}\Omega - \frac{\kappa}{2}\Omega_m^2 = 0
\end{equation}
whose solution is
\begin{equation}
	\Omega_{\pm}	= \frac{\Gamma_m\bar{\Delta} \pm \sqrt{\Gamma_m^2\bar{\Delta}^2 + 2\kappa\Omega_m^2 (\frac{\kappa}{2} +\Gamma_m)}}{\kappa+2\Gamma_m}
	= \boxed {\frac{\Gamma_m\bar{\Delta} \pm \Omega_m \sqrt{\frac{\Gamma_m^2}{\Omega_m^2}\bar{\Delta}^2 + \kappa^2 + 2\kappa\Gamma_m}}{\kappa+2\Gamma_m} }
\end{equation}
From the second equation we can then get for the critical power:
\begin{equation}
	\label{eq:P_crit}
	\frac{\alpha_{\pm}}{m} = \Gamma_m \Omega_{\pm} \left(\frac{\kappa}{2} + \frac{(\bar{\Delta}-\Omega_{\pm})^2}{\kappa/2} \right)
	\Rightarrow
	P_{\pm} = \frac{m \omega_l \Gamma_m \Omega_{\pm}}{G^2 \kappa_{\mathrm{ext}}} \left[ \left(\frac{\kappa}{2}\right)^2 + \bar{\Delta}^2 \right]
	\left(\frac{\kappa}{2} + \frac{(\bar{\Delta}-\Omega_{\pm})^2}{\kappa/2} \right)
\end{equation}
To better understand these results, we can apply a couple of approximations to the solutions that are usually true in optomechanics. Namely, we can assume $\Omega_m \gg \Gamma_m$ (high-Q mechanical resonator) and $\kappa \gg \Gamma_m$ (always true in USR). This then allows us to write $\kappa + 2\Gamma_m \approx \kappa$ and $\frac{\Gamma_m^2}{\Omega_m^2}\bar{\Delta}^2 + \kappa^2 + 2\kappa\Gamma_m = \kappa^2 \left(\frac{\Gamma_m^2}{\Omega_m^2}\frac{\bar{\Delta}^2}{\kappa^2} + 1 + \frac{2\Gamma_m}{\kappa} \right) \approx \kappa^2$. Then the equation for the critical frequencies simplifies to
\begin{equation}
	\Omega_{\pm} \simeq \frac{\Gamma_m \bar{\Delta}}{\kappa} \pm \Omega_m 
\end{equation}
Now we can clearly see that the value of two solutions consists of small corrections around $\pm \Omega_m$, as we would expect. Because $\alpha_{\pm} \propto \Gamma_m \Omega_{\pm}$, using the "$-$" solution would imply an unphysical negative value for the critical power ($\alpha \propto \bar{a}^2$ is always positive); so the only branch that we retain is the "$+$" one.

\subparagraph{Impact of the RWA approximation}
To quantify the impact of the sideband RWA in the unresolved-sideband regime, we directly compare the full expression \cref{eq:Aminus3} with its RWA limit \cref{eq:Aminus_RWA}, using representative parameters of our platform ($\Omega_m/2\pi \simeq \qty{2}{\mega\hertz}$, $Q\simeq\num{6e3}$, $M_{\mathrm{eff}}\simeq\qty{17}{\pico\gram}$, $\kappa/2\pi\simeq\qty{1}{\giga\hertz}$, $\lambda=\qty{920}{\nano\meter}$). We plot the reflected probe response $R(\Omega)=|a_{L_0}| \sqrt{\kappa_{\mathrm{ext}}}\,|A^-(\Omega)|$ (see derivation below), set $a_{L_0}=1$ for simplicity, and normalize all traces by the same
reference value (the maximum of $R$ at $G=0$).

\begin{figure}[t!]
	\centering
	\includegraphics[width=1.0\textwidth]{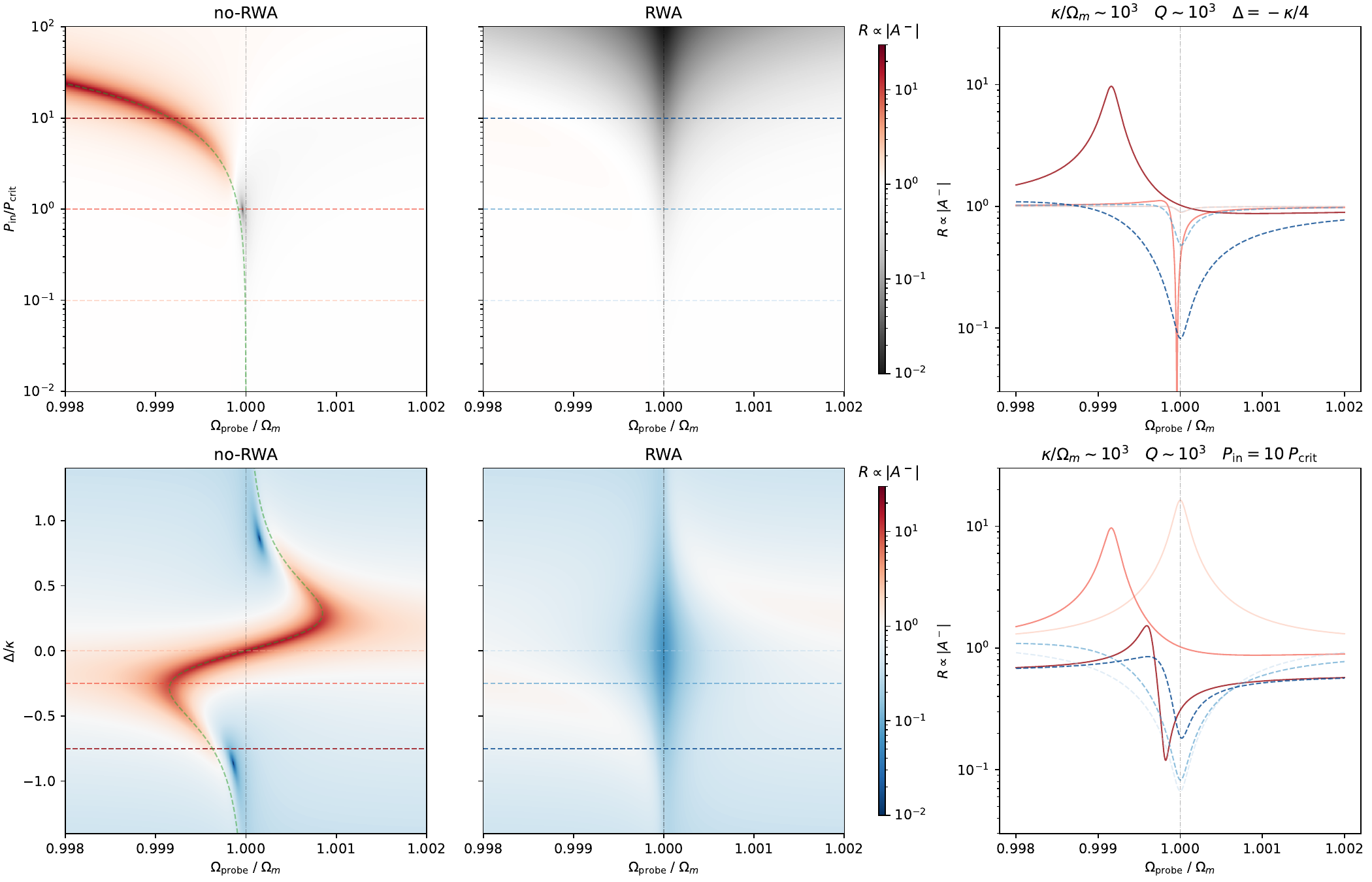}
	\caption{Comparison between the full USR probe response (\textbf{no-RWA}, \cref{eq:Aminus3}) and the sideband-RWA approximation (\textbf{RWA}, \cref{eq:Aminus_RWA}) for $\Omega_m/2\pi \sim \qty{2}{\mega\hertz}$, $Q\sim\num{6e3}$, $M_{\mathrm{eff}}\sim\qty{17}{\pico\gram}$ and $G/2\pi = \qty{1}{\giga\hertz\per\nano\meter}$. \textbf{Top row:} $R(\Omega)\propto \sqrt{\kappa_{\mathrm{ext}}}|A^-|$ versus $(\Omega_{\mathrm{probe}}/\Omega_m,P_{\mathrm{in}})$ at fixed $\Delta=-\kappa/4$, with line cuts at $P_{\mathrm{in}}/P_{\mathrm{crit}}=0.1,1,10$ (solid: no-RWA, dashed: RWA). \textbf{Bottom row:} $R(\Omega)$ versus $(\Omega_{\mathrm{probe}}/\Omega_m,\Delta)$ at fixed $P_{\mathrm{in}}=10\,P_{\mathrm{crit}}$, with line cuts at $\Delta/\kappa=0,-1/4,-3/4$. The dashed green traces mark $\Omega_m + \delta\Omega_m$ in the no-RWA maps.}
	\label{fig:USR_noRWA_VS_RWA}
\end{figure}

The top row of \cref{fig:USR_noRWA_VS_RWA} shows power sweeps at fixed detuning $\Delta=-\kappa/4$ and frequency-pull parameter $G/2\pi=\qty{1}{\giga\hertz\per\nano\meter}$ (corresponding to $g_0\simeq\qty{16}{\kilo\hertz}$). The power axis is normalized to the critical power $P_{\mathrm{crit}}=P_+$. The left panel is computed from the full model (no-RWA), the center panel from the RWA model, and the right panel shows line cuts at $P_{\mathrm{in}}/P_{\mathrm{crit}}=0.1,1,10$ (solid: no-RWA; dashed: RWA). The bottom row shows detuning sweeps (in units of $\kappa$) at fixed power $P_{\mathrm{in}}=10\,P_{\mathrm{crit}}$: left, full model; center, RWA model; right, line cuts at $\Delta/\kappa=0,-1/4,-3/4$ (same line-style convention). The dashed green guide in the no-RWA maps corresponds to the optical-spring-shifted resonance $\Omega_{\mathrm{eff}}=\Omega_m+\delta\Omega_m$, with
\begin{equation}
	\delta\Omega_m = g^2 \left[\frac{\Delta + \Omega_m}{(\Delta + \Omega_m)^2 + \kappa^2/4} + \frac{\Delta-\Omega_m}{(\Delta-\Omega_m)^2 + \kappa^2/4} \right]
\end{equation}
and $g^2=g_0^2 \bar{n}_{\mathrm{cav}}$.

Comparing no-RWA and RWA, three differences are evident. First, the full model predicts a true critical point where $A^-\to 0$ (perfect destructive interference), while the RWA model does not reproduce a complete cancellation under the same conditions and requires higher power to reach a similarly deep dip. Second, beyond the critical region, the full model rapidly evolves from a dip to a peak (relative probe enhancement with respect to the off-mechanical baseline), whereas the RWA prediction remains more monotonic. In resolved-sideband treatments, this monotonic evolution leads to the typical peak-splitting signature of strong coupling, while in the USR full model the evolution is qualitatively different (see below).
Third, the full model preserves a clear backaction-induced frequency drift of the dip/peak feature across power and detuning, especially visible in the detuning maps, which is instead absent in the RWA case. Finally, since \cref{eq:P_crit} depends on $\Delta$, varying the detuning at fixed input power also drives the system between the transparency-like and enhancement regimes.

\subparagraph{Impact of the oscillator thermal motion}
A second correction, relevant in our USR parameter range, comes from thermally induced cavity-frequency fluctuations. As discussed by Leijssen \textit{et al.} \cite{leijssenNonlinearCavityOptomechanics2017}, Brownian motion produces frequency jitter with rms amplitude
\begin{equation}
	\delta\omega_{\mathrm{rms}}=g_0\sqrt{2n_{\mathrm{th}}}, \qquad n_{\mathrm{th}}=\frac{k_B T}{\hbar\Omega_m},
\end{equation}
which broadens the optical response through a Gaussian contribution (Voigt picture). In the present model, we include this effect in an effective-linewidth form,
\begin{equation}
	\kappa_{\mathrm{eff}}(T)=\kappa+\delta\kappa(T),\qquad \delta\kappa(T)\equiv 2\sqrt{2\ln2}\,\delta\omega_{\mathrm{rms}},
\end{equation}
and use $\kappa_{\mathrm{eff}}$ consistently in both the cavity susceptibility and the intracavity photon number entering $\alpha=\hbar G^2\bar{a}^2$. Note that this treatment should be understood as a simple phenomenological implementation of the Voigt broadening, not as a full Lorentzian--Gaussian convolution treatment of the optical response.

\begin{figure}[t!]
	\centering
	\includegraphics[width=1.0\textwidth]{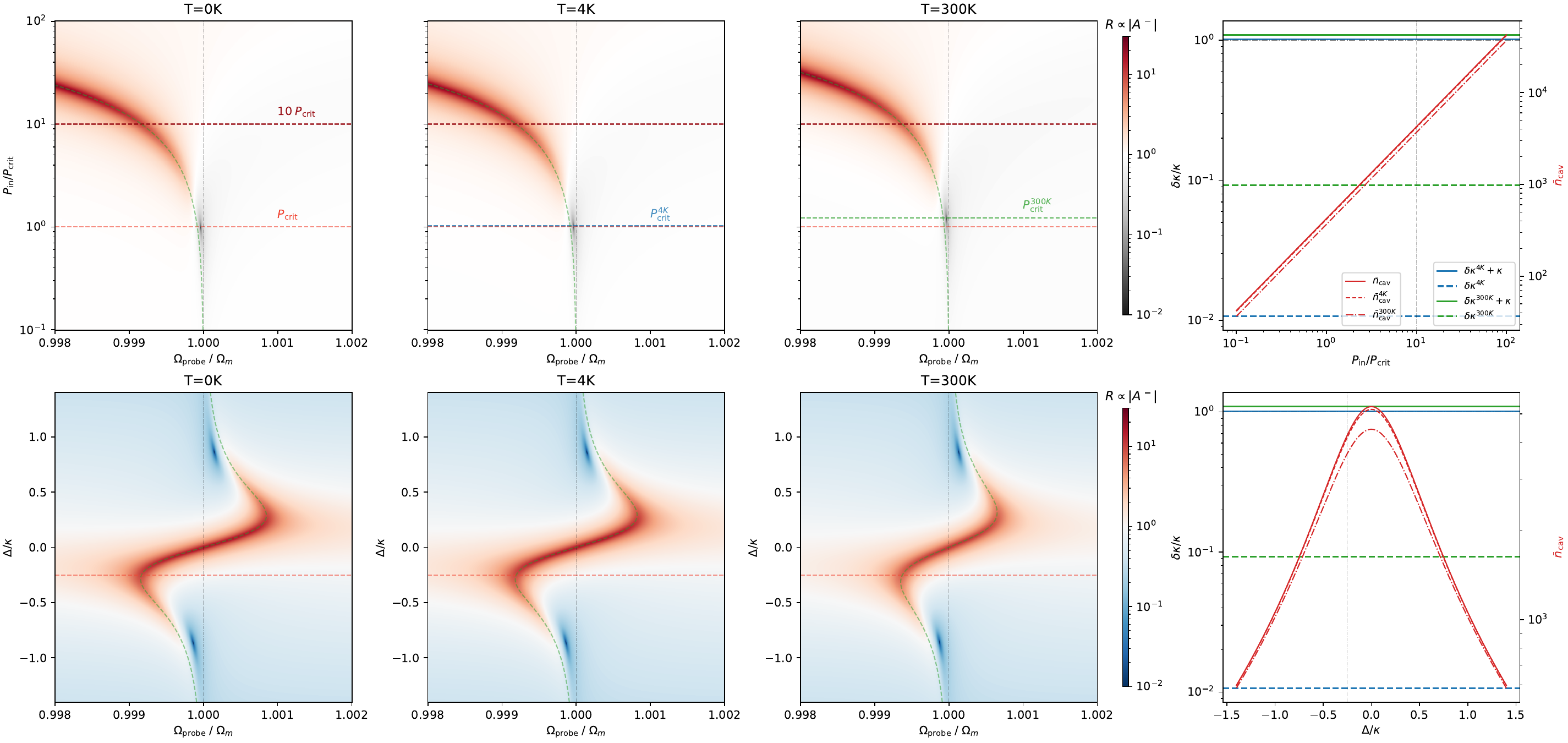}
	\caption{USR response including thermally induced cavity-linewidth broadening, following the mechanism discussed in \cite{leijssenNonlinearCavityOptomechanics2017}. Left three columns: normalized reflected probe response $R\propto|A^-|$ computed with $\kappa_{\mathrm{eff}}=\kappa+\delta\kappa(T)$ for $T=0,4,\qty{300}{\kelvin}$. Top row: $(\Omega_{\mathrm{probe}},P_{\mathrm{in}})$ maps at fixed $\Delta=-\kappa/4$; dashed horizontal markers indicate $P_{\mathrm{crit}}$ (including thermal broadening). Bottom row: $(\Omega_{\mathrm{probe}},\Delta)$ maps at fixed input power $P_{\mathrm{in}}=10P_{\mathrm{crit}}^{0K}$. Right column: corresponding $\delta\kappa/\kappa$ (left axis) and intracavity photon number $\bar{n}_{\mathrm{cav}}$ (right axis), versus power (top) and detuning (bottom). Increasing temperature broadens the effective cavity response, reduces $\bar n_{\mathrm{cav}}$ at fixed drive, shifts the OMIT critical power $P_{\mathrm{crit}}$ to higher values.}
	\label{fig:USR_OMIT_vs_dKappa}
\end{figure}

In \cref{fig:USR_OMIT_vs_dKappa}, using the same parameters as in \cref{fig:USR_noRWA_VS_RWA}, this gives $\delta\kappa/\kappa \simeq 1\times10^{-2}$ at \qty{4}{\kelvin} and $\delta\kappa/\kappa \simeq 1\times10^{-1}$ at \qty{300}{\kelvin}. The impact is twofold: (i) for fixed input power, $\bar{n}_{\mathrm{cav}}$ decreases because the cavity is effectively broader; (ii) the perfect-OMIT condition is shifted to higher power, so that $P_{\mathrm{crit}}$ increases (e.g. in our parameter set: $P_{\mathrm{crit}}^{300K}/P_{\mathrm{crit}}^{0K}\simeq1.2$). As a consequence, the dip-to-gain crossover is pushed to larger drive and the OMIT contrast is reduced at high temperature. For this parameter choice (which we employ for most of our measurements), the effect is moderate and acts as a correction at room temperature. However, using $G/2\pi\simeq\qty{11}{\giga\hertz\per\nano\meter}$ (the maximum achievable in our setup), the broadening $\delta\kappa$ becomes comparable to $\kappa$, effectively doubling the cavity linewidth (see below).

\Cref{fig:USR_OMIT_vs_dKappa} summarizes this trend. Top row: probe-frequency/power maps at fixed detuning ($\Delta=-\kappa/4$) for $T=0,4,\qty{300}{\kelvin}$; dashed horizontal lines mark the corresponding critical powers. Bottom row: probe-frequency/detuning maps at fixed input power ($P_{\mathrm{in}}=10P_{\mathrm{crit}}^{0K}$), again for $T=0,4,\qty{300}{\kelvin}$. The rightmost column reports the thermal broadening scale $\delta\kappa/\kappa$ (left axis) together with the corresponding $\bar{n}_{\mathrm{cav}}$ (right axis), explicitly showing how linewidth broadening and intracavity-population reduction jointly reshape the OMIT response.

\begin{figure}[t!]
	\centering
	\includegraphics[width=1.0\textwidth]{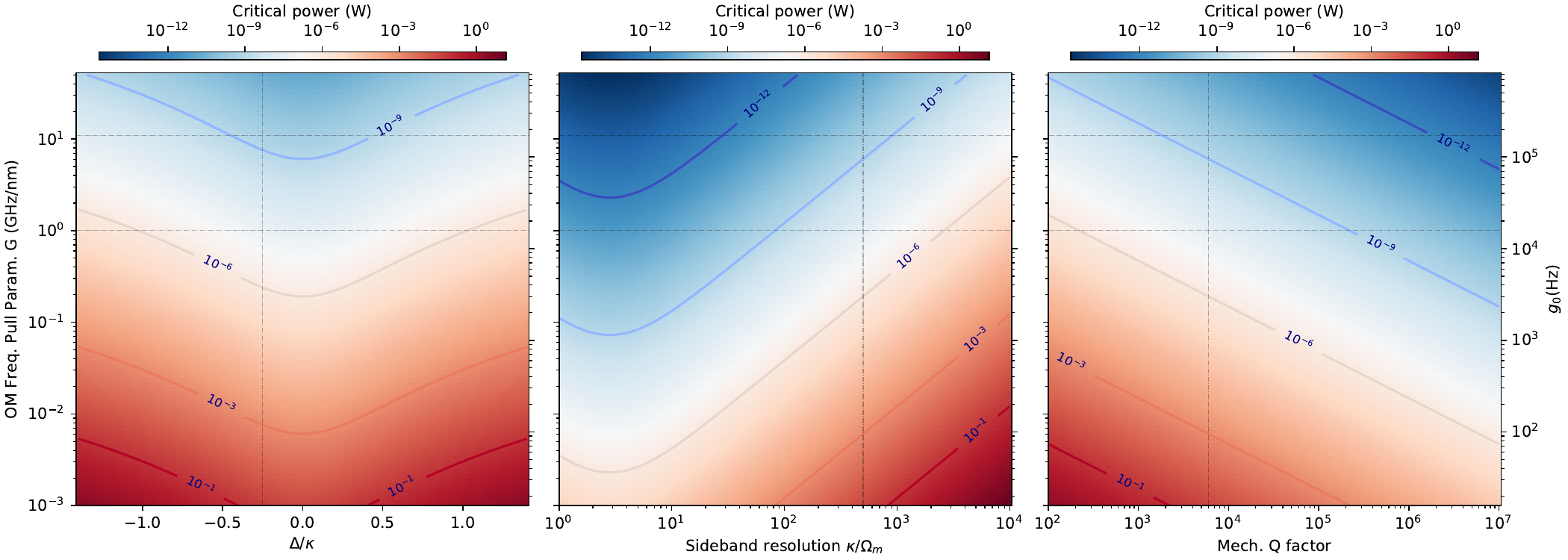}
	\caption{Critical-power maps in the USR at $T=0$, computed from \cref{eq:P_crit}. Left: $P_{\mathrm{crit}}(G,\Delta/\kappa)$ at fixed $\kappa/2\pi\simeq\qty{1}{\giga\hertz}$ and $Q\simeq\num{6e3}$. Center: $P_{\mathrm{crit}}(G,\kappa/\Omega_m)$ at fixed $\Delta=-\Omega_m$ ($Q$ as before). Right: $P_{\mathrm{crit}}(G,Q)$ at fixed $\Delta=-\Omega_m$ ($\kappa$ as before). Color and contour labels report absolute critical power (\unit{\watt}). Horizontal guide lines indicate nominal and maximum $G$ used in this work; vertical guide lines indicate the reference values of our system.}
	\label{fig:USR_Pcrit_VS_OMparam}
\end{figure}

\subparagraph{Critical Power vs. System Parameters}
The result in \cref{eq:P_crit} can be used to map a wide parameter space for perfect destructive interference, rather than a single operating point. \Cref{fig:USR_Pcrit_VS_OMparam} shows the predicted critical power (ignoring thermal effect) as a function of the frequency-pull parameter $G$ and three other important OM parameters: detuning $\Delta/\kappa$ (left), sideband resolution $\kappa/\Omega_m$ (center), and mechanical quality factor $Q$ (right). Contour labels indicate absolute power values in watts.

Three trends are immediate. First, increasing $G$ strongly lowers $P_{\mathrm{crit}}$ (consistent with the $1/G^2$ scaling in \cref{eq:P_crit}). Second, the sideband-resolution map exhibits a minimum for values comparable to the chosen detuning (here $\Delta=-\Omega_m$); moving away from this region increases the required power. Third, lower mechanical quality factor (larger $\Gamma_m$) penalizes the interference condition and raises $P_{\mathrm{crit}}$. The black guide lines mark the operating point used in previous figures (and in the experiment) and the maximum experimentally accessible coupling, providing a direct readout of the accessible parameter window. In this USR parameter set, the critical point can be reached at relatively low input power (from \unit{\nano\watt} to \unit{\micro\watt}), even for moderate $Q$. By contrast, many resolved-sideband implementations would require much higher optical power, often in the \unit{\milli\watt} range, to achieve comparable results (see below).

\begin{figure}[b!]
	\centering
	\includegraphics[width=1.0\textwidth]{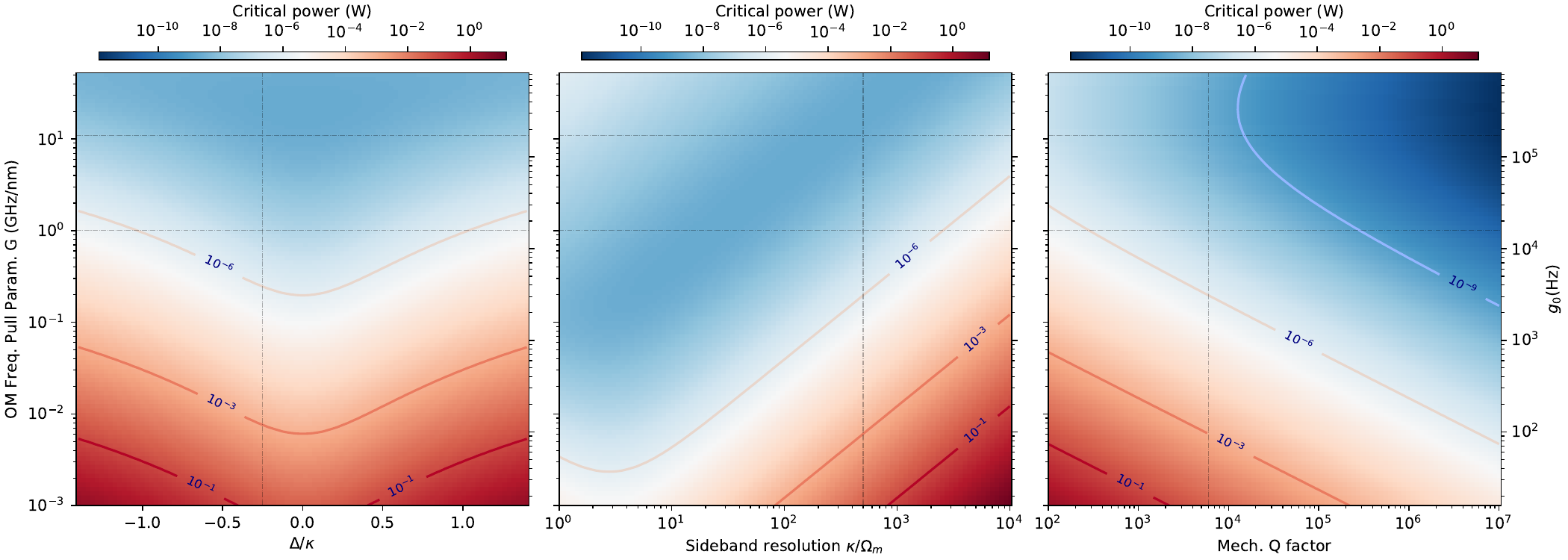}
	\caption{Critical-power maps in the USR at $T=\qty{300}{\kelvin}$, computed with the thermal correction $\kappa_{\mathrm{eff}}=\kappa+\delta\kappa(T)$. Panels and axes follow \cref{fig:USR_Pcrit_VS_OMparam}. Relative to $T=0$, the maps are shifted to higher power, with the largest increase in the high-$G$ region where thermally induced linewidth broadening is strongest.}
	\label{fig:USR_Pcrit_VS_OMparam_T300}
\end{figure}

To quantify how thermal broadening modifies these power maps, we repeat the same parameter sweeps at room temperature by using the effective linewidth model introduced above, $\kappa\rightarrow\kappa_{\mathrm{eff}}(T)=\kappa+\delta\kappa(T)$. The result is shown in \cref{fig:USR_Pcrit_VS_OMparam_T300}. Compared with \cref{fig:USR_Pcrit_VS_OMparam}, the overall structure is preserved but the absolute power scale shifts upward, reflecting reduced intracavity buildup at fixed input power. Because $\delta\kappa\propto g_0\sqrt{2n_{\mathrm{th}}}\propto G\sqrt{T}$ in this phenomenological model, the thermal penalty is strongest at large $G$. Furthermore, better sideband resolution makes thermal effects relevant sooner, which can require cryogenic operation for a given parameter set.

\begin{figure}[t!]
	\centering
	\includegraphics[width=1.0\textwidth]{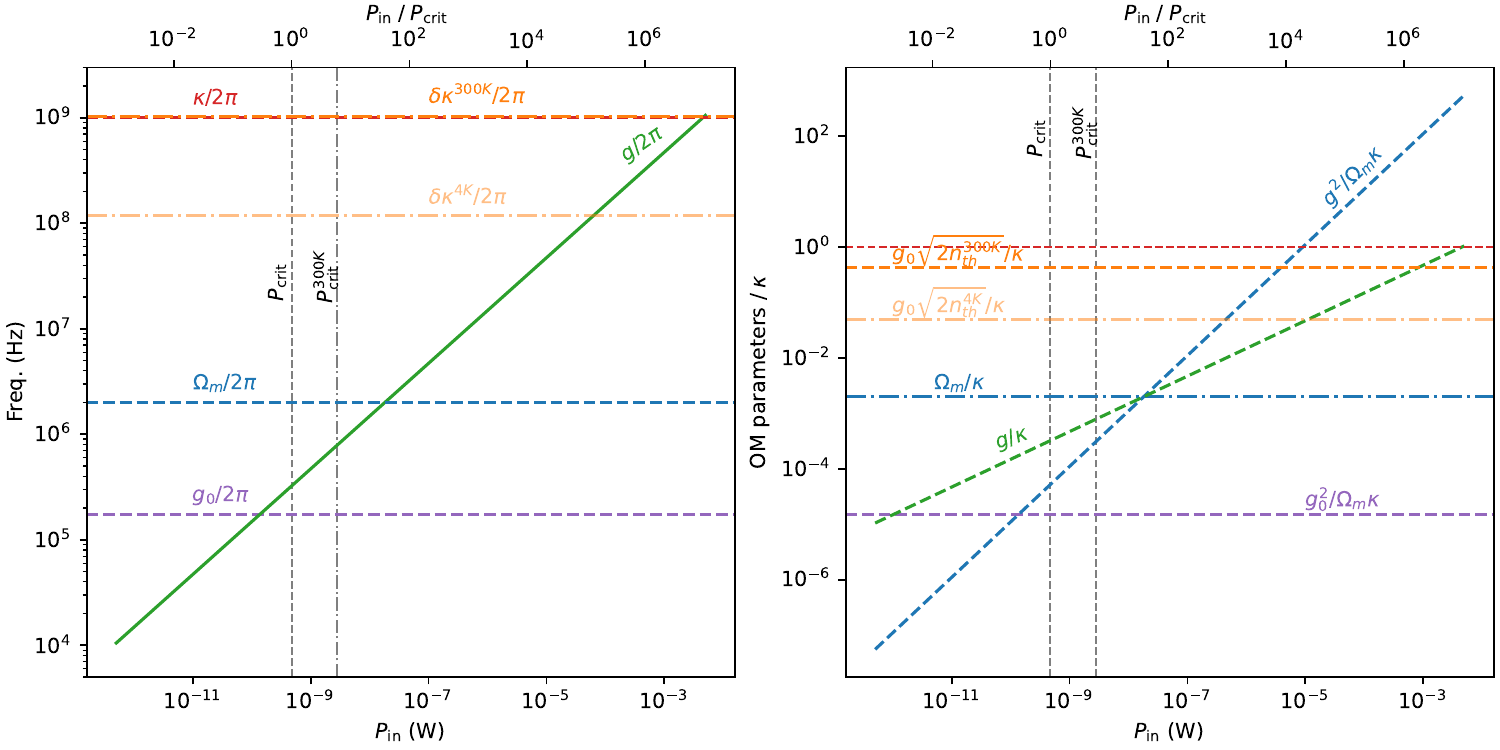}
	\caption{Power-dependent overview of key optomechanical parameters for the USR set. Left: absolute scales versus $P_{\mathrm{in}}$ (bottom axis) and $P_{\mathrm{in}}/P_{\mathrm{crit}}$ (top axis), including $g_0$, $g(P_{\mathrm{in}})$, $\Omega_m$, $\kappa$, $\delta\kappa^{4K}$, $\delta\kappa^{300K}$, and markers for $P_{\mathrm{crit}}^{0K}$ and $P_{\mathrm{crit}}^{300K}$. Right: dimensionless ratios normalized by $\kappa$, highlighting coupling, sideband-resolution, single-photon blockade parameter, and thermal-jitter indicators in a form suited for direct regime comparison.}
	\label{fig:USR_OM_params}
\end{figure}

\subparagraph{Overview of relevant optomechanical scales}
As a compact summary before comparing regimes, \cref{fig:USR_OM_params} collects the main optomechanical scales as a function of input power for the present USR parameter set. Assuming the maximum value achievable in our setup for the pull parameter $G/2\pi=\qty{11}{\giga\hertz\per\nano\meter}$, the left panel shows absolute frequencies, including the single-photon coupling $g_0$, field-enhanced coupling $g(P_{\mathrm{in}})$, $\Omega_m$, $\kappa$, and thermal linewidth broadening levels $\delta\kappa^{4K}$ and $\delta\kappa^{300K}$, together with vertical markers for $P_{\mathrm{crit}}^{0K}$ and $P_{\mathrm{crit}}^{300K}$.

The right panel rewrites the same information in normalized form, using $\kappa$ as the reference scale: $\Omega_m/\kappa$, $g/\kappa$, $g^2/(\Omega_m\kappa)$, $g_0^2/(\Omega_m\kappa)$, and the thermal-jitter metric $g_0\sqrt{2n_{\mathrm{th}}}/\kappa$ (4 K and 300 K), with the unity line as a visual threshold. This representation is convenient for cross-platform comparison and will be reused to contrast USR, small-USR, and RSR settings on equal footing.

\newpage
\subparagraph{RSR vs USR}
To isolate which features are specific to the unresolved-sideband regime, we now repeat the same analysis in a resolved-sideband reference set, using the experimental parameters from \cite{weisOptomechanicallyInducedTransparency2010} ($\Omega_m/2\pi=\qty{51.8}{\mega\hertz}$, $\Gamma_m=\qty{41}{\kilo\hertz}$, $M_{\mathrm{eff}}\simeq\qty{20}{\nano\gram}$, $\kappa/2\pi=\qty{15}{\mega\hertz}$, $\Delta=-\Omega_m$, $\lambda=\qty{1550}{\nano\meter}$), and compare directly with \cref{fig:USR_noRWA_VS_RWA,fig:USR_OMIT_vs_dKappa,fig:USR_Pcrit_VS_OMparam_T300,fig:USR_OM_params}.

\begin{figure}[!t]
	\centering
	\includegraphics[width=1.0\textwidth]{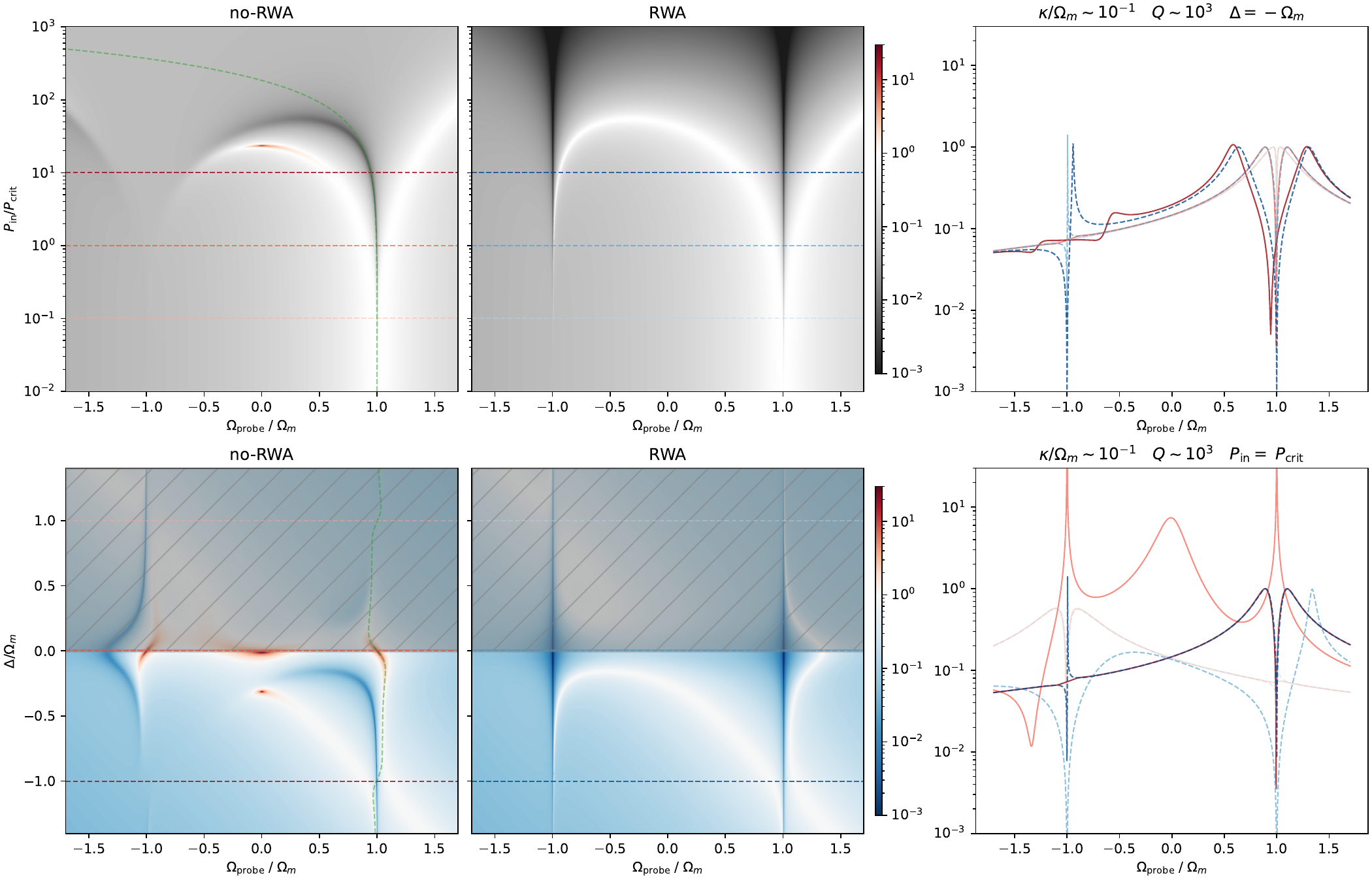}
	\caption{RSR comparison between the full probe response (\textbf{no-RWA}, \cref{eq:Aminus3}) and the sideband-RWA approximation (\textbf{RWA}, \cref{eq:Aminus_RWA}). \textbf{Top row:} normalized response $R\propto|A^-|$ versus $(\Omega_{\mathrm{probe}}/\Omega_m,P_{\mathrm{in}}/P_{\mathrm{crit}})$ at fixed $\Delta=-\Omega_m$. \textbf{Bottom row:} $R$ versus $(\Omega_{\mathrm{probe}}/\Omega_m,\Delta/\Omega_m)$ at fixed $P_{\mathrm{in}}=P_{\mathrm{crit}}$, with line cuts at $\Delta/\Omega_m=-1,0,+1$. The dashed green traces mark $\Omega_m + \delta\Omega_m$ in the no-RWA maps. The hatched region corresponds to blue detuning ($\Delta>0$), where the linearized steady-state model becomes dynamically unstable.}
	\label{fig:RSR_noRWA_VS_RWA}
\end{figure}
\begin{figure}[!t]
	\centering
	\includegraphics[width=0.93\textwidth]{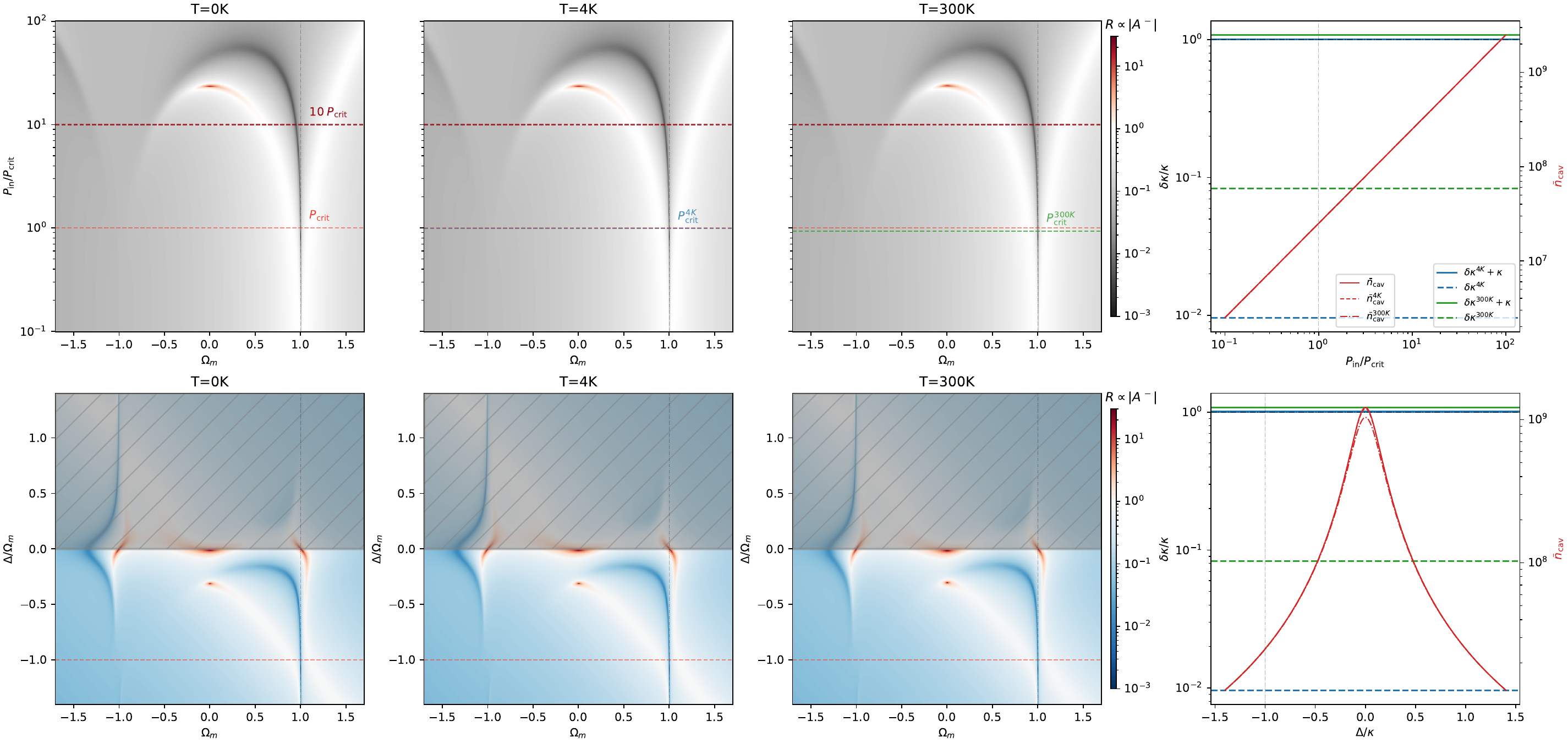}
	\caption{RSR response including thermally induced linewidth broadening, computed with $\kappa_{\mathrm{eff}}=\kappa+\delta\kappa(T)$ for $T=0,4,\qty{300}{\kelvin}$. Left three columns: normalized response maps like in \cref{fig:RSR_noRWA_VS_RWA}. Right column: corresponding $\delta\kappa/\kappa$ (left axis) and $\bar n_{\mathrm{cav}}$ (right axis), versus power (top) and detuning (bottom). Thermal broadening has a less pronounced impact in these RSR scans. The hatched region corresponds to blue detuning ($\Delta>0$), dynamically unstable.}
	\label{fig:RSR_OMIT_VS_dKappa}
\end{figure}

\Cref{fig:RSR_noRWA_VS_RWA} shows that, in the RSR, the full expression \cref{eq:Aminus3} and its RWA limit \cref{eq:Aminus_RWA} are much closer than in the USR case, at least for input power lower than $10 P_{\mathrm{crit}}$. In particular, around the red-sideband working point ($\Delta\simeq-\Omega_m$), the counter-rotating contribution has a weaker impact on the lineshape, so the no-RWA and RWA cuts nearly overlap over a broad power range. This is the expected sideband-filtering behavior: once $\kappa\ll\Omega_m$, the off-resonant channel is strongly suppressed and the RWA becomes quantitatively accurate. Again, the full model includes the effects due to the backaction which are absent from the RWA model, and it holds until the linear approximation is valid. In the detuning map, we can also see regions where gain effects emerge, particularly when approaching $\Delta=0$. Note, however, that in all RSR maps, the hatched half-plane ($\Delta>0$, blue detuning) marks a dynamically unstable region (parametric amplification/anti-damping), where the steady-state linear response used here is not reliable and the plotted values should not be interpreted quantitatively.

\Cref{fig:RSR_OMIT_VS_dKappa} adds thermal linewidth fluctuations using the same phenomenological model as above, $\kappa_{\mathrm{eff}}=\kappa+\delta\kappa(T)$. For this parameter set, the correction remains moderate (order $10^{-2}$ at 4 K and $10^{-1}$ at 300 K in relative linewidth), and it mainly shifts quantitative thresholds (critical-power lines and contrast) without changing the qualitative RSR behavior. Compared with the USR case, no rapid dip-to-enhancement inversion is observed in the same normalized scan range.

\begin{figure}[!b]
	\centering
	\includegraphics[width=0.93\textwidth]{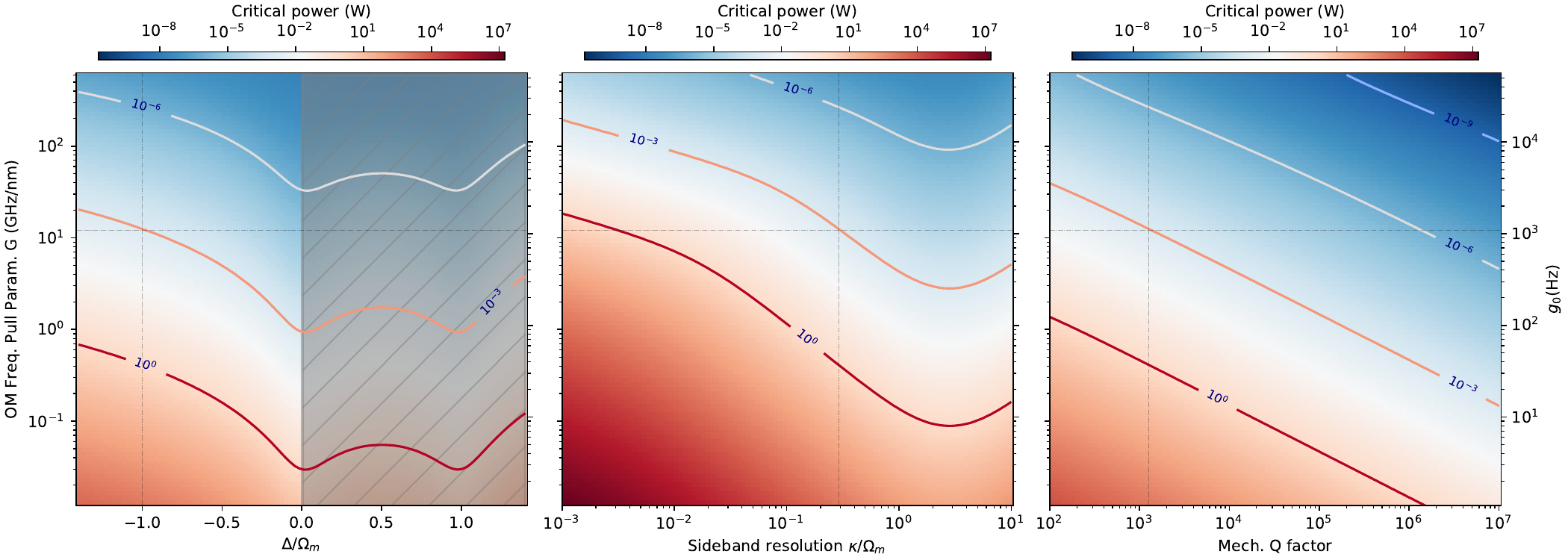}
	\caption{Critical-power maps in the RSR at $T=\qty{300}{\kelvin}$, computed with the thermal correction $\kappa_{\mathrm{eff}}=\kappa+\delta\kappa(T)$. Left: $P_{\mathrm{crit}}(G,\Delta/\Omega_m)$; center: $P_{\mathrm{crit}}(G,\kappa/\Omega_m)$; right: $P_{\mathrm{crit}}(G,Q)$. Colors and contour labels give absolute critical power (W). Guide lines indicate the reference operating point used in the RSR comparison (values from \cite{weisOptomechanicallyInducedTransparency2010}). The hatched region corresponds to blue detuning ($\Delta>0$), where the linearized steady-state model becomes dynamically unstable; results there are shown only for completeness and are not used for quantitative interpretation.}
	\label{fig:RSR_Pcrit_VS_OMparam_T300}
\end{figure}

\Cref{fig:RSR_Pcrit_VS_OMparam_T300} reports the critical-power landscape at room temperature. The same dependencies seen in USR are recovered (strong decrease with increasing $G$, penalty for lower $Q$, and sensitivity to sideband resolution), but the absolute power scale is significantly higher than in the USR maps, typically in the mW range around the chosen operating point. This highlights a key practical difference between regimes: in our USR configuration, critical conditions can be reached at much lower optical power. Again the sideband resolution map shows a minimum for values comparable to the detuning $\Delta$, and increases for higher resolved cases.

\begin{figure}[!t]
	\centering
	\includegraphics[width=0.8\textwidth]{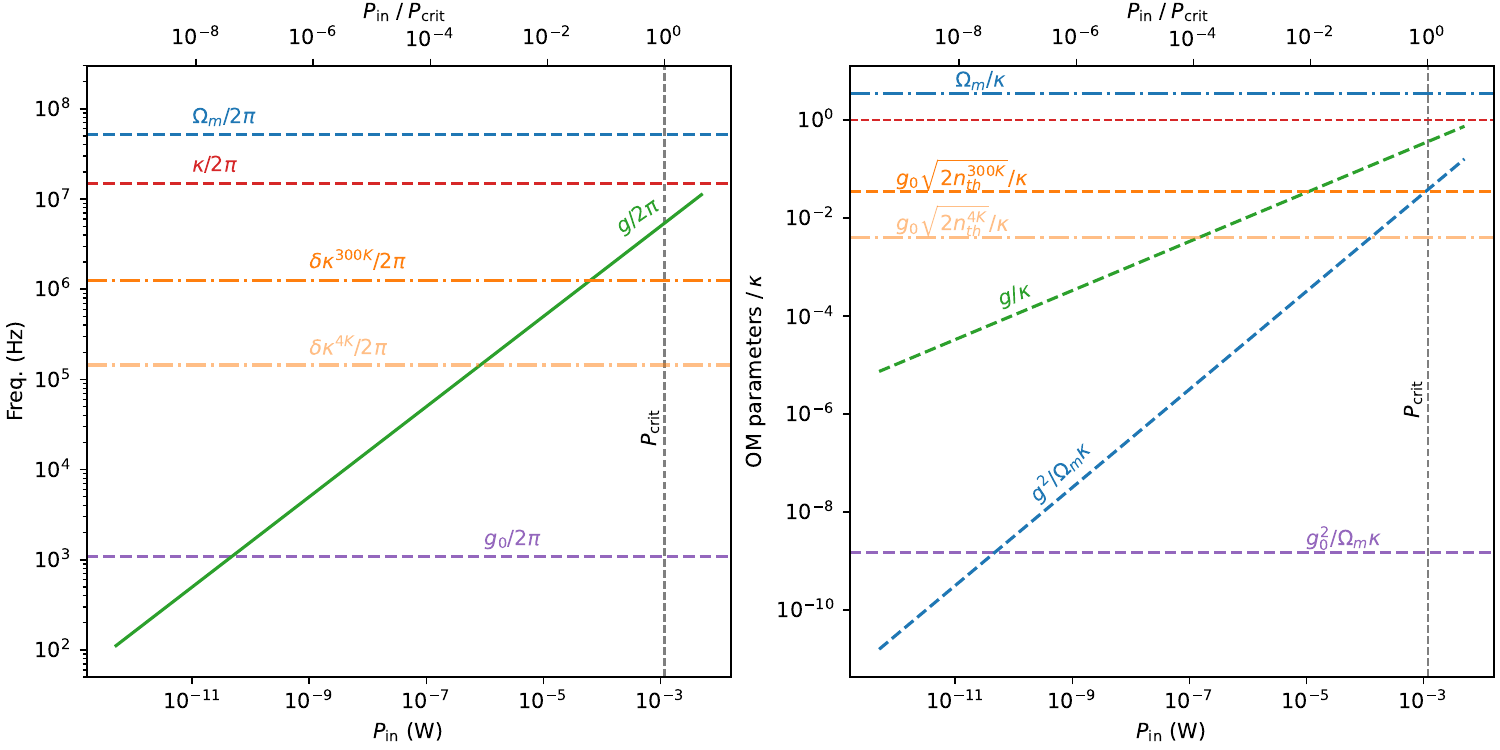}
	\caption{Power-dependent overview of key RSR optomechanical parameters. Left: absolute scales (including $g_0$, $g(P_{\mathrm{in}})$, $\Omega_m$, $\kappa$, and thermal broadening levels) versus $P_{\mathrm{in}}$ with a top axis in units of $P_{\mathrm{in}}/P_{\mathrm{crit}}$. Right: normalized quantities ($g/\kappa$, $g^2/\Omega_m\kappa$, $g_0^2/\Omega_m\kappa$, $\Omega_m/\kappa$, and $g_0\sqrt{2n_{\mathrm{th}}}/\kappa$), used to compare regime hierarchies on equal footing.}
	\label{fig:RSR_OM_params}
\end{figure}

Finally, \cref{fig:RSR_OM_params} summarizes the relevant scales versus input power for the RSR set. The left panel gives absolute rates/frequencies; the right panel gives normalized metrics ($g/\kappa$, $g^2/\Omega_m\kappa$, $g_0^2/\Omega_m\kappa$, $\Omega_m/\kappa$, and $g_0\sqrt{2n_{\mathrm{th}}}/\kappa$). Compared with the USR overview, the sideband resolution advantage is clear ($\Omega_m/\kappa>1$), but the much smaller single-photon coupling $g_0$ and the higher mechanical frequency $\Omega_m$ result in a much smaller value for the single-photon blockade parameter $g_0^2/\Omega_m\kappa$.

\subparagraph{Small USR}
To conclude the comparison, it is useful to consider an intermediate regime that is still unresolved-sideband, but only moderately so: $\kappa/\Omega_m \simeq 10$. We use parameters close to our experimental platform ($\Omega_m/2\pi\simeq \qty{5}{\mega\hertz}$, $Q\simeq\num{1e3}$, $M_{\mathrm{eff}}\simeq\qty{10}{\pico\gram}$, $\kappa/2\pi\simeq\qty{50}{\mega\hertz}$, $\lambda=\qty{920}{\nano\meter}$), with detuning $\Delta=-\Omega_m$. This intermediate point is particularly useful because it allows us to highlight features that can remain hidden in
more extreme parameter regimes.
\begin{figure}[b!]
	\centering
	\includegraphics[width=0.85\textwidth]{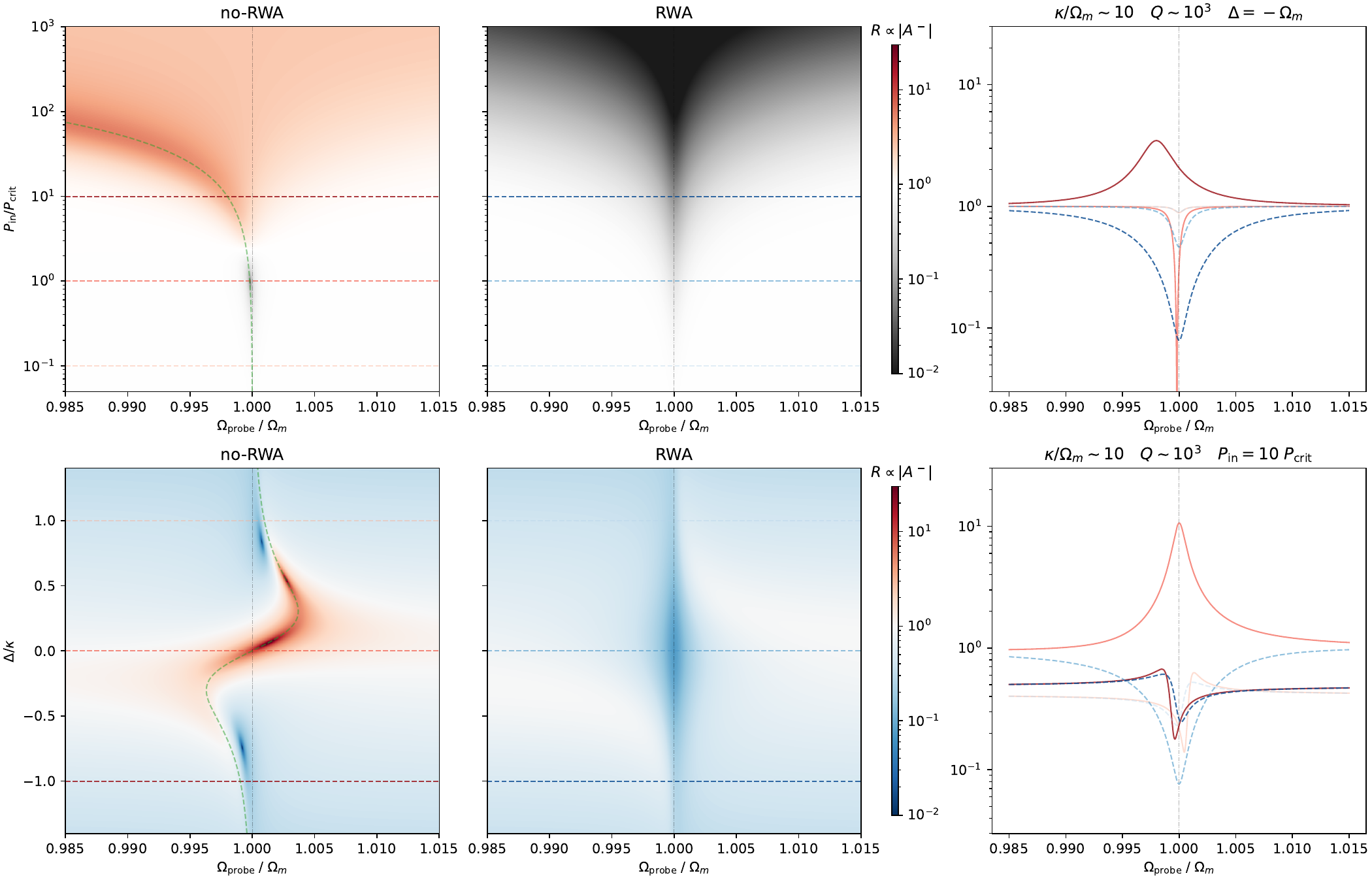}
	\caption{Small-USR comparison of no-RWA and RWA responses for $\kappa/\Omega_m\sim10$. Layout and normalization follow \cref{fig:USR_noRWA_VS_RWA}. The no-RWA detuning map shows a stronger blue-detuned enhancement ($\Delta>0$), consistent with anti-damping and OMIT gain acting together.}
	\label{fig:USRsmall_noRWA_VS_RWA}
\end{figure}

\Cref{fig:USRsmall_noRWA_VS_RWA} shows that this regime is still genuinely a USR, and the no-RWA and RWA models return clearly distinct results, especially in the detuning maps. In the no-RWA detuning map, the blue-detuned side ($\Delta>0$) exhibits a pronounced enhancement: cavity anti-damping and OMIT-induced probe enhancement act in the same direction, producing a stronger feature than in the red-detuned sector. This is precisely where the counter-rotating channel retained in \cref{eq:Aminus3} is most visible.

\begin{figure}[t!]
	\centering
	\includegraphics[width=1.0\textwidth]{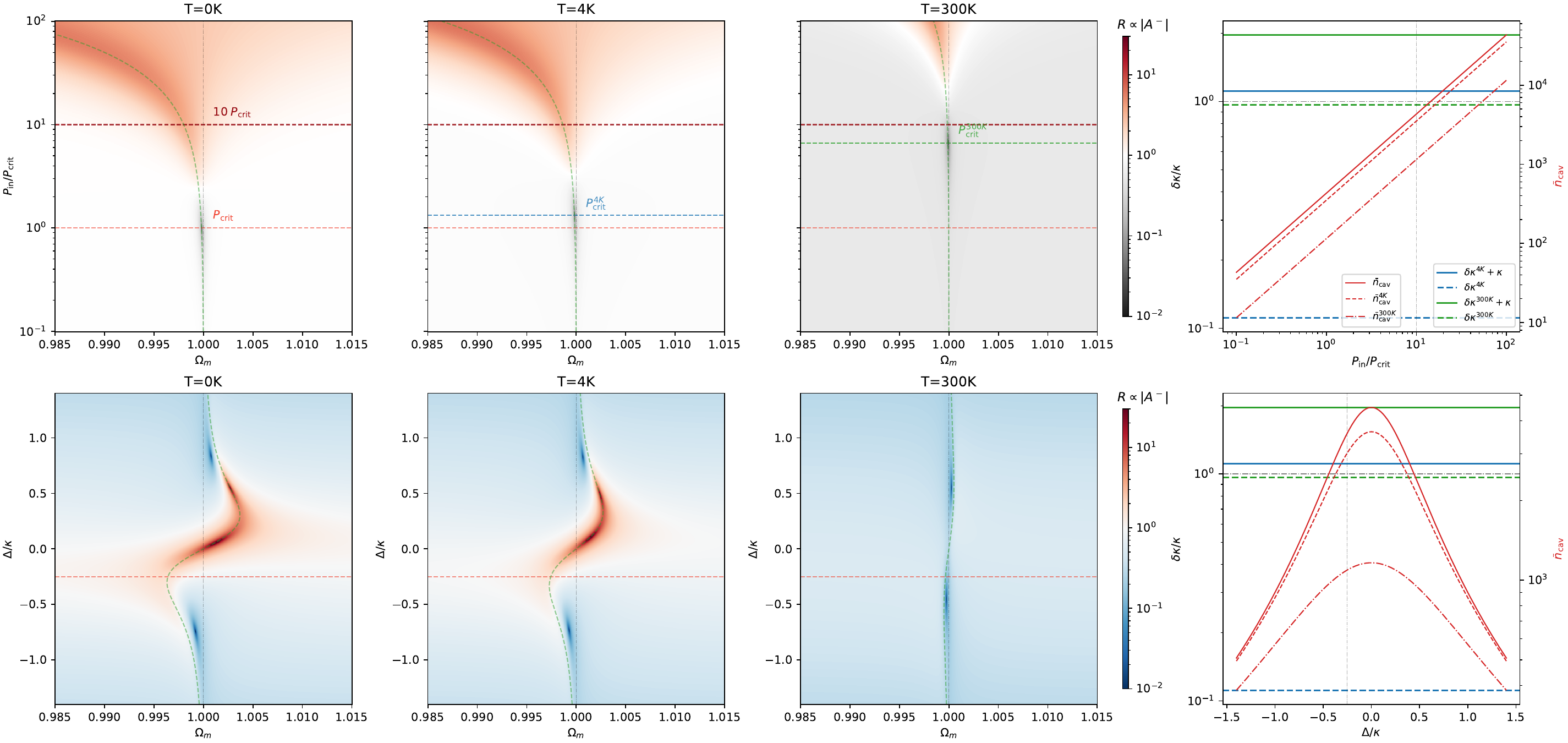}
	\caption{Small-USR response including thermal linewidth broadening ($\kappa_{\mathrm{eff}}=\kappa+\delta\kappa$) for $T=0,4,\qty{300}{\kelvin}$. Layout follows \cref{fig:USR_OMIT_vs_dKappa}. Compared with the deep-USR case, thermal effects are stronger (relative to $\kappa$) and shift the critical-power thresholds more significantly.}
	\label{fig:USRsmall_OMIT_VS_dKappa}
\end{figure}

\Cref{fig:USRsmall_OMIT_VS_dKappa} then shows that thermal broadening is much more critical in this small-USR case than in the previous USR example. With the same phenomenological model, the nominal-coupling set already gives $\delta\kappa/\kappa\sim10^{-1}$ at 4 K and order unity at 300 K; for larger accessible $G$, even 4 K can become strongly broadened, which points to dilution-temperature operation if one wants to access the high-coupling region quantitatively. This trend is reflected by the upward shift of the critical-power lines and the reduction of contrast.
\begin{figure}[b!]
	\centering
	\includegraphics[width=1.0\textwidth]{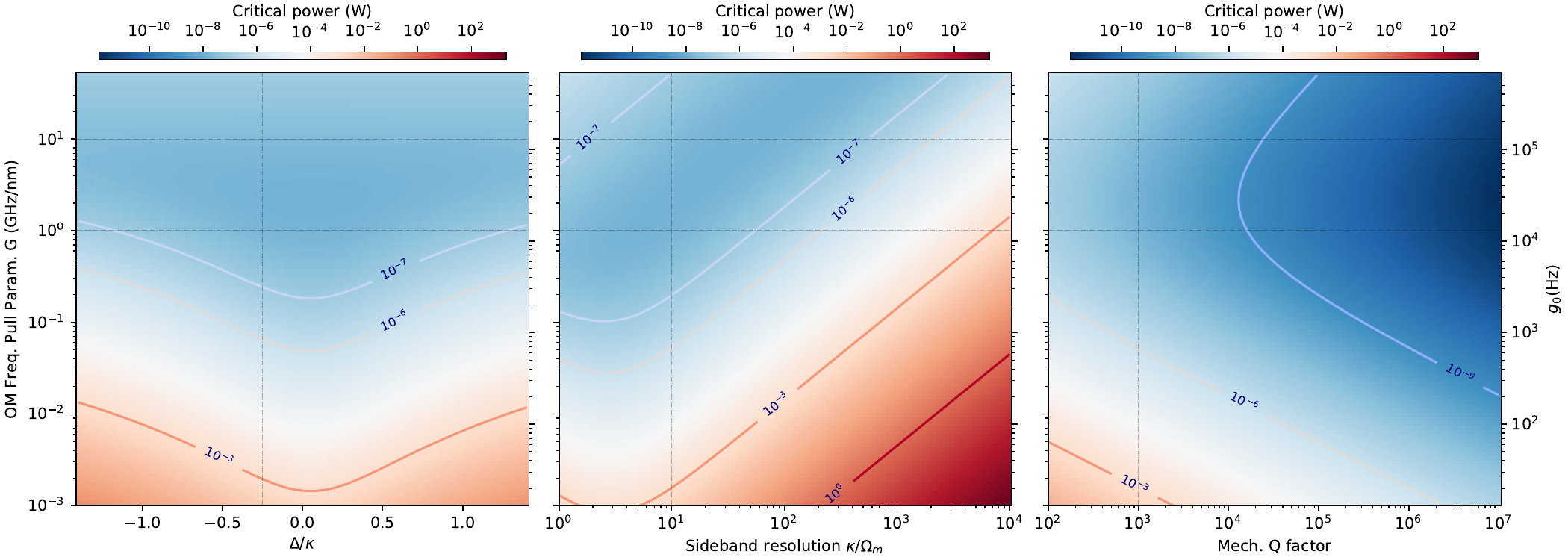}
	\caption{Critical-power maps for the small-USR set at $T=\qty{300}{\kelvin}$, with thermal correction $\kappa_{\mathrm{eff}}=\kappa+\delta\kappa(T)$. Panels and conventions follow \cref{fig:USR_Pcrit_VS_OMparam_T300}.}
	\label{fig:USRsmall_Pcrit_VS_OMparam_T300}
\end{figure}

\begin{figure}[t!]
	\centering
	\includegraphics[width=1.0\textwidth]{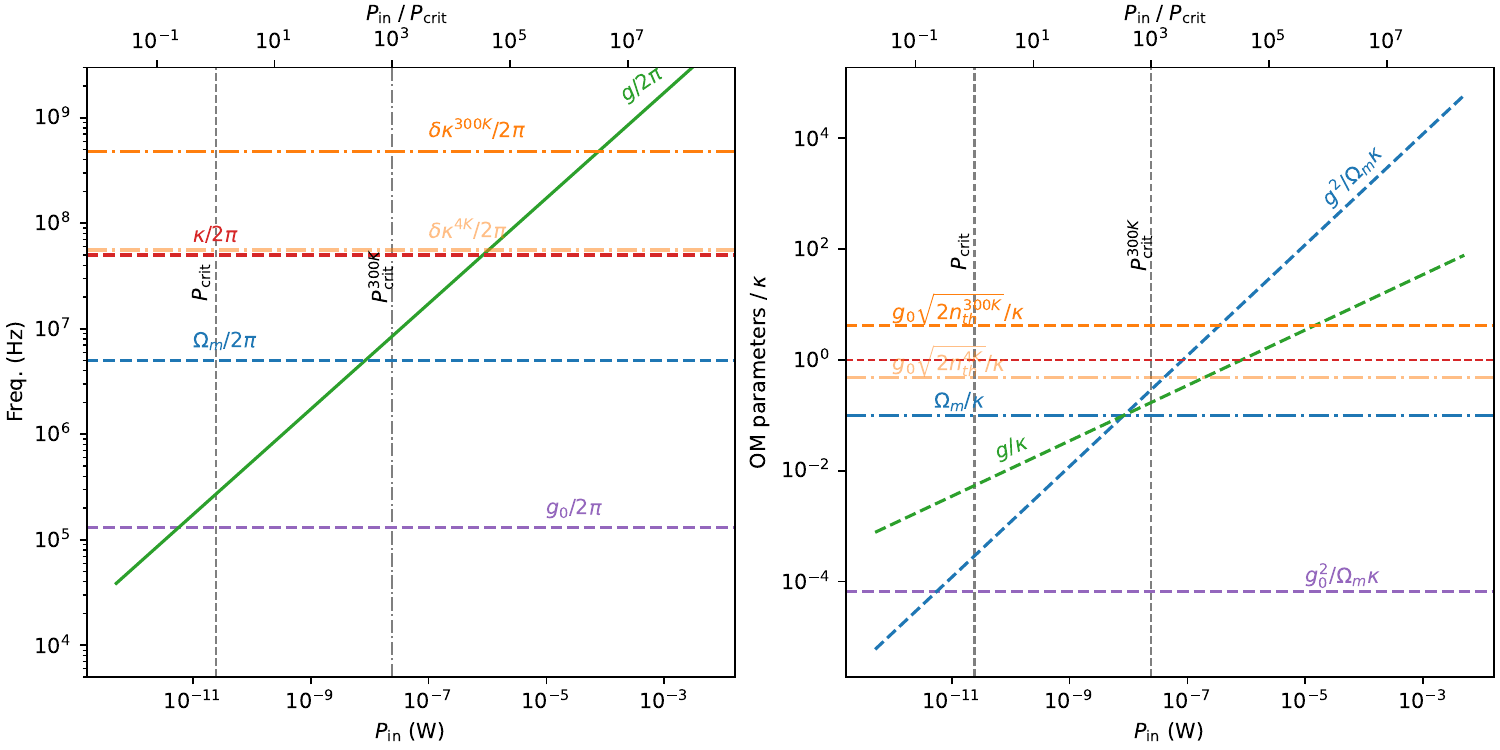}
	\caption{Power-dependent overview of key optomechanical scales for the small-USR set. Same quantities and conventions as \cref{fig:USR_OM_params}, highlighting the intermediate hierarchy between deep-USR and RSR.}
	\label{fig:USRsmall_OM_params}
\end{figure}

The room-temperature critical-power maps in \cref{fig:USRsmall_Pcrit_VS_OMparam_T300} keep the same qualitative structure as before, but show a stronger thermal penalty than in the deep-USR case. We note, at the same time, that the optical linewidth and coupling values employed here remain in an experimentally realistic range for fiber-cavity/hBN-type platforms \cite{sanchezarribasRadiationPressureBackaction2023}. Finally, \cref{fig:USRsmall_OM_params} summarizes the scale hierarchy: although this regime is experimentally favorable compared to strict RSR for reaching nonlinear signatures at moderate power, the single-photon-blockade metric $g_0^2/(\Omega_m\kappa)$ remains several orders of magnitude below unity, so true single-photon nonlinearity still requires substantially larger $g_0$ and/or lighter resonators (e.g. CNT-class devices).

\subsection{Spectrum of the reflected light, input-output relation and signal demodulation}
In the previous sections we have treated the intracavity dynamics and obtained the full expression for the probe-sideband amplitude $A^-$ in the unresolved-sideband regime. To connect this internal response to what is actually measured in the experiment, we now turn to the input–output description of the cavity field and to the spectrum of the light reflected from the cavity.

\subsubsection{Input-output formalism}
\label{subsec:input_output}
We first recall the standard input–output relations using the conventions from \cite{aspelmeyerCavityOptomechanics2014} and express the reflected field in terms of the incident field and the intracavity field $\hat{a}$. We then describe how the electro-optic modulator (EOM) is used to generate the various tones (lock, control/pump, and OMIT probe), and how the reflected signal contains the corresponding beat notes between carrier and sidebands. Finally, we show how demodulation (either in the Pound–Drever–Hall sense for cavity locking or in an IQ/heterodyne scheme for OMIT readout) selects the relevant quadratures of the reflected field, making explicit how the theoretical response $A^-(\Omega)$ appears in the experimentally measured signals.

Let's start by defining the input fields, the output fields and the field inside the cavity. According to our definition, the input field is:
\begin{subequations}
	\begin{empheq}[left=\empheqlbrace]{align}
		\hat{a}_{\mathrm{in}}(t) 		&= \left[ \bar{a}_{\mathrm{in}} + \delta\hat{a}_{\mathrm{in}}(t) \right] e^{-i\omega_l t} \\
		\delta\hat{a}_{\mathrm{in}}(t) 	&= a_p e^{-i\Omega t} \qquad \text{with} \qquad \Omega = \omega_p - \omega_l
	\end{empheq}
\end{subequations}
with $\bar{a}_{\mathrm{in}}$ the pump field and $\delta\hat{a}_{\mathrm{in}}$ the probe field. For the output field, according to input-output theory \cite{aspelmeyerCavityOptomechanics2014}, we have:
\begin{subequations}
	\begin{empheq}[left=\empheqlbrace]{align}
		\hat{a}^R_{\mathrm{out}}(t)	&= \hat{a}_{\mathrm{in}}(t) - \sqrt{\kappa_{\mathrm{ext}}}\hat{a}(t) \\
		\hat{a}^T_{\mathrm{out}}(t)	&= \sqrt{\kappa_{\mathrm{ext}}}\hat{a}(t)
	\end{empheq}
\end{subequations}
Finally, for the intracavity field, we can write, in the frame rotating at the laser frequency:
\begin{equation}
	\hat{a}(t) = \bar{a} + \delta\hat{a}(t)
\end{equation}
Focusing on the static field, we can write the following relations:
\begin{equation}
	\bar{a} = \bar{a}_{in} \sqrt{\kappa_{\mathrm{ext}}} \chi_{\mathrm{cav}}[\bar{\Delta}] = \frac{\sqrt{\kappa_{\mathrm{ext}}}}{\frac{\kappa}{2}-i\bar{\Delta}}\bar{a}_{in}
	\qquad \text{with} \qquad
	\chi_{\mathrm{cav}}[\bar{\Delta}] = \frac{1}{\frac{\kappa}{2}-i\bar{\Delta}}
\end{equation}
where $\chi_{\mathrm{cav}}$ is the optical cavity susceptibility. Then, for the reflected field, we can write:
\begin{equation}
	\bar{a}^R_{\mathrm{out}} 	= \bar{a}_{\mathrm{in}} - \sqrt{\kappa_{\mathrm{ext}}}\bar{a}
	= \bar{a}_{\mathrm{in}} \left( 1 - \kappa_{\mathrm{ext}} \chi_{\mathrm{cav}}[\bar{\Delta}] \right)
	= \bar{a}_{\mathrm{in}} \left( 1 - \frac{\kappa_{\mathrm{ext}}}{\frac{\kappa}{2}-i\bar{\Delta}} \right)
	= \bar{a}_{\mathrm{in}} \mathrm{K}^R_{\mathrm{cav}}[\bar{\Delta}]
\end{equation}
where with $\mathrm{K}^R_{\mathrm{cav}}$ we indicate the cavity reflection transfer function. Note also that for the average number of photons in the cavity we will use
\begin{equation}
	\bar{n}_{\mathrm{cav}} = |\!\left<\hat{a}\right>\!|^2 = |\bar{a}|^2 = \frac{\kappa_{\mathrm{ext}}}{\left(\frac{\kappa}{2}\right)^2 + \bar{\Delta}^2} \frac{P_{\mathrm{in}}}{\hbar \omega_l}
\end{equation}
with $P_{\mathrm{in}} = \hbar\omega_l |\bar{a}_{\mathrm{in}}|^2$ the cavity input power.

\subsubsection{EOM sideband generation and heterodyne detection}
In our experiment the electro-optic modulator (EOM) plays a dual role. First, it is used in the standard Pound–Drever–Hall (PDH) configuration to generate phase-modulation sidebands around the control laser, providing the error signal needed to lock the cavity close to resonance. Second, the same modulation chain forms the basis of our OMIT measurement: the tones created by the EOM (lock sidebands, control/pump tone, and the weak OMIT probe sideband) all propagate through the cavity and produce a set of well-defined beat notes in the reflected signal.

In this subsection we briefly review the description of phase modulation and sideband generation (following \cite{blackIntroductionPoundDrever2000} and related treatments), and introduce the notation that we will use for the carrier and its sidebands at the EOM output. We then outline how the reflected signal, containing the interference between the strong carrier and the various sidebands, is down-mixed and demodulated in a heterodyne to isolate the OMIT response at the probe 
frequency.

A phase‑modulating EOM is essentially a voltage‑controlled phase shifter: the instantaneous phase of the optical carrier is modulated at an RF frequency, which generates sidebands (additional optical tones) around the carrier. Let's consider a single‑tone phase modulation. If we take a monochromatic input field at the laser frequency $\omega_l$:
\begin{equation}
	a_{\phi_{\mathrm{in}}}(t) = a_{\phi_0}e^{-i\omega_l t}.
\end{equation}
The EOM driven by a voltage $V(t)$ imposes a phase $\phi(t) = \pi \frac{V(t)}{V_\pi}$, where $V_\pi$ is the half‑wave voltage (voltage for which the phase changes by $\pi$). For a sinusoidal drive at RF frequency $\Omega$,
\begin{equation}
	V(t) = V_0 \sin(\Omega t)
	\quad\Rightarrow\quad
	\phi(t) = \beta \sin(\Omega t),
\end{equation}
with modulation index $\beta = \pi \frac{V_0}{V_\pi}$. The output field is then
\begin{equation}
	a_{\phi_{\mathrm{out}}}(t) = a_{\phi_0}e^{-i(\omega_l t + \phi(t))} = a_{\phi_0}e^{-i\omega_l t}e^{-i\beta\sin(\Omega t)}.
\end{equation}
Using the Jacobi–Anger expansion
\begin{equation}
	e^{-i\beta\sin(\Omega t)} = \sum_{n=-\infty}^{+\infty} J_n(\beta) e^{-in\Omega t},
\end{equation}
we obtain
\begin{equation}
	\label{eq:EOM_1tone}
	a_{\phi_{\mathrm{out}}}(t) = a_{\phi_0} \sum_{n=-\infty}^{+\infty} J_n(\beta)e^{-i(\omega_l + n\Omega)t}.
\end{equation}
As a result, the EOM produces a carrier at $\omega_l$ (coefficient $J_0(\beta)$) and a comb of sidebands at $\omega_l + n\Omega$, with amplitudes $\propto J_n(\beta)$. For small modulation depth $\beta \ll 1$, we keep only the  carrier and first‑order sidebands. Using the relation $J_{-n}(z) = (-1)^n J_n(z)$:
\begin{equation}
	a_{\phi_{\mathrm{out}}}(t) \approx a_{\phi_0} e^{-i\omega_l t} \left[ J_0(\beta) + J_1(\beta)e^{-i\Omega t} - J_1(\beta)e^{+i\Omega t} \right].
\end{equation}
In this limit, the power in the first sidebands is $\sim J_1(\beta)^2 \approx (\beta/2)^2$, and the carrier is slightly depleted from $1$ to $J_0(\beta) \approx 1 - \beta^2/4$.

Generalizing, if now we consider the EOM driven by several RF tones (e.g. one for PDH locking, two for the OMIT pump/probe),
\begin{equation}
	V(t) = \sum_k V_k \sin(\Omega_k t), \quad \phi(t) = \sum_k \beta_k \sin(\Omega_k t),
\end{equation}
then
\begin{equation}
	a_{\phi_{\mathrm{out}}}(t) = a_{\phi_0} e^{-i\omega_l t} \prod_k e^{-i\beta_k \sin(\Omega_k t)},
\end{equation}
which expands into a set of sidebands at frequencies $\omega_l + \sum_k n_k \Omega_k$ with amplitudes given by
products of Bessel functions $\prod_k J_{n_k}(\beta_k)$. In practice (and for modest modulation indices), one
usually keeps only the carrier and the first‑order sidebands for each modulation frequency.

\subparagraph{Heterodyne detection of a PDH signal}
Let's now focus on the case of one single modulation tone for the detection (for example) of the PDH error signal. We can rewrite the EOM output signal from \cref{eq:EOM_1tone} in terms of our cavity input-output notation. If we identify $a_{\phi_{\mathrm{out}}}(t) \equiv \hat{a}_{\mathrm{in}}(t)$, we can write:
\begin{equation}
	\hat{a}_{\mathrm{in}}(t) = a_{L_0} e^{-i\omega_l t} + a_{L_1} e^{-i(\omega_l+\Omega) t} - a_{L_1} e^{-i(\omega_l-\Omega) t}
\end{equation}
where $a_{L_0}=a_{\phi_0} J_0(\beta)$ and $a_{L_1}=a_{\phi_0} J_1(\beta)$  are real given our modulation choice ($\sin(x)$ instead of $\cos(x)$) and relative Jacobi–Anger expansion.

We can now apply the input-output formalism to this signal and consider the case where the cavity is centered around one of the two sidebands (so $\bar{\Delta}+\Omega \simeq 0$):
\begin{equation}
	\begin{aligned}
		\hat{a}^R_{\mathrm{out}}(t)	&= \hat{a}_{\mathrm{in}}(t) \mathrm{K}^R_{\mathrm{cav}} = \\
		&= a_{L_0}\mathrm{K}^R_{\mathrm{cav}}[\bar{\Delta}] e^{-i\omega_l t}
		+ a_{L_1}\mathrm{K}^R_{\mathrm{cav}}[\bar{\Delta}+\Omega] e^{-i(\omega_l+\Omega) t}
		- a_{L_1}\mathrm{K}^R_{\mathrm{cav}}[\bar{\Delta}-\Omega] e^{-i(\omega_l-\Omega) t} = \\
		&\simeq a_{L_0} e^{-i\omega_l t}
		+ a_{L_1} \left( 1 - \frac{\kappa_{\mathrm{ext}}}{\frac{\kappa}{2}-i(\bar{\Delta}+\Omega)} \right) e^{-i(\omega_l+\Omega) t}
		- a_{L_1} e^{-i(\omega_l-\Omega) t} = \\
		&= \hat{a}_{\mathrm{in}}(t) - a_{L_1} \kappa_{\mathrm{ext}} \chi_{\mathrm{cav}}[\bar{\Delta}+\Omega] e^{-i(\omega_l+\Omega) t} 
		\equiv \hat{a}_{\mathrm{in}}(t) + \hat{a}^R_{\mathrm{cav}}(t)
	\end{aligned}									
\end{equation}
where we use the cavity sideband resolution to approximate $\mathrm{K}^R_{\mathrm{cav}} \approx 1$ outside the cavity. We can then compute the reflected power as
\begin{equation}
	\label{eq:Pout}
	P_{\mathrm{out}} \propto |\hat{a}^R_{\mathrm{out}}|^2 = (\hat{a}_{\mathrm{in}} + \hat{a}^R_{\mathrm{cav}})(\hat{a}^*_{\mathrm{in}} + {{}\hat{a}^R_{\mathrm{cav}}}^*)
\end{equation}
If we decompose this expression, the different terms can be grouped in different ways. First we have the constant terms: $|a_{L_0}|^2 + 2|a_{L_1}|^2 + |\hat{a}^R_{\mathrm{cav}}|^2$. After demodulation those terms rotate at frequency $\Omega$ and get discarded by the low-pass filter. So we can ignore them. Then we have the beatnotes between the local oscillator and the sidebands:
\begin{equation}
	a_{L_0} a_{L_1}^* e^{i\Omega t} + a_{L_0}^* a_{L_1} e^{-i\Omega t} - a_{L_0} a_{L_1}^* e^{-i\Omega t} - a_{L_0}^* a_{L_1} e^{i\Omega t}
\end{equation}
In order to demodulate the signal, we can multiply by $\cos(\Omega t + \phi) = \frac{e^{i(\Omega t + \phi)}+e^{-i(\Omega t + \phi)}}{2}$ and the low-pass filter the result:
\begin{multline}
	\frac{1}{2} \left\{ a_{L_0}a_{L_1}^* e^{i2\Omega t}e^{i\phi} + a_{L_0}a_{L_1}^* e^{-i\phi} + a_{L_0}^*a_{L_1} e^{i\phi} + a_{L_0}^*a_{L_1} e^{-i2\Omega t}e^{-i\phi} \right. \\
	\left. - a_{L_0}a_{L_1}^* e^{i\phi} - a_{L_0}a_{L_1}^* e^{-i2\Omega t}e^{-i\phi} - a_{L_0}^*a_{L_1} e^{i2\Omega t}e^{i\phi}
	- a_{L_0}^*a_{L_1} e^{-i\phi} \right\}
\end{multline}
After low-pass filtering:
\begin{multline}
	\frac{1}{2} \left\{ a_{L_0}a_{L_1}^* (e^{-i\phi}-e^{i\phi}) + a_{L_0}^*a_{L_1} (e^{i\phi}-e^{-i\phi}) \right\} = \\
	\frac{1}{2} \left\{ (a_{L_0}^*a_{L_1} - a_{L_0}a_{L_1}^*)(e^{i\phi}-e^{-i\phi}) \right\} = \\
	-2\Im[a_{L_0}^*a_{L_1}] \sin(\phi) = Q
\end{multline}
Recalling that both $a_{L_0}$ and $a_{L_1}$ are real values, we can see that this term goes to zero.

For the sake of completeness, we can also derive the other quadrature. We apply the same procedure but multiply by $-\sin(\Omega t + \phi) = -\frac{e^{i(\Omega t + \phi)}-e^{-i(\Omega t + \phi)}}{2i}$ instead:
\begin{multline}
	\frac{1}{2i} \left\{ -a_{L_0}a_{L_1}^* e^{i2\Omega t}e^{i\phi} + a_{L_0}a_{L_1}^* e^{-i\phi} - a_{L_0}^*a_{L_1} e^{i\phi} + a_{L_0}^*a_{L_1} e^{-i2\Omega t}e^{-i\phi} \right. \\
	\left. + a_{L_0}a_{L_1}^* e^{i\phi} - a_{L_0}a_{L_1}^* e^{-i2\Omega t}e^{-i\phi} + a_{L_0}^*a_{L_1} e^{i2\Omega t}e^{i\phi}
	- a_{L_0}^*a_{L_1} e^{-i\phi} \right\}
\end{multline}
After low-pass filtering:
\begin{multline}
	\frac{1}{2i} \left\{ a_{L_0}a_{L_1}^* (e^{-i\phi}+e^{i\phi}) - a_{L_0}^*a_{L_1} (e^{i\phi}+e^{-i\phi}) \right\} = \\
	\frac{1}{2i} \left\{ (a_{L_0}a_{L_1}^* - a_{L_0}^*a_{L_1})(e^{i\phi}+e^{-i\phi}) \right\} = \\
	2\Im[a_{L_0}^*a_{L_1}] \cos(\phi) = I
\end{multline}
As expected, this term is also zero.

An alternative way to obtain both quadratures at the same time is the dual-phase down-mixing, which mathematically means to multiply by $e^{-i(\Omega t +\phi)}= e^{-i\Omega t}[\cos(\phi)-i\sin(\phi)]$. Repeating the same procedure as before:
\begin{equation}
	a_{L_0}a_{L_1}^* e^{-i\phi} + a_{L_0}^*a_{L_1} e^{-i2\Omega t}e^{-i\phi} - a_{L_0}a_{L_1}^* e^{-i2\Omega t}e^{-i\phi} - a_{L_0}^*a_{L_1} e^{-i\phi}
\end{equation}
After low-pass filtering:
\begin{equation}
	(a_{L_0}a_{L_1}^* - a_{L_0}^*a_{L_1})e^{-i\phi} = 2\Im[a_{L_0}^*a_{L_1}] e^{-i\phi} = 2\Im[a_{L_0}^*a_{L_1}] [\cos(\phi)-i\sin(\phi)] = Z
\end{equation}
that as we can see is indeed the sum of the two quadrature: $Z=I+iQ$.

Going back to the remaining terms of \cref{eq:Pout}, we can safely ignore the cross terms between the two sidebands, which rotate at $2\Omega$ and focus on the interaction between the cavity signal and the input signal. The term of interest is now
\begin{multline}
	\hat{a}_{\mathrm{in}} {{}\hat{a}^R_{\mathrm{cav}}}^* + \hat{a}^*_{\mathrm{in}} \hat{a}^R_{\mathrm{cav}} = \\
	= -a_{L_0}a_{L_1}^* \kappa_{\mathrm{ext}} \chi^*_{\mathrm{cav}}[\bar{\Delta}+\Omega] e^{i\Omega t}
	- |a_{L_1}|^2 \kappa_{\mathrm{ext}} \chi^*_{\mathrm{cav}}[\bar{\Delta}+\Omega] \\
	+ |a_{L_1}|^2 \kappa_{\mathrm{ext}} \chi^*_{\mathrm{cav}}[\bar{\Delta}+\Omega] e^{i2\Omega t} + \mathrm{c.c.}
\end{multline}
to which applying the dual-phase down mixing as above and low-pass filtering we obtain:
\begin{equation}
	Z = -a_{L_0}a_{L_1}^* \kappa_{\mathrm{ext}} \chi^*_{\mathrm{cav}}[\bar{\Delta}+\Omega] e^{-i\phi}
\end{equation}
From there both quadratures can be obtained as $I=\Re[Z]$, $Q=\Im[Z]$, as well as the phase as $\arg[Z]$ and the amplitude as
\begin{equation}
	|Z| = \sqrt{Z\bar{Z}} = |a_{L_0}||a_{L_1}|\kappa_{\mathrm{ext}}|\chi_{\mathrm{cav}}[\bar{\Delta}+\Omega]|
\end{equation}
which we can see is proportional to the employed laser power (via $a_{L_0}$), to the power in the sidebands, to the external coupling losses and of course the cavity susceptibility.

\subsubsection{Heterodyne measurement of the OMIT signal}
\label{subsec:hetero_OMIT}
\begin{figure}[b!]
	\centering
	\includegraphics[width=1.0\textwidth]{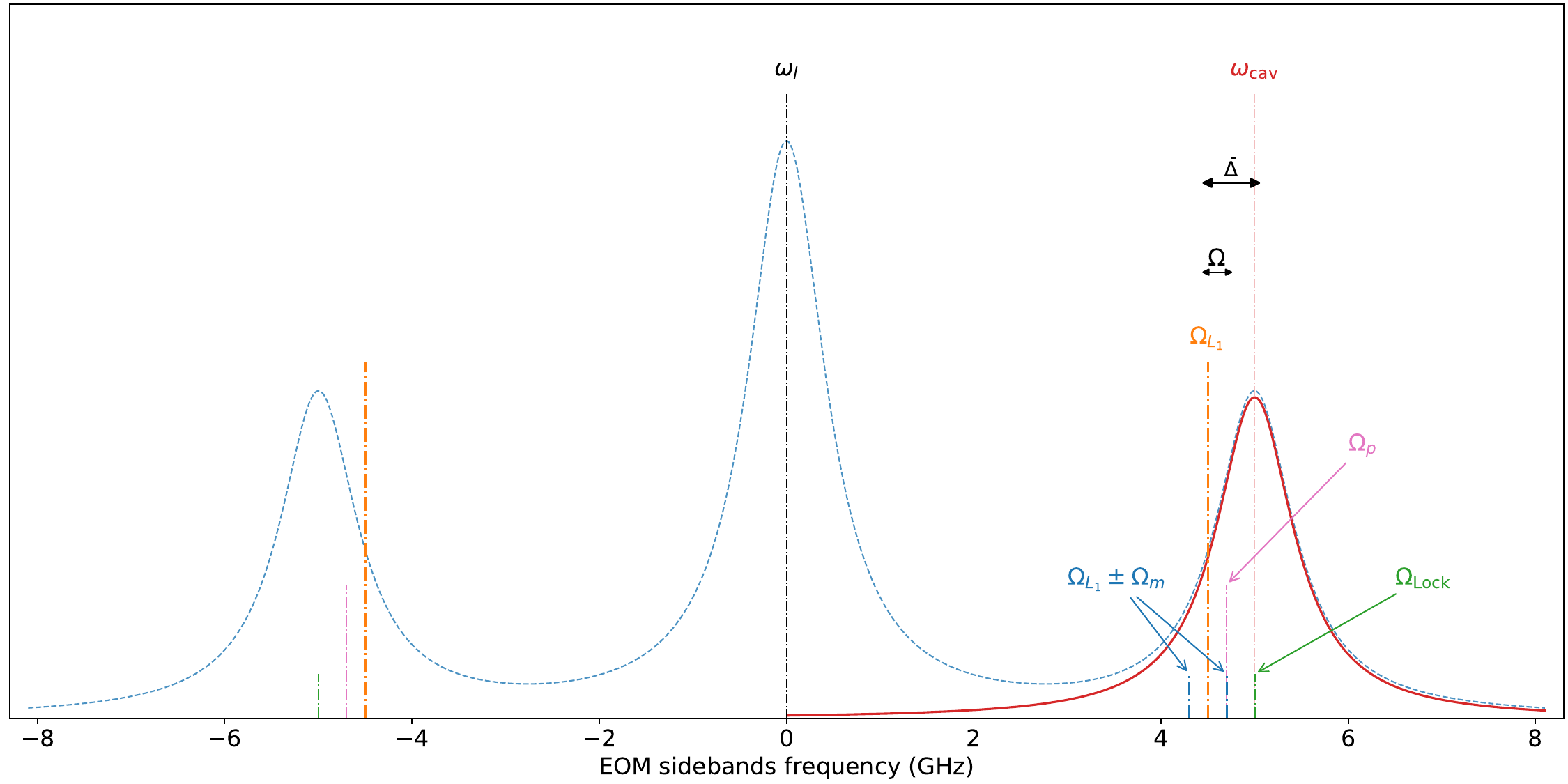}
	\caption{Graphical illustration of the cavity resonant peak(s) and of the optical sidebands generated and employed to lock the cavity and perform the OMIT measurements.}
	\label{fig:sidebands}
\end{figure}

In this last section, we want to discuss how we implement the OMIT measurement and in particular how we detect the reflection of the probe signal. We start by discussing the different tones that we generate with the EOM and the definition of their frequency, which we will adapt from our previous discussion and notation in order to match our experimental needs. These definitions are summarized in \cref{tab:sidebands} and a graphical representation is given in \cref{fig:sidebands}. As can be seen, we work with the cavity centered not around the main laser frequency $\omega_l$, but around one of the two groups of sidebands generated by the EOM. Those are shifted by around \qty{5}{\giga\hertz} with respect to the laser frequency, which acts as the local oscillator. The cavity is then locked at $\omega_{\mathrm{cav}} \simeq \omega_l + \Omega_{\mathrm{Lock}}$, using the demodulated phase of the PDH error signal, which allows for a larger locking bandwidth with respect to using one of the two quadratures. The pump laser tone used for the OMIT measurement is at $\omega_{L_1} = \omega_l + \Omega_{L_1}$, while the probe tone is at $\omega_p = \omega_l \pm \Omega_p$. The detuning is denoted as the difference between the cavity resonance and the pump tone, $\bar{\Delta} = \omega_{L_1} - \omega_{\mathrm{cav}} \simeq \Omega_{L_1} - \Omega_{\mathrm{Lock}}$. Finally, we define the difference between probe and pump frequency as $\Omega = \omega_p - \omega_{L_1} = \Omega_p - \Omega_{L_1}$.
\begin{table}[ht]
	\centering
	\begin{tabular}{l|c|c}
		\toprule
		Definition				& Abs. Freq.															& EOM RF Tone\\
		\midrule
		Laser Freq. (LO) 		& $\omega_l \equiv \omega_{L_0}$										& $\Omega_{L_0} = 0$\\
		Sidebands Lock Freq.	& $\omega_l \pm \Omega_{\mathrm{Lock}}$									& $\Omega_{\mathrm{Lock}}$ \\
		Cavity Freq.			& $\omega_{\mathrm{cav}} \simeq \omega_l + \Omega_{\mathrm{Lock}}$ 		& -- \\
		Pump Freq.				& $\omega_{L_1} = \omega_l \pm \Omega_{L_1}$							& $\Omega_{L_1}$ \\
		Probe Freq.				& $\omega_p = \omega_l \pm \Omega_p$									& $\Omega_p$ \\
		Mech. Sidebands			& $\omega_l + \Omega_{L_1} \pm \Omega_m$								& -- \\
		Cav. Detuning			& $\bar{\Delta} = \omega_{L_1} - \omega_{\mathrm{cav}}$					& $\bar{\Delta} \simeq \Omega_{L_1} - \Omega_{\mathrm{Lock}}$ \\
		Pump-Probe Detuning		& $\Omega = \omega_p - \omega_{L_1}$									& $\Omega = \Omega_p - \Omega_{L_1}$ \\
		\bottomrule
	\end{tabular}
	\caption{Optical tone and EOM sidebands definition.}
	\label{tab:sidebands}
\end{table}

In order to directly compare to the notations we used previously (compare particularly \cref{tab:agarwal}), it is sufficient to substitute $\omega_l$ with $\omega_{L_1}$. In other words, the main pump tone is no longer the laser frequency (which now acts only as the local oscillator), but the strongest of the sidebands that we indicate with $L_1$. In this experimental notation, this specifically means that the detuning $\Omega$ is indeed $\Omega = \omega_p - \omega_{L_1}$, rather than $\omega_p - \omega_l$.

We can now apply the same procedure as in the previous section. We start by defining the input field in our experimental context:
\begin{equation}
	\begin{aligned}
		\hat{a}_{\mathrm{in}}(t)	&= \left[ \hat{a}^{SSB}_{\mathrm{in}}(t) + \delta\hat{a}_{\mathrm{in}}(t) \right] = \\
		&= a_{L_0} e^{-i\omega_l t} + a_{L_1} e^{-i(\omega_l+\Omega_{L_1}) t} - a_{L_1} e^{-i(\omega_l-\Omega_{L_1}) t}
		+ a_p e^{-i(\omega_l+\Omega_p) t} - a_p e^{-i(\omega_l-\Omega_p) t} \\
		&= a_{L_0} e^{-i\omega_l t} + a_{L_1} e^{-i(\omega_l+\Omega_{L_1}) t} - a_{L_1} e^{-i(\omega_l-\Omega_{L_1}) t}
		+ a_p e^{-i(\omega_l+\Omega_{L_1}+\Omega) t} - a_p e^{-i(\omega_l-\Omega_{L_1}-\Omega) t}
	\end{aligned}									
\end{equation}
where "SSB" stands for single-sideband and $a_p$ is the amplitude of the probe signal. Similarly for the field inside the cavity we have:
\begin{equation}
	\hat{a}(t)	= \left[\hat{a}^{SSB}(t) + \delta\hat{a}(t) \right] 
	= \hat{a}^{SSB}(t) + \left[A^-e^{-i(\Omega_{L_1}+\Omega) t} + A^+e^{-i(\Omega_{L_1}-\Omega) t} \right]e^{-i\omega_l t}								
\end{equation}
We can then write for the reflected signal:
\begin{equation}
	\begin{aligned}
		\hat{a}^R_{\mathrm{out}}(t) &= \hat{a}_{\mathrm{in}}(t) - \sqrt{\kappa_{\mathrm{ext}}} \hat{a}(t) \\
		&\begin{multlined}[t]
			= \hat{a}^{SSB}_{\mathrm{in}}(t) - \sqrt{\kappa_{\mathrm{ext}}} \hat{a}^{SSB}(t) + a_p e^{-i(\omega_l+\Omega_p) t} - a_p e^{-i(\omega_l-\Omega_p) t} \\
			\shoveleft[0.3\linewidth] - \sqrt{\kappa_{\mathrm{ext}}} A^-e^{-i(\omega_l+\Omega_{L_1}+\Omega) t} - \sqrt{\kappa_{\mathrm{ext}}} A^+e^{-i(\omega_l+\Omega_{L_1}-\Omega) t}
		\end{multlined} \\
		&\begin{multlined}[t]
			= a_{L_0} (1- \kappa_{\mathrm{ext}} \chi_{\mathrm{cav}}[\bar{\Delta}]) e^{-i\omega_l t} \\
			\shoveleft[0.05\linewidth] + a_{L_1} (1- \kappa_{\mathrm{ext}} \chi_{\mathrm{cav}}[\bar{\Delta}+\Omega_{L_1}]) e^{-i(\omega_l+\Omega_{L_1}) t} \\
			\shoveleft[0.15\linewidth] - a_{L_1} (1- \kappa_{\mathrm{ext}} \chi_{\mathrm{cav}}[\bar{\Delta}-\Omega_{L_1}]) e^{-i(\omega_l-\Omega_{L_1}) t} \\
			\shoveleft[0.25\linewidth] + a_p (1- \kappa_{\mathrm{ext}} \bar{A}^-[\bar{\Delta}+\Omega_{L_1}+\Omega]) e^{-i(\omega_l+\Omega_{L_1}+\Omega) t} \\
			\shoveleft[0.35\linewidth] - a_p e^{-i(\omega_l-\Omega_{L_1}-\Omega) t}
			- a_p \kappa_{\mathrm{ext}} \bar{A}^+ [\bar{\Delta}+\Omega_{L_1}-\Omega] e^{-i(\omega_l+\Omega_{L_1}-\Omega) t}
		\end{multlined}	\\
		&\begin{multlined}[t]
			= a_{L_0}\mathrm{K}^R_{\mathrm{cav}}[\bar{\Delta}] e^{-i\omega_l t} \\
			\shoveleft[0.05\linewidth] + a_{L_1}\mathrm{K}^R_{\mathrm{cav}}[\bar{\Delta}+\Omega_{L_1}] e^{-i(\omega_l+\Omega_{L_1}) t} \\
			\shoveleft[0.15\linewidth] - a_{L_1}\mathrm{K}^R_{\mathrm{cav}}[\bar{\Delta}-\Omega_{L_1}] e^{-i(\omega_l-\Omega_{L_1}) t} \\
			\shoveleft[0.25\linewidth] + a_p (1- \kappa_{\mathrm{ext}} \bar{A}^-[\bar{\Delta}+\Omega_{L_1}+\Omega]) e^{-i(\omega_l+\Omega_{L_1}+\Omega) t} \\
			\shoveleft[0.35\linewidth] - a_p e^{-i(\omega_l-\Omega_{L_1}-\Omega) t}
			- a_p \kappa_{\mathrm{ext}} \bar{A}^+ [\bar{\Delta}+\Omega_{L_1}-\Omega] e^{-i(\omega_l+\Omega_{L_1}-\Omega) t}
		\end{multlined}	\\
		&\begin{multlined}[t]
			\simeq a_{L_0} e^{-i\omega_l t} + a_{L_1} e^{-i(\omega_l+\Omega_{L_1}) t} - a_{L_1} e^{-i(\omega_l-\Omega_{L_1}) t} \\
			\shoveleft[0.1\linewidth] - a_{L_1} \kappa_{\mathrm{ext}} \chi_{\mathrm{cav}}[\bar{\Delta}+\Omega_{L_1}] e^{-i(\omega_l+\Omega_{L_1}) t} \\
			\shoveleft[0.2\linewidth] + a_p (1- \kappa_{\mathrm{ext}} \bar{A}^-[\bar{\Delta}+\Omega_{L_1}+\Omega]) e^{-i(\omega_l+\Omega_{L_1}+\Omega) t} \\
			\shoveleft[0.3\linewidth] - a_p e^{-i(\omega_l-\Omega_{L_1}-\Omega) t}
			- a_p \kappa_{\mathrm{ext}} \bar{A}^+ [\bar{\Delta}+\Omega_{L_1}-\Omega] e^{-i(\omega_l+\Omega_{L_1}-\Omega) t}
		\end{multlined}	\\
		&\begin{multlined}[t]
			= \hat{a}^{SSB}_{\mathrm{in}}(t) + a_p e^{-i(\omega_l+\Omega_{L_1}+\Omega) t} - a_p e^{-i(\omega_l-\Omega_{L_1}-\Omega) t} \\
			\shoveleft[0.1\linewidth] - a_{L_1} \kappa_{\mathrm{ext}} \chi_{\mathrm{cav}}[\bar{\Delta}+\Omega_{L_1}] e^{-i(\omega_l+\Omega_{L_1}) t} \\
			\shoveleft[0.2\linewidth] - a_p \kappa_{\mathrm{ext}} \bar{A}^-[\bar{\Delta}+\Omega_{L_1}+\Omega] e^{-i(\omega_l+\Omega_{L_1}+\Omega) t} \\
			\shoveleft[0.3\linewidth] - a_p \kappa_{\mathrm{ext}} \bar{A}^+ [\bar{\Delta}+\Omega_{L_1}-\Omega] e^{-i(\omega_l+\Omega_{L_1}-\Omega) t} \\
		\end{multlined}	\\
		&\begin{multlined}[t]
			= \hat{a}_{\mathrm{in}}(t) - a_{L_1} \kappa_{\mathrm{ext}} \chi_{\mathrm{cav}}[\bar{\Delta}+\Omega_{L_1}] e^{-i(\omega_l+\Omega_{L_1}) t} \\
			\shoveleft[0.15\linewidth] - a_p \kappa_{\mathrm{ext}} \bar{A}^-[\bar{\Delta}+\Omega_{L_1}+\Omega] e^{-i(\omega_l+\Omega_{L_1}+\Omega) t} \\
			\shoveleft[0.3\linewidth] - a_p \kappa_{\mathrm{ext}} \bar{A}^+ [\bar{\Delta}+\Omega_{L_1}-\Omega] e^{-i(\omega_l+\Omega_{L_1}-\Omega) t} \\
		\end{multlined}	\\	
	\end{aligned}									
\end{equation}
where again we took advantage of the cavity sidebands' resolution and we defined $\bar{A}^{\pm} = \frac{A^{\pm}}{a_p \sqrt{\kappa_{\mathrm{ext}}}}$. At this point we can proceed as before to  compute what terms actually contribute to the demodulated signal. We start by computing the reflected power as $P_{\mathrm{out}} \propto |\hat{a}^R_{\mathrm{out}}|^2$. Leveraging the previous discussion, we can safely assume that all the DC terms of the type $|\hat{a}_{\mathrm{in}}(t)|^2$, $|a_{L_1} \dots|^2$ etc. go to zero after demodulation and low-pass filtering. Similarly, the cross terms of the type $a_{L_1} a_p^*$ and $|a_p|^2\dots$, which beat at frequencies different from the demodulation frequency, are filtered out.\\
Let's then consider the remaining terms. The first is:
\begin{equation}
	\begin{aligned}
		- \hat{a}_{\mathrm{in}}(t) a_{L_1}^* &\kappa_{\mathrm{ext}} \chi^*_{\mathrm{cav}}[\bar{\Delta}+\Omega_{L_1}] e^{+i(\omega_l+\Omega_{L_1}) t} + \text{c.c.} = \\
		&\begin{multlined}[t]
			= - a_{L_0} a_{L_1}^* \kappa_{\mathrm{ext}} \chi^*_{\mathrm{cav}}[\bar{\Delta}+\Omega_{L_1}] e^{+i \Omega_{L_1} t} \\
			\shoveleft[0.05\linewidth] - |a_{L_1}|^2 \kappa_{\mathrm{ext}} \chi^*_{\mathrm{cav}}[\bar{\Delta}+\Omega_{L_1}]
			+ |a_{L_1}|^2 \kappa_{\mathrm{ext}} \chi^*_{\mathrm{cav}}[\bar{\Delta}+\Omega_{L_1}] e^{+i 2\Omega_{L_1} t} \\
			\shoveleft[0.15\linewidth] - a_p a_{L_1}^* \kappa_{\mathrm{ext}} \chi^*_{\mathrm{cav}}[\bar{\Delta}+\Omega_{L_1}] e^{-i \Omega t}
			+ a_p a_{L_1}^* \kappa_{\mathrm{ext}} \chi^*_{\mathrm{cav}}[\bar{\Delta}+\Omega_{L_1}] e^{+i(2\Omega_{L_1} + \Omega) t} + \text{c.c.} \\
		\end{multlined}	\\
	\end{aligned}									
\end{equation}
It is straightforward to prove that none of these terms survive after demodulation.\\
The next term we need to consider is:
\begin{equation}
	\begin{aligned}
		- \hat{a}_{\mathrm{in}}(t) a_p^* &\kappa_{\mathrm{ext}} (\bar{A}^-)^*[\bar{\Delta}+\Omega_{L_1}+\Omega] e^{+i(\omega_l+\Omega_{L_1}+\Omega) t} + \text{c.c.} = \\
		&\begin{multlined}[t]
			= - a_{L_0} a_p^* \kappa_{\mathrm{ext}} (\bar{A}^-)^*[\bar{\Delta}+\Omega_{L_1}+\Omega] e^{+i(\Omega_{L_1}+\Omega) t} \\
			\shoveleft[0.05\linewidth] - a_{L_1} a_p^* \kappa_{\mathrm{ext}} (\bar{A}^-)^*[\bar{\Delta}+\Omega_{L_1}+\Omega] e^{+i\Omega t}
			+ a_{L_1} a_p^* \kappa_{\mathrm{ext}} (\bar{A}^-)^*[\bar{\Delta}+\Omega_{L_1}+\Omega] e^{+i(2\Omega_{L_1} + \Omega) t} \\
			\shoveleft[0.15\linewidth] - |a_p|^2 \kappa_{\mathrm{ext}} (\bar{A}^-)^*[\bar{\Delta}+\Omega_{L_1}+\Omega] 
			+ |a_p|^2 \kappa_{\mathrm{ext}} (\bar{A}^-)^*[\bar{\Delta}+\Omega_{L_1}+\Omega] e^{+i(2\Omega_{L_1} + 2\Omega) t} + \text{c.c.} \\
		\end{multlined}	\\
	\end{aligned}									
\end{equation}
As we will show in a moment, of all these terms only the first one, containing a $a_{L_0}$ factor, survives, all the others go to zero.\\
Finally the last term to consider is:
\begin{equation}
	\begin{aligned}
		- \hat{a}_{\mathrm{in}}(t) a_p^* &\kappa_{\mathrm{ext}} (\bar{A}^+)^*[\bar{\Delta}+\Omega_{L_1}-\Omega] e^{+i(\omega_l+\Omega_{L_1}-\Omega) t} + \text{c.c.} = \\
		&\begin{multlined}[t]
			= - a_{L_0} a_p^* \kappa_{\mathrm{ext}} (\bar{A}^+)^*[\bar{\Delta}+\Omega_{L_1}-\Omega] e^{+i(\Omega_{L_1}-\Omega) t} \\
			\shoveleft[0.05\linewidth] - a_{L_1} a_p^* \kappa_{\mathrm{ext}} (\bar{A}^+)^*[\bar{\Delta}+\Omega_{L_1}-\Omega] e^{-i\Omega t}
			+ a_{L_1} a_p^* \kappa_{\mathrm{ext}} (\bar{A}^+)^*[\bar{\Delta}+\Omega_{L_1}-\Omega] e^{+i(2\Omega_{L_1} - \Omega) t} \\
			\shoveleft[0.15\linewidth] - |a_p|^2 \kappa_{\mathrm{ext}} (\bar{A}^+)^*[\bar{\Delta}+\Omega_{L_1}-\Omega] e^{-i2\Omega t}
			+ |a_p|^2 \kappa_{\mathrm{ext}} (\bar{A}^+)^*[\bar{\Delta}+\Omega_{L_1}-\Omega] e^{+i2\Omega_{L_1} t} + \text{c.c.} \\
		\end{multlined}	\\
	\end{aligned}									
\end{equation}
Also in this case, none of these terms survive after demodulation and low-pass filtering.

To conclude, let's compute explicitly the value of the only surviving term. Using, as before, dual-phase down mixing and low-pass filtering we obtain:
\begin{multline}
	- a_{L_0} a_p^* \kappa_{\mathrm{ext}} (\bar{A}^-)^*[\bar{\Delta}+\Omega_{L_1}+\Omega] e^{+i(\Omega_{L_1}+\Omega) t} e^{-i\left[(\Omega_{L_1} + \Omega)t + \phi\right]} \\
	\quad \Rightarrow \quad
	-a_{L_0} a_p^* \kappa_{\mathrm{ext}} (\bar{A}^-)^*[\bar{\Delta}+\Omega_{L_1}+\Omega] e^{-i\phi} = I+iQ = Z
\end{multline}
Finally, given that we will perform our measurement using a network analyzer that measure the amplitude and phase of the demodulated signal, we get:
\begin{equation}
	R = |Z| = \sqrt{Z\bar{Z}} = \sqrt{|a_{L_0}|^2 |a_p|^2 \kappa_{\mathrm{ext}}^2 |\bar{A}^-|^2} = |a_{L_0}| \sqrt{\kappa_{\mathrm{ext}}} \, |A^-|
\end{equation}
and as we can see the measured signal is proportional to the LO sideband (which is proportional to the power injected in the cavity), to the external coupling losses and to $A^-$ as expected.

\subsubsection{Comparison with other OMIT readout schemes and sideband artifacts}
The intrinsic OMIT response discussed in the previous sections is encoded in the intracavity probe-sideband amplitude $A^-$. However, the experimentally measured OMIT signal depends not only on this internal response, but also on how the probe is generated and how the reflected or transmitted light is demodulated. Different measurement schemes can therefore introduce additional sideband contributions or transfer-function distortions that are not part of the bare optomechanical response itself.

A first distinction concerns whether the probe is generated as a separate optical tone or as modulation sidebands around a carrier. In the membrane-in-the-middle experiment of Karuza \textit{et al.}
\cite{karuzaOptomechanicallyInducedTransparency2013}, pump and probe are produced as separate beams with AOMs and the measured quantity is the beat between the transmitted pump and probe fields. In that case the analysis assumes that the opposite-frequency component is sufficiently far from cavity resonance to be neglected. By contrast, Buters \textit{et al.} \cite{butersStraightforwardMethodMeasure2017a} deliberately generate one control tone together with two symmetrically spaced probe tones using a single AOM and lock-in modulation. They show explicitly that when the cavity is not deeply sideband resolved, the second probe cannot be ignored and the standard single-probe OMIT formula must be modified.

A second distinction concerns phase-modulation-based readout, where both upper and lower optical sidebands are present by construction. This is particularly clear in the OMIT analysis of Bodiya \textit{et al.} \cite{bodiyaSubhertzOptomechanicallyInduced2019}, where the full transmitted-field model contains both sidebands and the measured response includes an unwanted lower-sideband contribution transmitted by the cavity because the sideband resolution is only modest. In addition, their measurement is performed from within the cavity control loop, so the observed response must be calibrated by dividing out the servo-loop transfer function. Similarly, in the amplitude-modulated oscillating-mode measurements of Sbarra \textit{et al.} \cite{sbarraMultiphysicsModelUltrahigh2021}, the demodulated lock-in quadratures depend explicitly on both $A^-$ and $A^+$, and can also be modified by a thermal degree of freedom when photothermal effects are significant.

Our experimental configuration avoids these specific complications because the cavity is not addressed directly on the laser carrier, but on a GHz-shifted EOM sideband branch. The laser carrier acts as a local oscillator, while the OMIT pump and probe are placed on the selected sideband group. As a result, the opposite PDH sideband is displaced by approximately $2\Omega_{\mathrm{Lock}}$, i.e. several GHz, and the probe image associated with the unused branch is likewise far from the cavity resonance relevant to the OMIT scan. In the expansion of the reflected power $P_{\mathrm{out}} \propto |\hat{a}^R_{\mathrm{out}}|^2$, the corresponding beat terms oscillate at frequencies that are rejected by the electrical demodulation and low-pass filtering. Under these conditions, the surviving demodulated complex response is directly proportional to $\sqrt{\kappa_{\mathrm{ext}}}A^-$, which is why the main-text OMIT measurements can be compared directly to the full theoretical expression for $A^-(\Omega)$ without introducing extra nearby-sideband correction terms.


\section{Frequency Comb}
This section expands on the frequency-comb observations reported in the main text. We first introduce a simple modulation-depth and visibility framework that separates the ideal comb span from the experimentally detectable sideband envelope, and use it to compare the present hBN fiber-cavity platform with other optomechanical systems. We then present supporting datasets that clarify the relation between the all-optical two-tone combs and mechanically driven combs.

\subsection{OM-comb mechanism and modulation-depth estimate}
To quantify the comb span, optomechanical frequency-comb generation can first be viewed as frequency modulation of the optical cavity resonance by periodic mechanical motion. We write the coherent component of the displacement as
\begin{equation}
	x(t)=x_0\cos(\Omega_m t),
\end{equation}
where $x_0$ is the peak displacement amplitude. The corresponding cavity-frequency modulation is
\begin{equation}
	\omega_{\mathrm{cav}}(t) = \omega_{\mathrm{cav}}^0-Gx_0\cos(\Omega_m t),
\end{equation}
with $G=-\partial\omega_{\mathrm{cav}}/\partial x$. The natural dimensionless modulation index is therefore
\begin{equation}
	\mathcal{M}_{\mathrm{OM}} \equiv \frac{Gx_0}{\Omega_m} = \frac{g_0}{\Omega_m} \frac{x_0}{x_{\mathrm{zpf}}},
	\label{eq:SI_M_OM}
\end{equation}
where $g_0=Gx_{\mathrm{zpf}}$. This quantity measures the peak cavity-frequency excursion in units of the mechanical frequency and gives the ideal upper limit for the number of mechanically spaced sidebands that can be generated. In the ideal phase-modulation limit, the optical field acquires the phase
\begin{equation}
	\phi(t) = \int^t Gx_0\cos(\Omega_m t')\,dt' = \mathcal{M}_{\mathrm{OM}}\sin(\Omega_m t),
\end{equation}
and the Jacobi--Anger expansion,
\begin{equation}
	e^{i\mathcal{M}_{\mathrm{OM}}\sin(\Omega_m t)} = \sum_{n=-\infty}^{+\infty} J_n(\mathcal{M}_{\mathrm{OM}})e^{in\Omega_m t},
\end{equation}
shows that the available sideband orders are controlled by $J_n(\mathcal{M}_{\mathrm{OM}})$. Since $J_n(\mathcal{M}_{\mathrm{OM}})$ decreases rapidly for $|n|\gtrsim\mathcal{M}_{\mathrm{OM}}$, the modulation index gives the ideal estimate of the comb span.

The cavity response determines how this ideal phase-modulation spectrum is converted into reflected or transmitted optical power. In a frame rotating at the optical drive, the intracavity field obeys
\begin{equation}
	\dot{a}(t)=	\left[i\left(\bar{\Delta}+Gx_0\cos\Omega_m t\right)-\frac{\kappa}{2}\right]a(t)	+\sqrt{\kappa_{\mathrm{ext}}}\,a_{\mathrm{in}},
	\label{eq:SI_comb_full_eom}
\end{equation}
where $\bar{\Delta}$ is the effective operating detuning, including any static displacement shift. Expanding the periodic steady state as
\begin{equation}
	a(t)=\sum_n a_n e^{in\Omega_m t},
\end{equation}
one obtains
\begin{equation}
	a_n = \sqrt{\kappa_{\mathrm{ext}}}\,a_{\mathrm{in}}	\sum_{m=-\infty}^{+\infty}
			\frac{J_{n-m}(\mathcal{M}_{\mathrm{OM}})J_m(-\mathcal{M}_{\mathrm{OM}})}{\kappa/2+i(m\Omega_m-\bar{\Delta})}.
	\label{eq:SI_full_comb_sidebands}
\end{equation}
This is the large-amplitude sideband solution used in optomechanical-comb analyses such as Hu \textit{et al.} \cite{huGenerationOpticalFrequency2021}, written in the present notation. The Bessel functions encode the modulation depth, while the denominator gives the cavity weighting of each mechanically spaced component. 

The generated sidebands with $n\neq0$ are proportional to the intracavity Fourier coefficients $a_n$, up to constant detection factors. We can therefore use $|a_n|^2$ as an estimate of the sideband-power envelope. This convention is sufficient for comparing comb spans, because the overall proportionality factor is absorbed into the experimentally determined dynamic range. For a given dynamic range $D_{\mathrm{dB}}$, the largest visible positive order in the full model can then be estimated from
\begin{equation}
	N_{\mathrm{full}} =\max\left\{n>0: 10\log_{10}\left(\frac{|a_n|^2}{A_{\mathrm{full}}}\right) \ge -D_{\mathrm{dB}} \right\}.
	\label{eq:SI_Nfull}
\end{equation}
Here $A_{\mathrm{full}}$ is the chosen normalization of the calculated sideband-power envelope, for example the maximum positive-order value of $|a_n|^2$. This expression is the appropriate quantity when $\kappa$ is comparable to $\Omega_m$ or when the cavity weighting changes appreciably from one sideband order to the next.

In the present unresolved-sideband measurements, the experimentally visible comb span lies in a regime where the cavity response is close to instantaneous on the mechanical period. The envelope can then be approximated by the adiabatic cavity transduction. Setting $\dot{a}\simeq0$ in \cref{eq:SI_comb_full_eom} gives
\begin{equation}
	a(t) \propto \frac{1}{\kappa/2-i[\bar{\Delta}+Gx_0\cos(\Omega_m t)]}.
	\label{eq:SI_comb_adiabatic_field}
\end{equation}
For $\bar{\Delta}\simeq0$, the Fourier components of this periodic response have the geometric power envelope
\begin{equation}
	|a_n|^2 \simeq \rho^{2|n|},
	\label{eq:SI_rho_envelope}
\end{equation}
where
\begin{equation}
	\rho = \frac{|Gx_0|}{\sqrt{(\kappa/2)^2+(Gx_0)^2}+\kappa/2}	= \frac{\eta}{\sqrt{1+\eta^2}+1},
		\qquad
			\eta = \frac{|Gx_0|}{\kappa/2}.
	\label{eq:SI_rho_eta}
\end{equation}
Here $\eta$ is the transduction strength. This expression explains the approximately triangular decay observed on a logarithmic power scale in our measured comb spectra.

The transduction envelope provides a simple estimate of the largest comb order detectable above the measurement floor. If the spectrum is measured as a power spectral density, the line heights can be compared to
\begin{equation}
	S_n = S_{\mathrm{floor}} + A\rho^{2|n|},
	\label{eq:SI_comb_envelope_fit}
\end{equation}
where $S_{\mathrm{floor}}$ is the local noise floor and $A$ is the extrapolated excess power of the comb envelope at $n=0$. The effective dynamic range is
\begin{equation}
	D_{\mathrm{dB}} = 10\log_{10} \left(\frac{A}{S_{\mathrm{floor}}}\right).
	\label{eq:SI_comb_dynamic_range}
\end{equation}
Since the envelope decreases by $-10\log_{10}(\rho^2)$ dB per mechanical sideband order, the largest detectable positive comb order in this approximation is
\begin{equation}
	N_{\rho} \simeq \frac{D_{\mathrm{dB}}}{-10\log_{10}(\rho^2)}.
	\label{eq:SI_Nrho}
\end{equation}
This estimate assumes that all line heights are extracted with the same resolution bandwidth or frequency-bin convention.

The ideal modulation index and the transduction estimate describe two different limitations. The displacement required to reach comb order $N$ in the ideal phase-modulation sense is
\begin{equation}
	x_{\mathcal{M}}(N) = \frac{N\Omega_m}{G}.
	\label{eq:SI_x_M_req}
\end{equation}
At the same time, a general visibility requirement can be defined from the full sideband envelope. For a target order $N$ and dynamic range $D_{\mathrm{dB}}$, we define
\begin{equation}
	x_{\mathrm{full}}(N,D_{\mathrm{dB}}) = \min\left\{x_0: N_{\mathrm{full}}(x_0,D_{\mathrm{dB}})\ge N \right\}.
	\label{eq:SI_x_full_req}
\end{equation}
This quantity is obtained by evaluating \cref{eq:SI_full_comb_sidebands,eq:SI_Nfull} as a function of $x_0$, or equivalently of $\mathcal{M}_{\mathrm{OM}}=Gx_0/\Omega_m$. It gives the displacement required for the $N$-th mechanically spaced sideband to remain visible once the cavity weighting of the comb envelope is included. In the adiabatic unresolved-sideband limit, $x_{\mathrm{full}}$ reduces to a simple analytic expression. Setting $N=N_{\rho}$ gives, for a target dynamic range $D_{\mathrm{dB}}$,
\begin{equation}
	\rho_N = 10^{-D_{\mathrm{dB}}/(20N)},
\end{equation}
and therefore
\begin{equation}
	x_{\rho}(N,D_{\mathrm{dB}}) = \frac{\kappa}{G} \frac{\rho_N}{1-\rho_N^2}.
	\label{eq:SI_x_rho_req}
\end{equation}
When the adiabatic approximation applies, $x_{\rho}$ is a closed-form approximation to $x_{\mathrm{full}}$. The ideal modulation index and the visibility estimate therefore describe two distinct constraints. For a given coherent displacement, the observable positive comb order is set approximately by the smaller of the ideal span $\mathcal{M}_{\mathrm{OM}}$ and the detectable order $N_{\mathrm{full}}$. Conversely, to observe a target order $N$, the required displacement is set by the larger of $x_{\mathcal{M}}$ and $x_{\mathrm{full}}$ (or $x_{\rho}$ in the adiabatic unresolved-sideband limit). This distinction is useful because $G$, $\Omega_m$, $\kappa$, and $g_0$ enter the two requirements differently: large $G$ lowers both displacement scales, broad $\kappa$ avoids sideband filtering but weakens transduction contrast, and large $g_0$ reduces the optical power needed to generate the motion.

The analysis so far treats the mechanical oscillation amplitude $x_0$ as a given parameter. This is the appropriate description for the piezo-driven measurements, where the mechanical motion is imposed externally. It also provides the kinematic part of an all-optical comb model: once radiation pressure has generated a coherent oscillation, the optical comb span and detected envelope are still governed by $\mathcal{M}_{\mathrm{OM}}$, $N_{\mathrm{full}}$, and the cavity linewidth. The remaining question is therefore how efficiently the optical field can generate a large $x_0$.

This picture connects directly to previous theoretical work on optomechanical frequency combs. The large-amplitude treatment reviewed in \cite{aspelmeyerCavityOptomechanics2014} gives sideband amplitudes involving the same dimensionless argument $GA/\Omega_m$, multiplied by the cavity response. Miri \textit{et al.} \cite{miriOptomechanicalFrequencyCombs2018} go beyond this kinematic picture by treating comb formation as a self-consistent dynamical instability of a driven single-mode optomechanical cavity: radiation pressure destabilizes the stationary state, drives periodic mechanical motion, and thereby generates a comb of optical sidebands spaced by the mechanical frequency. A key conclusion is that comb generation does not require sideband-resolved operation; in the unresolved-sideband regime, the cavity linewidth can cover many mechanically spaced sidebands and the comb can remain less prone to chaos in parts of parameter space.

The two-tone mechanism proposed by Gu \textit{et al.} \cite{guOpticalFrequencyComb2024} is the one adopted in our measurements. In this case, the cavity is driven by two optical tones whose frequency difference is close to the mechanical frequency,
\begin{equation}
	|\omega_1-\omega_2| \simeq \Omega_m .
\end{equation}
In the unresolved-sideband regime, this drive can produce a self-organized nonlinear resonance that enhances the mechanical oscillation amplitude and broadens the comb. In the notation of Gu \textit{et al.}, the mechanical amplitude is written as a dimensionless quantity $A_m\simeq x_0/x_{\mathrm{zpf}}$, so their comb-span parameter is equivalent to $\mathcal{M}_{\mathrm{OM}}$.

A simple resonant linear estimate illustrates the two-tone drive scaling. If two intracavity tones with photon numbers $n_1$ and $n_2$ beat at $\Omega_m$, the intracavity photon number contains an oscillating term with peak amplitude $2\sqrt{n_1n_2}$. Since the radiation-pressure force is $F=\hbar G n_{\mathrm{cav}}$, the force component at the beat frequency is
\begin{equation}
	F_{\Omega} \simeq 2\hbar G\sqrt{n_1n_2}.
	\label{eq:SI_two_tone_force}
\end{equation}
The displacement response follows from the mechanical susceptibility,
\begin{equation}
	x(\Omega) =	\chi_m(\Omega)F(\Omega),
		\qquad
			\chi_m(\Omega) = \frac{1}{M_{\mathrm{eff}}(\Omega_m^2-\Omega^2-i\Gamma_m\Omega)}.
\end{equation}
At resonance, $|\chi_m(\Omega_m)|=1/(M_{\mathrm{eff}}\Gamma_m\Omega_m)$, giving the peak coherent displacement
\begin{equation}
	x_0 \simeq \frac{F_{\Omega}}{M_{\mathrm{eff}}\Gamma_m\Omega_m}.
\end{equation}
Using $g_0=Gx_{\mathrm{zpf}}$ and $x_{\mathrm{zpf}}=\sqrt{\hbar/(2M_{\mathrm{eff}}\Omega_m)}$, this becomes
\begin{equation}
	\frac{x_0}{x_{\mathrm{zpf}}} \simeq \frac{4g_0\sqrt{n_1n_2}}{\Gamma_m},
\end{equation}
and therefore
\begin{equation}
	\mathcal{M}_{\mathrm{OM}} \simeq \frac{4g_0^2\sqrt{n_1n_2}}{\Gamma_m\Omega_m}.
	\label{eq:SI_MOM_drive_scaling}
\end{equation}
This estimate is only the linear resonant-drive limit and does not include nonlinear saturation, optical-spring shifts, or frequency locking. It nevertheless shows why large $g_0$ strongly reduces the optical power required to reach a given modulation index.

The same frequency-pull scale also appears in the thermal-broadening discussion above. Thermal Brownian motion produces an rms cavity-frequency jitter
\begin{equation}
	\delta\omega_{\mathrm{rms}} = Gx_{\mathrm{rms}} \simeq g_0\sqrt{2n_{\mathrm{th}}},
\end{equation}
in the high-temperature limit. This is the origin of the effective thermal linewidth contribution $\delta\kappa$ used in the OMIT theory section. The comb case differs because the motion is coherent and periodic rather than random: a coherent displacement $x(t)=x_0\cos(\Omega_m t)$ produces sidebands with peak frequency excursion $Gx_0$, while random thermal motion produces cavity broadening. For a coherent sinusoid, the rms frequency excursion would be $Gx_0/\sqrt{2}$.

We first apply these estimates to the mechanically driven fundamental hBN mode shown in \cref{fig:FigS11_Freq_Comb_Modes}. At the operating point used for that measurement, $\kappa/2\pi\simeq\qty{1}{\giga\hertz}$ and $G/2\pi\simeq\qty{2.75}{\giga\hertz\per\nano\meter}$, comparable to the value extracted from the backaction calibration. The thermally driven motion, $x_0\simeq\qty{38}{\pico\meter}$, gives $\mathcal{M}_{\mathrm{OM}}\simeq50$ and $N_{\mathrm{full}}\simeq3$ for the measured dynamic range $D_{\mathrm{dB}}\simeq58$, consistent with the weak comb structure observed without piezo drive. At the largest piezo drive, $x_0\simeq\qty{0.6}{\nano\meter}$ gives $\mathcal{M}_{\mathrm{OM}}\simeq8\times10^2$, while finite cavity transduction limits the expected visible order to $N_{\mathrm{full}}\simeq20$ for $D_{\mathrm{dB}}\simeq48$, again matching the measured envelope.

The same estimates illustrate the potential of operating closer to the maximum static coupling measured in the present device. For $G/2\pi\simeq\qty{11}{\giga\hertz\per\nano\meter}$ and $x_0\simeq\qty{1}{\nano\meter}$, one obtains $\mathcal{M}_{\mathrm{OM}}\simeq5.5\times10^3$. With the present linewidth, $\kappa/2\pi\simeq\qty{1}{\giga\hertz}$, the full cavity-weighted estimate gives $N_{\mathrm{full}}\simeq 10^2$ detectable positive-order lines for a dynamic range near \qty{50}{\decibel}. Reducing the linewidth to $\kappa/2\pi\simeq\qty{100}{\mega\hertz}$, a scale already reached in related fiber-cavity experiments, would increase the expected visible order to $N_{\mathrm{full}}\simeq 10^3$ while remaining deeply in the unresolved-sideband regime for the present $\Omega_m/2\pi\simeq\qty{2}{\mega\hertz}$ mode. As a higher repetition-rate projection, we also consider a smaller hBN drum with diameter $\sim\qty{5}{\micro\meter}$ and thickness $\sim\qty{35}{\nano\meter}$, using the $G$ and $g_0$ values estimated from \cref{fig:FigS9_OMcoupling_Theo}. With a mechanical frequency $\Omega_m/2\pi\simeq\qty{50}{\mega\hertz}$, this case still gives about $10^2$ visible positive-order lines for nanometer-scale motion. These numbers show that the large frequency-pull parameter and design flexibility of the hBN fiber-cavity platform can make broad mechanically spaced combs accessible with sub-nanometer coherent motion, provided that stable operation can be maintained near high-coupling positions.

For an all-optical two-tone drive, the relevant question is how efficiently radiation pressure can generate the required coherent motion. Using the resonant linear estimate of \cref{eq:SI_MOM_drive_scaling}, two intracavity tones with $n_1\simeq n_2\simeq15$ photons, corresponding to about $\qty{10}{\nano\watt}$ per tone for a near-resonant symmetric cavity with $\kappa/2\pi\simeq\qty{1}{\giga\hertz}$, would already give $\mathcal{M}_{\mathrm{OM}}\sim2\times10^3$ at the maximum measured coupling $g_0/2\pi\simeq\qty{180}{\kilo\hertz}$, using the measured mechanical linewidth of the fundamental hBN mode. This value highlights why increasing $g_0$ is particularly valuable for optically generated combs.

To place these numbers in context, \cref{tab:SI_comb_platform_comparison} compares representative optomechanical platforms using the same set of figures of merit. The table is meant to show how each platform would perform under the same simple frequency-modulation metric, and the entries combine representative or maximum reported parameters. We list the mechanical frequency and quality factor, the optical linewidth, the dispersive pull parameter $G$, and the single-photon coupling $g_0$. From these values we compute $\mathcal{M}_{\mathrm{OM}}$ and $N_{\mathrm{full}}$ for a common coherent displacement $x_0=\qty{1}{\nano\meter}$ and dynamic range $D_{\mathrm{dB}}=\qty{50}{\decibel}$. We also give $x_{\mathcal{M}}(100)$, the full visibility requirement $x_{\mathrm{full}}(100,\qty{50}{\decibel})$, and the corresponding per-tone optical power $P_{\mathrm{req}}$ estimated from the resonant linear two-tone drive. The all-optical power estimates are deliberately optimistic because they neglect saturation, detuning errors, optical-spring shifts, heating, and lock stability.

The present hBN fiber-cavity system is already competitive in this comparison, despite not being designed as a comb generator. The current linewidth, which was a deliberate design choice for stable unresolved-sideband OMIT measurements, is not a fundamental limitation of the architecture; reducing it to values already achieved in related fiber-cavity platforms would strongly improve the visibility requirement while preserving USR operation. The comparison with other fiber-cavity MIM systems clarifies this point. Rochau \textit{et al.} demonstrated an ultrahigh-finesse SiN fiber cavity with linewidths around $\kappa/2\pi\simeq\qty{60}{\mega\hertz}$ and pull factors up to $G/2\pi\simeq\qty{1}{\giga\hertz\per\nano\meter}$ \cite{rochauDynamicalBackactionUltrahighFinesse2021}. The narrow linewidth gives a favorable visibility scale, $x_{\mathrm{full}}\simeq\qty{0.54}{\nano\meter}$, but the smaller $g_0$ means that the optical drive efficiency relies strongly on their higher mechanical quality factor. S\'anchez Arribas \textit{et al.} reached linewidths near $\qty{50}{\mega\hertz}$ and pull factors of a few $\unit{\giga\hertz\per\nano\meter}$ in an hBN fiber cavity \cite{sanchezarribasRadiationPressureBackaction2023}. This gives a very small visibility displacement, $x_{\mathrm{full}}\simeq\qty{0.19}{\nano\meter}$, but again with a much smaller $g_0$ than in the present device, leading to larger estimated optical powers. These comparisons show that narrow coatings can compensate smaller $G$ for detecting externally driven combs, whereas the large $g_0$ of the present hBN drum is especially advantageous for all-optical comb generation.

Direct-laser-written polymer membranes provide another open-cavity comparison. Tenbrake \textit{et al.} report pull factors comparable to our maximum value, $G/2\pi\simeq\qty{11}{\giga\hertz\per\nano\meter}$, and $g_0/2\pi$ of order $\qty{30}{\kilo\hertz}$ \cite{tenbrakeDirectLaserwrittenOptomechanical2024}. However, the inferred optical linewidth is several GHz and the mechanical quality factor is modest. As a result, the ideal modulation index for \qty{1}{\nano\meter} motion is large, but the visible order remains only $N_{\mathrm{full}}\simeq32$, with $x_{\mathrm{full}}\simeq\qty{2.8}{\nano\meter}$ and a much higher two-tone power requirement. In this sense, the hBN fiber cavity combines a similar frequency-pull scale with substantially better optical-drive scaling.

The sliced photonic-crystal nanobeam of Leijssen \textit{et al.} represents an extreme nanophotonic limit rather than a direct MIM comparison \cite{leijssenNonlinearCavityOptomechanics2017}. Its pull parameter and single-photon coupling, $G/2\pi\simeq\qty{0.8}{\tera\hertz\per\nano\meter}$ and $g_0/2\pi\simeq\qty{25}{\mega\hertz}$, are much larger than ours, giving extremely favorable optical-power estimates. However, the effective cavity linewidth is also very large, especially at room temperature because of thermally induced cavity fluctuations. Even at \qty{3}{\kelvin}, the table gives $N_{\mathrm{full}}\simeq401$ for \qty{1}{\nano\meter} motion,
while at room temperature the effective linewidth reduces this to $N_{\mathrm{full}}\simeq66$. This illustrates the tradeoff that a very large $G$ can overcome a broad linewidth, but only if thermally induced broadening and stability can be controlled.

Integrated photonic and phononic-crystal systems provide a complementary limit. Guo and Gr\"oblacher combine large $G$, large $g_0$, and an exceptionally high mechanical quality factor, leading to a very favorable optical-power estimate \cite{guoIntegratedOpticalreadoutHighQ2022a}. Nevertheless, the broad optical linewidth limits the visible sideband order in the same transduction metric, giving $N_{\mathrm{full}}\simeq24$ for \qty{1}{\nano\meter} motion. This is an example where the ideal modulation index is enormous, $\mathcal{M}_{\mathrm{OM}}\simeq2\times10^4$, but the detected comb span is limited by cavity transduction. Conversely, the soft-clamped membrane-in-the-middle system of Planz \textit{et al.} has an ultranarrow optical linewidth and an extraordinary mechanical quality factor near $10^9$, but the long cavity gives a small pull parameter, $G/2\pi\simeq\qty{50}{\mega\hertz\per\nano\meter}$ \cite{planzMembraneinthemiddleOptomechanicsSoftclamped2023}. The required physical displacement therefore remains at the nanometer scale, and the system is competitive in power mainly because of its highly engineered mechanical quality factor. The comparison is useful because the present hBN device reaches comparable or better comb-generation metrics with a much lower current $Q_m$, owing to its much larger optomechanical pull.

The last rows compare directly with recent optomechanical-comb experiments. Hu \textit{et al.} demonstrated hundreds of comb lines in a silica microcavity with $\Omega_m/2\pi\simeq\qty{50}{\mega\hertz}$ \cite{huGenerationOpticalFrequency2021}. Their demonstrated mechanism is thermal and blue-detuned self-oscillation rather than the two-tone mechanism used here, so the power column should not be compared literally. Still, the table shows that a smaller hBN drum with a comparable mechanical frequency could reach similar or better modulation and visibility metrics with much lower estimated two-tone power, because of the larger $g_0$. Mercad\'e \textit{et al.} and Gou \textit{et al.} operate at GHz mechanical frequencies in integrated optomechanical resonators \cite{mercadeMicrowaveOscillatorFrequency2020,gou_chip-scale_2025}. These devices are useful high-frequency benchmarks: their large $G$ can still produce sizable modulation, but the high $\Omega_m$ increases the displacement required for a given modulation index, and the optical drive power becomes large unless compensated by high $g_0$ and high $Q_m$. The present platform instead targets the low-MHz, low-repetition-rate regime, where broad combs can be generated from much smaller optical powers.

Overall, the table shows that the present experiment is not merely an incidental observation of a few nonlinear sidebands. The measured hBN fiber-cavity parameters place the platform in a favorable region for low-repetition-rate optomechanical combs: the current device already shows comb generation, while realistic changes to mirror coating, membrane geometry, and mechanical-$Q$ engineering could increase the detectable span by one to two orders of magnitude. The distinctive advantage is the combination of open-cavity tunability, large $G$, comparatively large $g_0$, and unresolved-sideband operation. Other platforms can outperform one of these metrics, but usually by sacrificing another: long MIM cavities have small $G$, nanophotonic cavities have less reconfigurability and often broad effective linewidths, and several demonstrated comb systems rely on auxiliary nonlinearities or very high mechanical quality factors. This makes hBN fiber cavities a promising route toward broadband, low-power, mechanically spaced combs in a reconfigurable MIM-like architecture.

\begin{table}[!hb]
	\centering
	\begin{tabular}{l|r|c|c|c|c|c|c|c|c|c}
		\toprule
		System									&$\Omega_m/2\pi$	&$Q_m$			&$\kappa/2\pi$		&$G/2\pi$				&$g_0/2\pi$			&$\mathcal{M}_{\rm OM}$		&$N_{\rm full}$		&$x_{\mathcal{M}}$	&$x_{\rm full}$		&$P_{\rm req}$\\
												&					&				&					&						&					&$(\qty{1}{\nm})$			&$(\qty{1}{\nm})$	&(100)				&(100,50dB)			&(per tone)\\
		\midrule
		This work, max. $G$						&\qty{2}{\MHz}		&\num{5e3}		&\qty{1}{\GHz}		&\qty{11}{\GHz/\nm}		&\qty{180}{\kHz}	&5500						&127 				&\qty{18}{\pm}		&\qty{780}{\pm} 	&\qty{18}{\nW}\\
		This work, max. $G$, small $\kappa$		&\qty{2}{\MHz} 		&\num{5e3}		&\qty{100}{\MHz}	&\qty{11}{\GHz/\nm}		&\qty{180}{\kHz}	&5500						&1260 				&\qty{18}{\pm}		&\qty{79}{\pm} 		&\qty{0.18}{\nW}\\
		This work, small drum					&\qty{50}{\MHz}     &\num{5e3}		&\qty{100}{\MHz}	&\qty{10}{\GHz/\nm}		&\qty{240}{\kHz}	&202						&218				&\qty{500}{\pm}		&\qty{430}{\pm}		&\qty{15}{\nW}\\
		\midrule
		
		Rochau 2021 \cite{rochauDynamicalBackactionUltrahighFinesse2021}, max. $G$					&\qty{0.93}{\MHz}	&\num{2e5}		&\qty{62}{\MHz}		&\qty{1}{\GHz/\nm}		&\qty{5.8}{\kHz}	&1100						&186 				&\qty{93}{\pm}		&\qty{540}{\pm} 	&\qty{0.47}{\nW}\\
		
		Sanchez 2023 \cite{sanchezarribasRadiationPressureBackaction2023}, max. $G$					&\qty{1.6}{\MHz}	&\num{3e3}		&\qty{54}{\MHz}		&\qty{2.5}{\GHz/\nm}	&\qty{1.2}{\kHz}	&1600						&526				&\qty{64}{\pm}		&\qty{190}{\pm} 	&\qty{940}{\nW}\\
		
		Tenbrake 2024 \cite{tenbrakeDirectLaserwrittenOptomechanical2024}							&\qty{3}{\MHz}		&\num{6e2}		&\qty{3.6}{\GHz}	&\qty{11}{\GHz/\nm}		&\qty{33}{\kHz}		&3700						&32 				&\qty{27}{\pm}		&\qty{2.8}{\nm} 	&\qty{100}{\uW}\\
		
		Leijssen 2017 \cite{leijssenNonlinearCavityOptomechanics2017}, \qty{3}{\K}				&\qty{3.3}{\MHz}	&\num{3e4}		&\qty{23}{\GHz}		&\qty{800}{\GHz/\nm}	&\qty{25}{\MHz}		&240000						&401			 	&\qty{0.41}{\pm}	&\qty{250}{\pm} 	&\qty{0.08}{\nW}\\
		Leijssen 2017 \cite{leijssenNonlinearCavityOptomechanics2017}, \qty{300}{\K}			&\qty{3.3}{\MHz}	&\num{3e4}		&\qty{140}{\GHz}	&\qty{800}{\GHz/\nm}	&\qty{25}{\MHz}		&240000						&66					&\qty{0.41}{\pm}	&\qty{1.5}{\nm}		&\qty{2.9}{\nW}\\
		
		Guo 2022 \cite{guoIntegratedOpticalreadoutHighQ2022a}								&\qty{1}{\MHz}		&\num{1e7}		&\qty{10}{\GHz}		&\qty{20}{\GHz/\nm}		&\qty{260}{\kHz}	&20000						&24 				&\qty{5}{\pm}		&\qty{4.3}{\nm} 	&\qty{0.13}{\nW}\\
		\midrule
		Planz 2023 \cite{planzMembraneinthemiddleOptomechanicsSoftclamped2023}								&\qty{1.32}{\MHz}	&\num{1e9}		&\qty{2}{\MHz}		&\qty{50}{\MHz/\nm}		&\qty{1.2}{\Hz}		&38							&47					&\qty{2.6}{\nm}		&\qty{2.3}{\nm} 	&\qty{36}{\nW}\\
		\midrule
		Hu 2021 \cite{huGenerationOpticalFrequency2021}									&\qty{50.1}{\MHz}	&\num{5.5e3}	&\qty{60}{\MHz}		&\qty{6.35}{\GHz/\nm}   &\qty{470}{\Hz}		&127						&141				&\qty{790}{\pm}		&\qty{690}{\pm}		&\qty{660}{\uW}\\
		Mercad\'e 2020 \cite{mercadeMicrowaveOscillatorFrequency2020}							&\qty{3.9}{\GHz}  	&\num{2.4e3}	&\qty{39}{\GHz}		&\qty{245}{\GHz/\nm}	&\qty{660}{\kHz}	&63							&60					&\qty{1.6}{\nm}		&\qty{1.6}{\nm}		&\qty{11}{\mW}\\
		Gou 2025 \cite{gou_chip-scale_2025}								&\qty{1.65}{\GHz}	&\num{1.4e4}	&\qty{3.1}{\GHz}	&\qty{80.8}{\GHz/\nm}	&\qty{41.6}{\kHz}	&49							&59					&\qty{2.0}{\nm}		&\qty{1.8}{\nm}		&\qty{2.3}{\mW}\\
		\bottomrule
	\end{tabular}
	\caption{Comparison of representative optomechanical platforms for mechanically spaced comb generation. The estimates use $D_{\mathrm{dB}}=\qty{50}{\decibel}$. $\mathcal{M}_{\rm OM}$ and $N_{\rm full}$ are evaluated for a coherent displacement $x_0=\qty{1}{\nano\meter}$. $x_{\mathcal{M}}(100)$ is the ideal displacement required to reach modulation index $100$, while $x_{\rm full}(100,50dB)$ is the displacement required for the 100th positive sideband to remain above the specified dynamic range using the full cavity-weighted sideband envelope. $P_{\rm req}$ is the corresponding per-tone input power estimated from the linear resonant two-tone drive, using the larger of the ideal and visibility requirements.}
	\label{tab:SI_comb_platform_comparison}
\end{table}

\subsection{Supporting Data}
We now return to the experimental datasets supporting the main-text comb interpretation. The all-optical two-tone comb shown in the Letter displays the essential signature of optomechanical comb generation, namely a set of lines spaced by $\Omega_m$, but its span is smaller and its weaker harmonics are more visible than in the strongest piezo-driven spectra. The measurements below show that such spectra are qualitatively similar to mechanically driven combs obtained either at intermediate drive strength or under slightly off-resonant drive conditions, and that cleaner and denser comb states can be reached in both optical and mechanical configurations. 

\begin{figure}[!t]
	\includegraphics[width=1.0\linewidth]{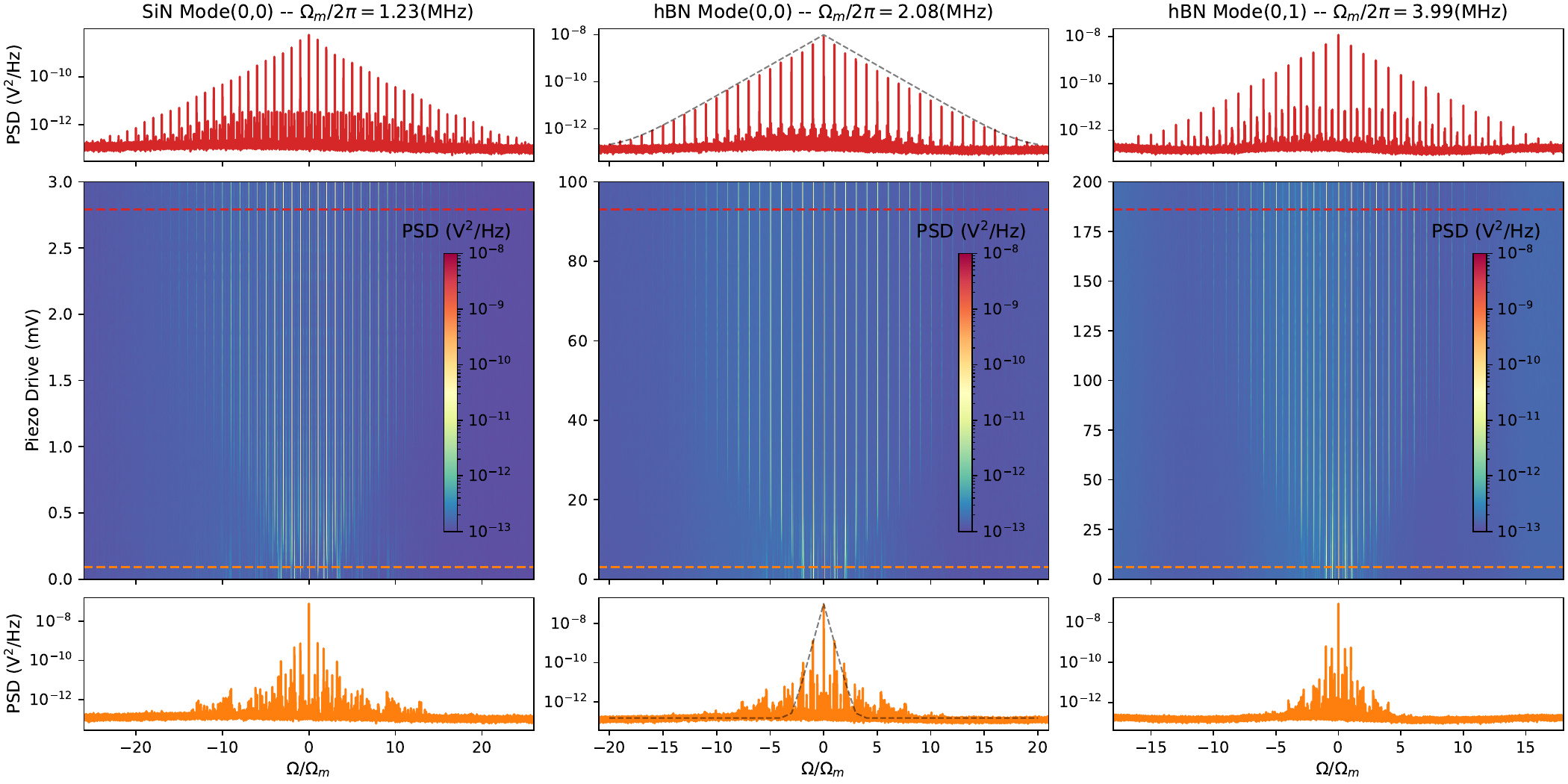}
	\caption{Mechanically driven frequency-comb generation for three different vibrational modes: the fundamental Si$_3$N$_4$ mode, the fundamental hBN mode, and a higher-order hBN mode. In each column, the central panel shows the	reflected power spectral density as a function of piezo-drive amplitude, while the spectra above and below correspond to the drive values marked by the dashed horizontal lines. For the fundamental hBN spectra, black dashed lines indicate the geometric envelope of \cref{eq:SI_rho_envelope}}
	\label{fig:FigS11_Freq_Comb_Modes}
\end{figure}
\Cref{fig:FigS11_Freq_Comb_Modes} shows mechanically driven comb generation obtained by applying a coherent piezoelectric drive to three different vibrational modes of the hybrid membrane system: the fundamental mode of the Si$_3$N$_4$ membrane, the fundamental mode of the hBN drum, and a higher-order hBN mode around \qty{4}{\mega\hertz}. In all three cases, the cavity is driven by a single optical tone while the mechanical mode is excited resonantly with the piezo actuator. The central panels show the reflected power spectral density as a function of piezo-drive amplitude, with the horizontal axis expressed in units of the corresponding mechanical frequency $\Omega_m$. The spectra shown above and below each map correspond to the drive values indicated by the dashed horizontal lines.

For the fundamental hBN mode, the spectra shown above and below the map also include black dashed guides corresponding to the geometric envelope of \cref{eq:SI_rho_envelope}. These guides are not meant to capture the exact height of every individual tone, but they reproduce the approximately triangular logarithmic envelope expected from adiabatic cavity transduction. This provides an experimental check of the visibility-limited picture introduced above: the ideal modulation index can be large, while the number of detected lines is set by the cavity-weighted envelope and the available dynamic range.

The common qualitative behavior across the three columns is that, at weak drive, the spectrum is dominated by the central pump and a small number of sidebands, whereas at stronger drive it evolves into a broad frequency comb with many harmonics spaced by $\Omega_m$. Between these two limits, an intermediate regime appears in which only a limited number of dominant comb lines are visible together with weaker additional harmonics. This intermediate regime is particularly relevant for interpreting the main-text all-optical two-tone comb, whose less regular structure more closely resembles this partially developed mechanically driven state than the cleanest high-drive piezo-driven spectra. The details of the comb span and threshold depend on the mode under consideration, but the same general transition is observed for all three cases. \Cref{fig:FigS11_Freq_Comb_Modes} therefore shows that the mechanically driven comb regime is not restricted to a specific mode and is instead a general nonlinear feature of the coupled cavity--membrane system.

\begin{figure}[!t]
	\includegraphics[width=1\linewidth]{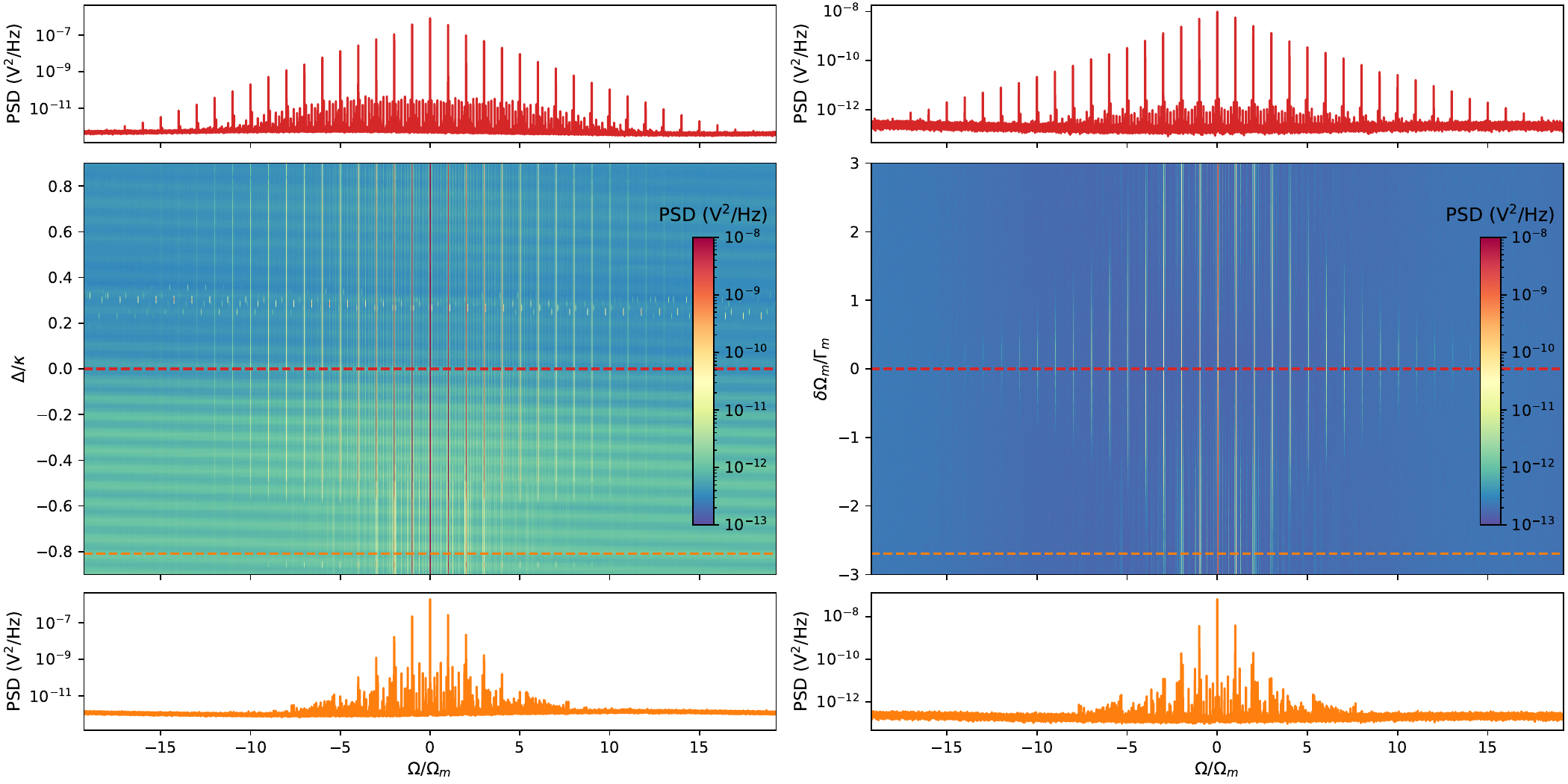}
	\caption{Left: stronger two-tone optical comb measurement as a function of cavity detuning, showing a sudden transition to a much denser and cleaner comb state at a particular detuning. Right: mechanically driven comb measurement in which the piezo-drive frequency is swept around the mechanical resonance. The comb is strongest near resonance and rapidly weakens as the drive is detuned by several mechanical linewidths.}
	\label{fig:FigS12_Freq_Comb_Det}
\end{figure}

\Cref{fig:FigS12_Freq_Comb_Det} provides two additional datasets that further support this interpretation. The left panel shows a stronger two-tone optical measurement of the same type as the all-optical comb data presented in the main text. Here the cavity is driven by two coherent optical tones whose frequency separation is matched to the independently measured mechanical resonance frequency. Over a small detuning range, the observed spectrum contains only a moderate number of comb lines. However, at a particular detuning, the spectrum changes abruptly and a much denser and cleaner comb state appears. This sharp transition is suggestive of entry into a more strongly synchronized or locking-enhanced regime, although the present data do not by themselves constitute a direct demonstration of phase locking.

The right panel of \cref{fig:FigS12_Freq_Comb_Det} returns to the mechanically driven configuration, but now the drive frequency is swept around the mechanical resonance while keeping the optical configuration fixed. The vertical axis is expressed as the drive-frequency detuning from the mechanical resonance in units of the mechanical linewidth $\Gamma_m$. The comb is strongest when the piezo drive is close to resonance and rapidly weakens as the drive is detuned by several mechanical linewidths. As the drive is moved slightly away from resonance, the spectrum again contains fewer dominant comb lines together with more visible weaker harmonics, resembling both the intermediate-drive regime of \cref{fig:FigS11_Freq_Comb_Modes} and the less developed all-optical two-tone comb shown in the main text. This behavior shows directly that, in the present measurements, efficient comb generation relies on resonant enhancement of the mechanical motion and provides strong support for a mechanically mediated origin of the comb spacing.

Taken together, \cref{fig:FigS11_Freq_Comb_Modes,fig:FigS12_Freq_Comb_Det} reinforce and clarify the interpretation of the main-text comb data. The observed comb spectra can be accessed both optically, through two-tone pumping, and mechanically, through direct resonant piezo actuation; they appear for several distinct vibrational modes; and their detailed structure depends strongly on the degree of resonant mechanical excitation. The main-text two-tone comb is therefore most naturally interpreted as a partially developed mechanically mediated comb state, likely limited by the effective drive strength and operating stability of the all-optical configuration, while the stronger optical dataset of \cref{fig:FigS12_Freq_Comb_Det} shows that significantly cleaner comb states can also be reached in that geometry. Overall, these observations are consistent with resonantly enhanced optomechanical frequency-comb generation in the hBN fiber-cavity platform.

An additional attractive perspective is that, in upcoming work, we show that the mechanical resonance frequency of hBN drum resonators can be tuned with temperature by up to about 300\% of its original value. In the present fiber-cavity platform, this could enable widely tunable optomechanical frequency combs with a comb spacing directly set by the mechanical resonance.

\newpage
\bibliographystyle{apsrev4-2}
\bibliography{OMIT_SI}